\documentclass[trackchanges,twocolumn]{aastex63}

\usepackage{graphicx}
\usepackage{txfonts}
\usepackage{xcolor}
\usepackage[caption=false]{subfig}
\usepackage{natbib}

\received{2020 October 27}
\revised{2020 December 1}
\accepted{2020 December 1}

\shorttitle{Stellar outbursts prior to interacting SNe}
\shortauthors{Strotjohann et al.}

\graphicspath{{./}{figs/}}

\begin{document}

\title{Bright, months-long stellar outbursts announce the explosion of interaction-powered supernovae}


\author[0000-0002-4667-6730]{Nora L. Strotjohann}
\affiliation{Benoziyo Center for Astrophysics, The Weizmann Institute of Science, Rehovot 76100, Israel}
\email{nora.linn.strotjohann@gmail.com}

\author{Eran O. Ofek}
\affiliation{Benoziyo Center for Astrophysics, The Weizmann Institute of Science, Rehovot 76100, Israel}

\author{Avishay Gal-Yam}
\affiliation{Benoziyo Center for Astrophysics, The Weizmann Institute of Science, Rehovot 76100, Israel}

\author{Rachel Bruch}
\affiliation{Benoziyo Center for Astrophysics, The Weizmann Institute of Science, Rehovot 76100, Israel}

\author{Steve Schulze}
\affiliation{Benoziyo Center for Astrophysics, The Weizmann Institute of Science, Rehovot 76100, Israel}

\author[0000-0003-3894-8422]{Nir Shaviv}
\affiliation{The Racah Institute of Physics, The Hebrew University of Jerusalem, Jerusalem 91904, Israel}

\author[0000-0003-1546-6615]{Jesper Sollerman}
\affiliation{Department of Astronomy, The Oskar Klein Centre, Stockholm University, AlbaNova, 10691 Stockholm, Sweden}

\author{Alexei V. Filippenko}
\affiliation{Department of Astronomy, University of California, Berkeley, CA 94720-3411, USA}
\affiliation{Miller Senior Fellow, Miller Institute for Basic Research in Science, University of California, Berkeley, CA 94720, USA}

\author{Ofer Yaron}
\affiliation{Benoziyo Center for Astrophysics, The Weizmann Institute of Science, Rehovot 76100, Israel}

\author[0000-0002-4223-103X]{Christoffer~Fremling}
\affiliation{Division of Physics, Mathematics and Astronomy, California Institute of Technology, Pasadena, CA 91125, USA}

\author{Jakob Nordin}
\affiliation{Institut fur Physik, Humboldt-Universitat zu Berlin, Newtonstr. 15, 12489, Berlin, Germany}

\author[0000-0002-7252-3877]{Erik C. Kool}
\affiliation{Department of Astronomy, The Oskar Klein Centre, Stockholm University, AlbaNova, 10691 Stockholm, Sweden}

\author[0000-0001-8472-1996]{Dan A. Perley}
\affiliation{Astrophysics Research Institute, Liverpool John Moores University, Liverpool Science Park, 146 Brownlow Hill, Liverpool L35RF, UK}

\author[0000-0002-9017-3567]{Anna Y. Q.~Ho}
\affiliation{Division of Physics, Mathematics and Astronomy, California Institute of Technology, Pasadena, CA 91125, USA}
\affiliation{Department of Astronomy, University of California, Berkeley, CA 94720-3411, USA}
\affiliation{Miller Institute for Basic Research in Science, University of California, Berkeley, CA 94720, USA}

\author{Yi Yang}
\affiliation{Benoziyo Center for Astrophysics, The Weizmann Institute of Science, Rehovot 76100, Israel}

\author[0000-0001-6747-8509]{Yuhan Yao}
\affiliation{Division of Physics, Mathematics and Astronomy, California Institute of Technology, Pasadena, CA 91125, USA}

\author{Maayane T. Soumagnac}
\affiliation{Department of Astronomy, University of California, Berkeley, CA 94720-3411, USA}

\author[0000-0002-9154-3136]{Melissa L. Graham}
\affiliation{DiRAC Institute, Department of Astronomy, University of Washington, 3910 15th Avenue NE, Seattle, WA 98195, USA}

\author[0000-0002-3821-6144]{Cristina~Barbarino}
\affiliation{Department of Astronomy, The Oskar Klein Centre, Stockholm University, AlbaNova, 10691 Stockholm, Sweden}

\author[0000-0003-3433-1492]{Leonardo~Tartaglia}
\affiliation{Department of Astronomy, The Oskar Klein Centre, Stockholm University, AlbaNova, 10691 Stockholm, Sweden}
\affiliation{INAF - Osservatorio Astronomico di Padova, Vicolo dell'Osservatorio 5, 35122 Padova, Italy}

\author[0000-0002-8989-0542]{Kishalay De}
\affil{Division of Physics, Mathematics and Astronomy, California Institute of Technology, Pasadena, CA 91125, USA}

\author{Daniel A. Goldstein}
\affiliation{Division of Physics, Mathematics and Astronomy, California Institute of Technology, Pasadena, CA 91125, USA}

\author[0000-0002-6877-7655]{David O. Cook}
\affiliation{IPAC, California Institute of Technology, 1200 E. California Blvd, Pasadena, CA 91125, USA}

\author[0000-0001-5955-2502]{Thomas G. Brink}
\affiliation{Department of Astronomy, University of California, Berkeley, CA 94720-3411, USA}

\author[0000-0002-5748-4558]{Kirsty Taggart}
\affiliation{Astrophysics Research Institute, Liverpool John Moores University, Liverpool Science Park, 146 Brownlow Hill, Liverpool L35RF, UK}

\author{Lin Yan}
\affiliation{Caltech Optical Observatories, California Institute of Technology, Pasadena, CA 91125, USA}

\author[0000-0001-9454-4639]{Ragnhild Lunnan}
\affiliation{Department of Astronomy, The Oskar Klein Centre, Stockholm University, AlbaNova, 10691 Stockholm, Sweden}

\author{Mansi Kasliwal}
\affiliation{Division of Physics, Mathematics and Astronomy, California Institute of Technology, Pasadena, CA 91125, USA}

\author{Shri R. Kulkarni}
\affiliation{Division of Physics, Mathematics and Astronomy, California Institute of Technology, Pasadena, CA 91125, USA}

\author{Peter E. Nugent}
\affiliation{Department of Astronomy, University of California, Berkeley, CA 94720-3411, USA}
\affiliation{Computational Science Department, Lawrence Berkeley National Laboratory, 1 Cyclotron Road, MS 50B-4206, Berkeley, CA 94720, USA}

\author[0000-0002-8532-9395]{Frank J. Masci}
\affiliation{IPAC, California Institute of Technology, 1200 E. California Blvd, Pasadena, CA 91125, USA}

\author[0000-0002-6099-7565]{Philippe Rosnet}
\affiliation{Universit\'e Clermont Auvergne, CNRS/IN2P3, LPC, Clermont-Ferrand, France}

\author{Scott M. Adams}
\affiliation{Division of Physics, Mathematics and Astronomy, California Institute of Technology, Pasadena, CA 91125, USA}

\author{Igor Andreoni}
\affiliation{Division of Physics, Mathematics and Astronomy, California Institute of Technology, Pasadena, CA 91125, USA}

\author{Ashot Bagdasaryan}
\affiliation{Division of Physics, Mathematics and Astronomy, California Institute of Technology, Pasadena, CA 91125, USA}

\author[0000-0001-8018-5348]{Eric C. Bellm}
\affiliation{DiRAC Institute, Department of Astronomy, University of Washington, 3910 15th Avenue NE, Seattle, WA 98195, USA}

\author{Kevin Burdge}
\affiliation{Division of Physics, Mathematics and Astronomy, California Institute of Technology, Pasadena, CA 91125, USA}

\author[0000-0001-5060-8733]{Dmitry A. Duev}
\affiliation{Division of Physics, Mathematics and Astronomy, California Institute of Technology, Pasadena, CA 91125, USA}

\author{Alison Dugas}
\affiliation{Division of Physics, Mathematics and Astronomy, California Institute of Technology, Pasadena, CA 91125, USA}
\affiliation{Institute for Astronomy, University of Hawai'i, 2680 Woodlawn Drive, Honolulu, HI 96822, USA}

\author[0000-0001-9676-730X]{Sara Frederick}
\affiliation{Department of Astronomy, University of Maryland, College Park, MD 20742, USA}

\author{Samantha Goldwasser}	
\affiliation{Benoziyo Center for Astrophysics, The Weizmann Institute of Science, Rehovot 76100, Israel}

\author{Matthew Hankins}
\affiliation{Division of Physics, Mathematics and Astronomy, California Institute of Technology, Pasadena, CA 91125, USA}

\author{Ido Irani}
\affiliation{Benoziyo Center for Astrophysics, The Weizmann Institute of Science, Rehovot 76100, Israel}

\author{Viraj Karambelkar}
\affiliation{Division of Physics, Mathematics and Astronomy, California Institute of Technology, Pasadena, CA 91125, USA}

\author[0000-0002-6540-1484]{Thomas Kupfer}
\affiliation{Texas Tech University, Department of Physics \& Astronomy, Box 41051, 79409, Lubbock, TX, USA}

\author{Jingyi Liang}
\affiliation{Benoziyo Center for Astrophysics, The Weizmann Institute of Science, Rehovot 76100, Israel}

\author[0000-0002-0466-1119]{James D. Neill}
\affiliation{Division of Physics, Mathematics and Astronomy, California Institute of Technology, Pasadena, CA 91125, USA}

\author{Michael Porter}
\affiliation{Caltech Optical Observatories, California Institute of Technology, Pasadena, CA 91125, USA}

\author[0000-0002-0387-370X]{Reed L. Riddle}	
\affiliation{Caltech Optical Observatories, California Institute of Technology, 1200 E. California Blvd, Pasadena, CA 91125, USA}

\author{Yashvi Sharma}
\affiliation{Division of Physics, Mathematics and Astronomy, California Institute of Technology, Pasadena, CA 91125, USA}

\author[0000-0002-5096-9464]{Phil Short}
\affiliation{Institute for Astronomy, University of Edinburgh, Royal Observatory, Blackford Hill, Edinburgh EH9 3HJ, UK}

\author{Francesco Taddia}
\affiliation{Department of Astronomy, The Oskar Klein Centre, Stockholm University, AlbaNova, 10691 Stockholm, Sweden}

\author[0000-0003-0484-3331]{Anastasios Tzanidakis}
\affiliation{Division of Physics, Mathematics and Astronomy, California Institute of Technology, Pasadena, CA 91125, USA}

\author[0000-0002-2626-2872]{Jan van~Roestel}
\affiliation{Division of Physics, Mathematics and Astronomy, California Institute of Technology, Pasadena, CA 91125, USA}

\author{Richard Walters}
\affiliation{Division of Physics, Mathematics and Astronomy, California Institute of Technology, Pasadena, CA 91125, USA}

\author[0000-0002-1945-2299]{Zhuyun Zhuang}
\affiliation{Division of Physics, Mathematics and Astronomy, California Institute of Technology, Pasadena, CA 91125, USA}



\begin{abstract}
Interaction-powered supernovae (SNe) explode within an optically-thick circumstellar medium (CSM) that could be ejected during eruptive events. To identify and characterize such pre-explosion outbursts we produce forced-photometry light curves for 196 interacting SNe, mostly of Type IIn, detected by the Zwicky Transient Facility between early 2018 and June 2020. Extensive tests demonstrate that we only expect a few false detections among the $70,000$ analyzed pre-explosion images after applying quality cuts and bias corrections.
We detect precursor eruptions prior to 18 Type IIn SNe and prior to the Type Ibn SN\,2019uo. Precursors become brighter and more frequent in the last months before the SN and month-long outbursts brighter than magnitude $-13$ occur prior to 25\% (5--69\%, 95\% confidence range) of all Type IIn SNe within the final three months before the explosion. With radiative energies of up to $10^{49}\,\text{erg}$, precursors could eject $\sim1\,\text{M}_\odot$ of material. Nevertheless, SNe with detected precursors are not significantly more luminous than other SNe IIn and the characteristic narrow hydrogen lines in their spectra typically originate from earlier, undetected mass-loss events. The long precursor durations require ongoing energy injection and they could, for example, be powered by interaction or by a continuum-driven wind. Instabilities during the neon and oxygen burning phases are predicted to launch precursors in the final years to months before the explosion; however, the brightest precursor is 100 times more energetic than anticipated.
\end{abstract}

\section{Introduction}
\label{sec:intro}


Despite the detection of more than $2000$ core-collapse supernovae (SNe) per year, the processes leading to their explosions are still not entirely understood (see, e.g., \citealt{janka2016, mueller2016, glas2019}) and remain unobservable as they happen deep within the cores of stars in distant galaxies.
However, at least for some progenitor stars, the impending core collapse seems to have direct implications for the stellar envelope. Bright optical flares have been observed in the years leading up to the SN explosion and may offer another means of probing the conditions near the surface of progenitor stars, which are, with exception of the nearest events, too faint to be detected by any telescope.

The first pre-explosion outburst was detected two years prior to the explosion of the Type Ibn SN\,2006jc \citep{pastorello2007, foley2007}. Most precursors were observed prior to Type IIn SNe (see, e.g., \citealt{ofek2013, mauerhan2013, fraser2013, margutti2013, tartaglia2016b, elias-rosa2016, ofek2016, thoene2017, nyholm2017, pastorello2018, reguitti2019}), as well as prior to a broad-lined Type Ic SN \citep{ho2019}, and possibly a SN IIb \citep{strotjohann2015}. This suggests that numerous types of progenitor stars can produce such flares. However, since most flares have been detected prior to Type IIn SNe, the progenitors of these relatively rare explosions are either more likely to generate such flares, or the generated flares are brighter. A systematic study by \citet{ofek2014} showed that precursor eruptions prior to Type IIn SNe are the rule rather than the exception. In a similar search, \citet{bilinski2015} did not find any precursors and claim that the rate is lower, but they relied on a small SN sample.

Type IIn SNe are characterized by relatively narrow hydrogen emission lines (see, e.g., \citealt{filippenko1997,gal-yam2017, smith2017}), which indicate the presence of a slowly moving circumstellar medium (CSM) surrounding the SN ejecta. This material originates from the star itself and is expelled in the years to decades before the explosion either during precursor eruptions or by a stellar wind. The SN ejecta crash into the CSM and a fraction of the ejecta kinetic energy is converted to high-energy photons (see, e.g., \citealt{katz2011, murase2011, murase2014}). If the CSM is optically thick to gamma-ray and X-ray photons, a part of the radiation may be converted to the UV-optical regime. Some Type IIn SNe therefore reach much brighter optical peak magnitudes than noninteracting SNe \citep{kiewe2012, stritzinger2012, gal-yam2019} and their diverse light-curve shapes can be explained by different CSM geometries, which might consist of several shells (see, e.g., ~\citealt{margutti2013, nyholm2017}) or be aspherical \citep{patat2011, soumagnac2019, soumagnac2020}, as is  commonly observed for planetary nebulae in our Galaxy.

The progenitor stars of a few nearby Type IIn SNe were identified in archival images and are consistent with being luminous blue variables (LBVs; see, e.g., \citealt{gal-yam2007, gal-yam2009, foley2011, kochanek2011}). These bright and massive stars are named after their hot surface temperatures and their high-amplitude luminosity variability. They launch strong winds, which remove part of their hydrogen envelope. They were therefore traditionally considered stars in a transitional phase which evolve from a main-sequence star into a hydrogen-stripped Wolf-Rayet star \citep{humphreys1994}. Another possibility is that they develop from a main-sequence star that gains mass and angular momentum from a binary companion, which turns it into an LBV rather than a red supergiant (\citealt{smith2017b} see also \citealt{justham2014}). However, there is also a class of lower-luminosity transients with Type-IIn-like spectra that originate from heavily obscured stars with masses of $\sim10\,M_{\odot}$ (see e.g. \citealt{prieto2008, kochanek2011b, szcygiel2012}). It is hence possible that not all Type IIn SNe originate from massive LBVs.

In addition to Type IIn SNe, evidence for interaction has also been observed for several other SN classes. Type Ibn SNe explode within a helium-rich CSM and their rapid light-curve evolution might indicate that the CSM is confined to a small radius (see, e.g., \citealt{pastorello2016, gal-yam2017,hosseinzadeh2017}). The spectra of Type II superluminous supernovae (SLSNe-II) often look similar to the ones of Type IIn SNe and their large radiative energy is usually attributed to strong CSM interaction (see, e.g.,~\citealt{gal-yam2019}). So called flash-spectroscopy SNe exhibit narrow emission features during for the first few days after their explosion; these could originate from a confined CSM shell that is flash ionized by radiation from the shock breakout and is the quickly swept up by the expanding ejecta (see, e.g., \citealt{gal-yam2014, khazov2016, yaron2017, smith2017,bruch2020}). Type Ia-CSM SNe are thermonuclear explosions of white dwarfs that explode inside a hydrogen-rich CSM, potentially produced by a binary companion star \citep{hamuy2003, dilday2012, silverman2013, gal-yam2017}. In the following, we use the expression ``interaction-powered SNe'' to refer to all these subclasses.


A first systematic search for precursor eruptions was done by \citet{ofek2014} for a sample of 16 nearby Type IIn SNe using data from the Palomar Transient Factory (PTF; \citealt{law2009, rau2009}). It established that most Type IIn progenitor stars undergo one or several precursor eruptions in the last 2.5\,yr before the SN and that the rate increases in the last 4 months before the explosion. However, the study was limited by the small SN sample and by the relatively sparse sampling of the pre-explosion light curves. The majority of the observations were obtained in the Mould-$R$ band, such that the precursor colors could not be determined.

Here, we build on the work by \citet{ofek2014} and use data from the Zwicky Transient Facility (ZTF; \citealt{bellm2019, graham2019}) to systematically search for precursor eruptions prior to interacting SNe, mostly of Type IIn SNe. Compared to PTF, ZTF has a $\sim\!15$ times faster survey speed: with its large field of view of $47\,\text{deg}^2$, it monitors nearly the complete sky at declinations larger than $-30^\circ$ and smaller than $80^\circ$ \citep{bellm2019}. Since the commissioning of the ZTF camera in fall 2017, the survey has detected more than 200 interacting SNe for which nearly $10^5$ pre-explosion images are available in the $g$, $r$, and $i$ bands. We here search unbinned and binned light curves for pre-explosion activity. 
Owing to the abundant photometric data provided by the ZTF survey and the larger SN sample, we expect to detect more precursor eruptions and measure the precursor rate more precisely. Thus, we extend the previous search to fainter, shorter, and less-common precursors and expect that the eruptions are better observed with data in multiple bands.

The paper is structured as follows. Section~\ref{sec:fp} describes the analysis and quality cuts which allow us to reduce the rate of false-positive detections. The detected precursors are described in Sec.~\ref{sec:precursors} and the luminosity-dependent precursor rates are measured in Sec.~\ref{sec:rates}. In Sec.~\ref{sec:sne}, we show that the material ejected during most of the detected precursors cannot account for the characteristic narrow hydrogen lines in the spectra of Type IIn SNe. One exception is the Type Ibn SN\,2019uo, described Sec.~\ref{sec:sne_sn2019uo}, for which the observed interaction can be explained by the precursor 320 days before the explosion. In Sec.~\ref{sec:nature} we consider which mechanisms might power the precursor luminosity and whether wave-driven mass loss could launch the observed precursors. Our findings are summarized in Sec.~\ref{sec:conclusion}.

\section{Methods}
\label{sec:fp}

\begin{deluxetable*}{l l l c c c c c l}
\tablecaption{SNe with detected pre-explosion activity \label{tab:sample}}
\tablewidth{0pt}
\tablehead{
\colhead{IAU name} & \colhead{ZTF name} & \colhead{SN Type} & \colhead{R.A. (J2000)} & \colhead{Dec. (J2000)} & \colhead{$z$} & \colhead{$t_0$} & \colhead{Separation} & \colhead{Comment}\\    
\colhead{} & \colhead{} & \colhead{} & \colhead{(deg)} & \colhead{(deg)} & \colhead{} & \colhead{(JD)} & \colhead{(arcsec)} & }
\startdata
SN\,2018eru & ZTF\,18ablqehq & IIn & $ 185.115828 $ & $ 41.79289029 $ & $ 0.03069 $ & $ 2458316.6 $ & $ 11 $& \\
SN\,2018gho & ZTF\,18abucxcj & IIn & $ 246.8412533 $ & $ 39.1091986 $ & $ 0.033 $ & $ 2458366.4 $ & $ 2.8 $& \\
SN\,2018hxe & ZTF\,18abwlupf & IIn & $ 221.0426466 $ & $ 62.89518 $ & $ 0.134 $ & $ 2458370.7 $ & $ 0.37 $& \\
SN\,2018kag & ZTF\,18acwzyor & IIn & $ 133.951979 $ & $ 3.64152020 $ & $ 0.02736 $ & $ 2458466.5 $ & $ 6.3 $& \\
SN\,2019uo & ZTF\,19aadnxbh & Ibn & $ 180.6525136 $ & $ 41.0616364 $ & $ 0.020454 $ & $ 2458501.2 $ & $ 27 $& \\
SN\,2019bxq & ZTF\,19aamkmxv & IIn & $ 254.4938 $ & $ 78.6037 $ & $ 0.0139 $ & $ 2458555.8 $ & $ 0.01 $& \\
SN\,2019cmy & ZTF\,19aanpcep & IIn & $ 227.2118369 $ & $ 40.7137261 $ & $ 0.0314 $ & $ 2458567.9 $ & $ 5.8 $& \\
SN\,2019iay & ZTF\,19abandzh & IIn & $ 200.27061809 $ & $ 8.1684897 $ & $ 0.0406 $ & $ 2458656.6 $ & $ 7.9 $& \\
SN\,2019meh & ZTF\,19abclykm & SLSN-II & $ 321.8227253 $ & $ 64.4164373 $ & $ 0.0935 $ & $ 2458657.3 $ & $ 0.04 $& bg AGN $^{a}$\\
SN\,2019gjs & ZTF\,19abiszoe & IIn & $ 224.7381972 $ & $ 20.0529308 $ & $ 0.043 $ & $ 2458690.7 $ & $ 10 $& \\
SN\,2019mom & ZTF\,19ablojrw & IIn & $ 28.9021955 $ & $ 53.5918978 $ & $ 0.0488 $ & $ 2458690.9 $ & $ - $& \\
SN\,2019njv & ZTF\,19abpidqn & IIn & $ 304.98829689 $ & $ 15.37745280 $ & $ 0.01458 $ & $ 2458706.9 $ & $ 2.9 $& \\
SN\,2019fmb & ZTF\,19aavyvbn & IIn & $ 186.68196 $ & $ 56.0757834 $ & $ 0.016 $ & $ 2458715.8 $ & $ 17 $& $t_0$ uncertain \\
SN\,2019sae & ZTF\,19acahbxd & IIn & $ 41.2693283 $ & $ 26.0714348 $ & $ 0.048 $ & $ 2458728.8 $ & $ 9.9 $& \\
SN\,2019aafe & ZTF\,19abzfxel & IIn & $ 349.0686877 $ & $ 48.4284178 $ & $ 0.075 $ & $ 2458740.8 $ & $ - $& \\
SN\,2019vkl & ZTF\,19acukucu & IIn & $ 29.1283206 $ & $ 18.4399406 $ & $ 0.064 $ & $ 2458808.6 $ & $ - $& \\
SN\,2019vts & ZTF\,19acxmnkc & IIn & $ 98.7642676 $ & $ 50.434783 $ & $ 0.0395 $ & $ 2458816.9 $ & $ 24 $& \\
SN\,2019qny & ZTF\,19adannbl & IIn & $ 53.24978869 $ & $ -2.778198 $ & $ 0.048 $ & $ 2458827.0 $ & $ 5.7 $& \\
SN\,2020iq & ZTF\,20aabcemq & IIn & $ 43.8321616 $ & $ -11.4134991 $ & $ 0.096 $ & $ 2458832.6 $ & $ 2.5 $& \\
SN\,2019yzx & ZTF\,19adcbxkw & Ia-CSM & $ 142.6721432 $ & $ 21.4558832 $ & $ 0.057 $ & $ 2458840.0 $ & $ 2.0 $& \\
SN\,2019zrk & ZTF\,20aacbyec & IIn & $ 174.9475073 $ & $ 19.9296524 $ & $ 0.0362 $ & $ 2458889.0 $ & $ 14 $& \\
SN\,2020dcs & ZTF\,20aaocqkr & IIn & $ 183.3561586 $ & $ 37.6993902 $ & $ 0.023958 $ & $ 2458894.9 $ & $ 3.0 $& \\
SN\,2020dfh & ZTF\,20aasivpe & IIn & $ 265.506694 $ & $ 3.2008709 $ & $ 0.0293 $ & $ 2458903.1 $ & $ 6.3 $& \\
SN\,2020edh & ZTF\,20aaswzdm & IIn & $ 259.0999066 $ & $ 40.8081331 $ & $ 0.033 $ & $ 2458914.9 $ & $ 3.4 $& 
\enddata
\tablecomments{The R.A. and Dec. values represent the median coordinates of at least 10 ZTF detections. The discovery time $t_0$ is either the first detection time announced on TNS or a smaller value if the transient flux is visible earlier in ZTF data. The penultimate column lists the separation from the center of the host galaxy, to judge whether active galactic nucleus (AGN) activity might contribute to the pre-explosion variability. Here, we only list SNe for which pre-explosion activity is detected (see Sec.~\ref{sec:precursors}). The full table, containing all 227 considered SNe described in Sec.~\ref{sec:fp_sample}, is available online.\\
$^a$The detected variability likely originates from AGN activity in the center of the host galaxy and not from the progenitor star (see Sec.~\ref{sec:fp_tests} and~\ref{sec:pre_detections} for details).}
\end{deluxetable*}


The following subsections introduce the sample selection (Sec.~\ref{sec:fp_sample}), the forced photometry pipeline (Sec.~\ref{sec:fp_pipeline}), and the tests we perform on the pipeline (Sec.~\ref{sec:fp_bgsample}). Next, we explain how images with astrometric errors are rejected (Sec.~\ref{sec:fp_astrometry}) and how we correct the baseline offsets and rescale underestimated error bars (Sec.~\ref{sec:fp_rescaling}). Finally, we describe how observations are combined in bins (Sec.~\ref{sec:fp_binned}) and estimate the expected number of false detections in Sec.~\ref{sec:fp_tests}.

\subsection{Sample Selection}
\label{sec:fp_sample}

The ZTF survey produces about 1 million alerts per night \citep{patterson2019} which are then scored by a deep-learning algorithm to identify genuine astrophysical transients \citep{duev2019}. The resulting alert stream is filtered either by the AMPEL broker \citep{nordin2019, soumagnac2018} or the GROWTH ``Marshal'' \citep{kasliwal2019} based on different science goals, such as the detection of young SNe \citep{gal-yam2019b, bruch2020} or bright transients \citep{fremling2020}. In most science programs potentially interesting objects are identified by astronomers who request spectroscopy or other follow-up observations. Transients brighter than magnitude $\sim19$ are usually first classified based on spectra from the SED Machine \citep{ben-ami2012, blagorodnova2018, rigault2019} and higher-resolution spectra might be obtained later.

The commissioning phase of the ZTF survey started in fall 2017, while the survey officially began in spring 2018 after commissioning and building reference images. To select a sample of interaction-powered SNe with ZTF pre-explosion observations, we query both the Transient Name Server (TNS\footnote{\url{https://wis-tns.weizmann.ac.il/}}) and the private ZTF database, the GROWTH Marshal using the ZTFquery code \citep{rigault2018}, for transients discovered since 2018 January 1 and until 2020 June 24. We only consider SNe at locations that are observable by ZTF, with declinations larger than $-30^\circ$. Our sample includes all objects that are classified as SNe of Type IIn, Ibn, Ia-CSM, or SLSNe-II by members of the ZTF team or on TNS (see, e.g., \citealt{perley2020} for details). In addition, we include objects that show flash-spectroscopy features in early-time spectra, which were identified by \citet{bruch2020}. This brings the total sample to 239 SNe.

An accurate localization is required to perform forced photometry (see, e.g.,~\citealt{yao2019}), and we therefore only consider objects with at least ten ZTF detections. We find that this ensures that the position is within $0.15''$ of the best position for 90\% of the SNe in the ZTF coordinate system\footnote{A precision of $\lesssim0.15''$ is the required threshold for forced photometry (Frank Masci, priv. comm.).}. Out of 239 SNe, 12 objects have fewer than ten ZTF detections and are discarded. The remaining 227 SNe are listed in the online version of Table~\ref{tab:sample}.

To confirm both the SN classification and the redshift, we visually inspect spectra from the ZTF Marshal as well as the TNS. We discard in total 18 objects which we cannot verify are interacting transients. For most of these objects no good spectra are available or the observed narrow lines might originate from the host galaxy. For objects that are classified as SLSNe-II, we check whether they surpass a peak magnitude of $-21$ in any band. SNe with fainter peak magnitudes are here considered regular Type IIn SNe. 

Forced photometry is obtained for all 209 remaining SNe and we apply the quality cuts as described in the following sections. After all cuts, pre-explosion observations are available for 196 SNe.
This remaining sample consists of 131 Type IIn SNe, 26 SLSNe-II, 20 SNe with flash-spectroscopy signatures, 12 Type Ibn SNe, and 7 SNe~Ia-CSM. Table~\ref{tab:sample} lists the SNe for which pre-explosion activity is detected (see Sec.~\ref{sec:precursors}) and a full version of this table containing all initially considered 227 SNe is available online.

\subsection{The Forced Photometry Pipeline}
\label{sec:fp_pipeline}

We perform forced photometry using the pipeline described by \citet{yao2019} on difference images obtained from IPAC via IRSA\footnote{\url{https://irsa.ipac.caltech.edu/Missions/ztf.html}}. Details of the ZTF image reduction are given by \citet{masci2019} and image subtraction is based on the method developed by \citet{zackay2016}.
We have access to images from the ZTF partnership survey (40\% of the observation time) and images that became available during the third data release\footnote{\url{https://www.ztf.caltech.edu/page/dr3}}, which includes images from the public survey (also 40\% of the time) until December 2019 and Caltech data (20\% of the time) until December 2018. Forced photometry on more recent public or Caltech data cannot be done as the full images are not yet available.

The forced-photometry pipeline was implemented by \citet{yao2019}. It relies on the IPAC difference images and the measured point-spread functions (PSFs). An image cutout around the SN position is produced and the background is measured within an annulus with an inner radius of 10 pixels and an outer radius of 15 pixels, where 1 pixel corresponds of $1.01''$ on the sky. The median background flux is subtracted from the cutout and the $7\times7$ pixels around the SN position are used for the PSF fit. To quantify the uncertainty in the flux, the normalization of the PSF is fitted with a Markov chain Monte Carlo algorithm. While \citet{yao2019} used 250 random walkers for the fit, we lower the number to 50 walkers to reduce the computation time. For 50 walkers the fitting algorithm introduces an uncertainty that is smaller than $2\%$ of the typical error in the measured flux. We hence find that 50 walkers provide sufficient accuracy.

Based on the procedure of \citet{yao2019} as well as our own findings, we exclude some data points from the light curves. Our exclusion criteria are as follows.
\begin{enumerate}
\item Images obtained early in the survey with an unknown quadrant ID for which the reference image cannot be identified.
\item Flagged difference images which might suffer from issues during the image subtraction.
\item Observations with seeing $>4''$. The PSF fit is only done on the inner $7\times7$ pixels and might not be accurate for a very broad PSF.
\item Images affected by bad pixels at the SN position (inner $7 \times 7$ pixels).
\item Early $g$-band observations obtained between JD 2458120 and 2458140, which are not well calibrated. 
\item Difference images with a background standard deviation $>25$ in units of detector data number (see~\citealt{yao2019}) which indicate problems during the image subtraction.
\item Data points with flux errors that are seven times larger than the median flux error for this SN to remove images for which the PSF fit did not converge.
\end{enumerate}
These initial quality cuts remove $\sim10\%$ of the data (see also Sec.~\ref{tab:cuts}). We are left with 85,333 pre-explosion data points which are listed in Table~\ref{tab:fp_fluxes}. 
All fluxes are corrected for Milky Way extinction using the python package \emph{sfdmap}, which is based on the dust map of \citet{schlegel1998} recalibrated to the values of \citet{schlafly2011} and the \citet{cardelli1989} extinction law.

\begin{deluxetable*}{l l c c c c c c c c}
\tablecaption{Forced photometry pre-explosion light curves}
\label{tab:fp_fluxes}
\tablewidth{0pt}
\tablehead{
\colhead{SN name} & \colhead{ZTF name} & \colhead{JD} & \colhead{band} & \colhead{ref. im.} & \colhead{flux} & \colhead{flux err.} & \colhead{sys. err} & \colhead{red. $\chi^2$} & \colhead{red. $\chi^2_{\text{star}}$} \\
\colhead{} & \colhead{} & \colhead{} & \colhead{} & \colhead{} & \colhead{$10^{-10}$} & \colhead{$10^{-10}$} & \colhead{$10^{-10}$} & \colhead{} & \colhead{}
}
\startdata
SN\,2018atq & ZTF\,18aahmhxu & 2458076.93147 & $r$ & 5751232 & $-0.0448$ & $8.38$ & $1.26$ & 0.56 & 0.85 \\
SN\,2018atq & ZTF\,18aahmhxu & 2458079.03350 & $r$ & 5751232 & $6.49$ & $11.7$ & $1.26$ & 0.96 & 1.33 \\
SN\,2018atq & ZTF\,18aahmhxu & 2458089.03796 & $r$ & 5751232 & $6.66$ & $8.41$ & $1.26$ & 1.15 & 1.30 \\
SN\,2018atq & ZTF\,18aahmhxu & 2458091.02603 & $r$ & 5751232 & $-39.4$ & $24.4$ & $1.26$ & 0.44 & 1.25 \\
SN\,2018atq & ZTF\,18aahmhxu & 2458091.04729 & $r$ & 5751232 & $11.5$ & $29.3$ & $1.26$ & 0.87 & 1.32
\enddata
\tablecomments{The fifth column specifies which reference image was used (e.g. for the first rows the image for the ZTF field 575, CCD 12, quadrant 3 and filter 2, the $r$ band; see also \citealt{yao2019}). All fluxes have been corrected for the zeropoint and are given as a dimensionless ratio (see Eq.~8 in \citealt{yao2019}). This flux ratio is also known as "maggie" \citep{finkbeiner2004}. The third to last column lists the noise level in the reference image which is a systematic error on the measured flux. The two last columns show the reduced $\chi^2$ of the PSF fit at the SN location as well at the location of a nearby faint star (see Sect.~\ref{sec:fp_astrometry}). The full version of the table is available online.}
\end{deluxetable*}

\subsection{Background Samples}
\label{sec:fp_bgsample}

We quantify the expected rate of false detections by performing forced photometry in locations where no precursors are expected. The four background samples are
\begin{enumerate}
\item empty positions in the sky close to the SN position, but outside of the host galaxy;
\item faint \emph{Gaia} stars with $g$-band magnitudes between $20.5$ and $18.5$ close to the SN position, to identify misaligned images;
\item the SN position mirrored across the center of its host galaxy; and
\item the positions of Type IIn SNe discovered during the PTF survey before 2015.
\end{enumerate}
The tests are designed such that they start from a case for which image subtraction is easy (an empty position in the image) and progress to increasingly more realistic, but challenging environments for our pipeline.
The first three tests are done for the exact same images that also contain the SN positions; hence, they have the same observing conditions, reference images, and subtractions. The two last tests are considered the most realistic ones as they are performed in host galaxies or locations where Type IIn SNe explode. The second background sample is used to identify and exclude images with astrometric errors.

The positions for the background samples are generated as follows. For empty locations we randomly pick several locations at a distance of 50 pixels (i.e., $50.6''$) from the SN position. Faint stars or unresolved galaxies are selected from the \emph{Gaia} catalog. To reject extended sources we require an astrometric excess noise of less than $1\,\text{milliarcsec}$ and the $g$-band magnitude is limited to values between $20.5$ to $18.5$ to ensure that the luminosity is similar to that of a faint precursor. Moreover, the separation from the SN position is required to be at least 20 pixels, such that the SN light does not fall within the annulus region for which the background level is calculated (see Sec.~\ref{sec:fp_pipeline}). To identify the SN host galaxies, we query the NED database for objects close to the SN position. We reject those identified as stars, the SN itself, and infrared sources, many of which are also stars \citep{cutri2013}.

The selected empty locations, faint stars, and host-galaxy candidates are then displayed on top of the reference image for visual inspection. When selecting empty positions and stars, we check that they are isolated, located outside of the host galaxy, and are not affected by artefacts in the reference images, such as dead columns, stellar spikes, or the edge of the image. Among the host-galaxy candidates we select the most likely host. For most images, a known galaxy is consistent with the visible center of the host in the reference image, but for a few objects we select a UV source. If several NED sources are close to the center of the host we compare with multicolor SDSS images to identify the most likely center. We caution that we might not identify the true host center in all cases. These positions are primarily used to build a background sample, so we do not require a high accuracy. With this method we locate the presumable centers of 160 host galaxies. The hosts of the remaining SNe are not listed in the NED database, mostly because they are faint. Some of them are even undetected in the ZTF reference images, especially for SLSNe.

The SN position is then mirrored on the location of the identified host galaxy and we verify that the two positions are sufficiently separated. The PSF fit is done for the inner $7 \times 7$ pixels --- that is, the pixel containing the SN position and the three neighboring ones. However, if the seeing disk is large, the PSF of the SN could be broader. We therefore require a separation of at least 10 pixels between the actual and mirrored positions. Only 59 out of 160 SNe with identified host galaxies show a sufficiently large separation (see also Table~\ref{tab:sample}). To increase the sample size, we select in addition SNe of Type IIn that were discovered during the PTF survey. We query the TNS database for publicly available SNe detected prior to 2015. A slowly developing Type IIn SN might still be detectable after $\sim3$\,yr, but an inspection of the ZTF light curves shows that this is not the case for any of the selected objects. Moreover, we add six objects analyzed by \citet{ofek2014} for which the SN was not observed by PTF. This brings the sample to a total of 104 objects out of which ZTF data are available for 100.

We produce forced photometry light curves for all selected positions to test the pipeline. The sample of \emph{Gaia} stars is used in Sec.~\ref{sec:fp_astrometry} to reject misaligned images with astrometric residuals produced during the image subtraction. The other samples are used in Sec.~\ref{sec:fp_rescaling} to inspect the data quality and in Sec.~\ref{sec:fp_tests} to estimate the rate of false-positive detections. Table~\ref{tab:cuts} shows the impact of the derived cuts and corrections on the number of (false) detections and on the total number of data points. The sample of \emph{Gaia} stars is omitted in the table, because variable stars may result in actual detections.

\begin{deluxetable*}{c l c c c c}
\tablecaption{Cuts on data quality\label{tab:cuts}}
\tablewidth{0pt}
\tablehead{
\colhead{} & \colhead{step} & \multicolumn4c{$\#$ precursors / $\#$ data points} \\
\colhead{} & \colhead{} & \colhead{empty pos.} & \colhead{mirrored pos.} & \colhead{PTF SNe} & \colhead{real data}
}
\startdata
0	&	before cuts	&	4 / 176815	&	3 / 45092	&	116 / 48250	&	415 / 95442	\\
1	&	known reference image	&	4 / 175888	&	3 / 44491	&	116 / 48067	&	415 / 94515	\\
2	&	difference image not flagged	&	4 / 169522	&	3 / 43007	&	94 / 46328	&	399 / 91000	\\
3	&	seeing $\leqslant4''$	&	3 / 166119	&	3 / 42155	&	82 / 45292	&	382 / 88850	\\
4	&	no bad pixels within $7\times7$ pixels	&	3 / 166119	&	3 / 42155	&	82 / 45292	&	382 / 88850	\\
5	&	no early $g$-band images	&	3 / 165351	&	3 / 41988	&	75 / 45038	&	365 / 88078	\\
6	&	std. of bkg. $<25$	&	3 / 163808	&	3 / 40877	&	75 / 44349	&	362 / 86141	\\
7	&	err. on flux $<7$ times median err.	&	3 / 162485	&	3 / 40575	&	75 / 44054	&	361 / 85333	\\
8	&	red. $\chi^2<1.4$ for nearby star	&	3 / 150637	&	3 / 37774	&	73 / 40740	&	265 / 78946	\\
9	&	red. $\chi^2<1.4$ at SN position	&	2 / 148884	&	3 / 36888	&	5 / 37058	&	204 / 73105	\\
10	&	$\geqslant$ 20 pre-expl. observations	&	2 / 136338	&	3 / 33600	&	4 / 36300	&	189 / 70420	\\
11	&	offset correction	&	2 / 136338	&	9 / 33600	&	11 / 36300	&	189 / 70420	\\
12	&	error-bar scaling	&	2 / 136338	&	1 / 33600	&	3 / 36300	&	136 / 70420	\\
\hline
13	&	ref. sys. error / final unbinned	&	2 / 136338	&	0 / 33600	&	3 / 36300	&	152 / 70420	\\
14	&	1-day bins	&	1 / 63791	&	1 / 15682	&	4 / 16979	&	124 / 32993	\\
15	&	7-day bins	&	0 / 25528	&	2 / 6456	&	4 / 7616	&	84 / 14193	\\
16	&	90-day bins	&	0 / 3983	&	2 / 1045	&	2 / 1281	&	37 / 2093	
\enddata
\tablecomments{Number of remaining data points and (false) $5\sigma$ detections after each step of the analysis, as described in Sec.~\ref{sec:fp_pipeline} (steps 1 to 7), Sec.~\ref{sec:fp_astrometry} (step 8 and 9), Sec.~\ref{sec:fp_rescaling} (steps 10 to 12) and in Sec.~\ref{sec:fp_binned} (step 13 and 14). Our actual search (last column) yields a much larger number of precursor detections than the three background samples.
The initially large number of detections for the PTF sample is due to AGN activity in the host galaxy of SN\,2011cc (see Sec.~\ref{sec:fp_tests}) and a few detections of this AGN persist after all cuts. The false detections for the empty and mirrored positions are all caused by a faulty reference image.}
\end{deluxetable*}

\subsection{Astrometric Errors}
\label{sec:fp_astrometry}

The large number of analyzed observations requires tight cuts on the data quality to avoid false-positive detections.
Some of the reference or difference images might suffer from misalignments such that residuals are created in the image-subtraction process. Alignment errors result from several factors, including atmospheric scintillations (e.g.,~\citealt{osborn2015, ofek2019}).
To identify and remove affected images, we perform forced photometry at the position of a relatively faint star or an unresolved galaxy close to the SN position as described in Sec.~\ref{sec:fp_bgsample}. We choose faint stars because they roughly represent the surface brightness of bright galaxies. If the images are well aligned, no detection is expected for a nonvariable star, or for a variable star the residual should be well described by the PSF.

Images with astrometric residuals are identified via the reduced $\chi^2$ of the PSF fit. We find that requiring a reduced $\chi^2 <1.4$ at the position of the star removes most false detections. The corresponding difference images are flagged and are not used when searching for precursors at the SN position. In addition to misalignments, there could be more localized residuals or artefacts. We therefore require that the reduced $\chi^2$ at the SN position is also smaller than $1.4$. As shown in Table~\ref{tab:cuts}, these two cuts remove in total $12\%$ of the data. The reduced $\chi^2$ values for each data point are given in Table~\ref{tab:fp_fluxes}.

\subsection{Offsets and Rescaling Flux Errors}
\label{sec:fp_rescaling}

As a next step, we verify that the pre-explosion light curves are centered around zero flux and that the estimated flux errors account for the observed flux scatter. 
When visually inspecting pre-explosion light curves, we find that the median fluxes are sometimes systematically offset from zero. In some cases, the offset could be due to light in the reference image either from the transient or from a precursor. However, we also see such offsets for the background samples. They can be as large as the typical error bar of the unbinned fluxes. We therefore do a baseline correction for all light curves. Consequently, we cannot identify precursors during the reference period or very long-lasting precursors that affect all data points (see also Appendix~\ref{sec:sn2019cmy}). Moreover, we find that the size of the error bars is overestimated or underestimated by typically 10--20\%. For a handful of locations, the errors even have to be increased by as much as 50\% to account for the observed scatter. 

These biases are corrected for each reference image separately. To do this precisely, we require at least 20 pre-explosion observations with the same reference image. If fewer observations are available, the corresponding data points are discarded (step 10 in Table~\ref{tab:cuts}). After applying all cuts, we find that no pre-explosion observations are left for 13 out of 209 SNe (see Sec.~\ref{sec:fp_sample}); most of them were detected in the beginning of the survey (see online version of Table~\ref{tab:sample}). Our final sample hence consists of 196 SNe.

We find that using the median pre-explosion flux to correct the baseline does not work for all SNe, because some of them have long-lasting precursors that contain close to half of the data points. We therefore calculate the \emph{iterative median} which is more robust. We first combine same-night observations in bins to avoid  individual nights with many observations dominating the result. Next, the median is calculated for the binned data points and the data point with the largest deviation from the median (regardless of the size of the error bars) is removed from the sample. This last step is repeated; we recalculate the median for the remaining points and remove the most distant data point, until only 30\%, but at least 20, of the data points are left. The median of these remaining points is used as the baseline correction. We find that this algorithm reliably identifies the zero flux level and removes the impact of any precursors during the reference period.

When searching for precursors at the SN positions, we select all objects with positive or negative $5\sigma$ detections and check whether we can redo the baseline correction for a time range that excludes the potential precursor, preferentially after the SN has faded. If this is possible we recalculate the baseline correction, this time using a simple median. This step leads to additional precursor detections for SNe with few pre-explosion observations (e.g., SN\,2018eru and SN\,2018kag) and improves the baseline correction for SNe for which a large fraction of the data points are part of the precursor, such as SN\,2019fmb (see Sec.~\ref{sec:pre_detections}). We also find that observations obtained after SN\,2019cmy had faded are systematically lower than pre-explosion observations. As discussed in Appendix~\ref{sec:sn2019cmy}, we are not sure whether this drop in flux is due to a systematic error or an extremely bright progenitor star. In this paper, we exclude the late-time observations and only discuss the short precursor detected relative to the flux level of the pre-explosion light curve (see Sec.~\ref{sec:pre_detections}).

Next, we scale up the flux errors if they are underestimated, which is again done for every reference image separately. As before, the result might be biased by precursors which can inflate the error bars and remain undetected as a consequence. We therefore split the pre-explosion light curves for each reference image into equal segments of 15 or more data points. We calculate the \emph{local robust standard deviation} for each segment by determining the 15.9\% and 84.1\% percentile and dividing its difference by 2. The median standard deviation for all segments is used to judge whether the error bars are sufficiently large to account for the observed noise level. If the standard deviation is larger than 1 (i.e., the error bars cannot fully account for the observed size of the $1\sigma$-region), the error bars are multiplied with the robust standard deviation of the median segment. No scaling is done if the standard deviation is smaller than 1  (i.e., the errors are overestimated compared to the observed scatter).

\subsection{Binned Light Curves and the Systematic Error of the Reference Image}
\label{sec:fp_binned}

To increase our sensitivity to faint precursors we also search binned light curves. 
The bins are chosen such that same-night observations are always combined in the same bin and the edge of the last pre-explosion bin is at the end of the night in which the SN is discovered. We ensure that data points before and after the estimated explosion date are never combined in the same bin by binning the two parts of the light curve separately. For each bin, we use the median observation date as the observation time of the bin and calculate the weighted mean flux and its uncertainty.

When combining a large number of observations in one bin, the uncertainty in the flux can become very small. However, the ZTF reference images only consist of about 15 coadded observations; hence, the noise level in the reference image has to be considered. For this purpose we convert the limiting magnitude of the reference image to a flux which is given in Table~\ref{tab:fp_fluxes}. This systematic error is added in quadrature to the uncertainty of the unbinned or binned fluxes. It is typically ten times smaller than the uncertainty in the flux measured in a single image and thus only becomes relevant if many observations are coadded in a bin. When combining flux measurements that have different reference images we use the median systematic error.

\begin{figure}[tb]
\centering
\includegraphics[width=\columnwidth]{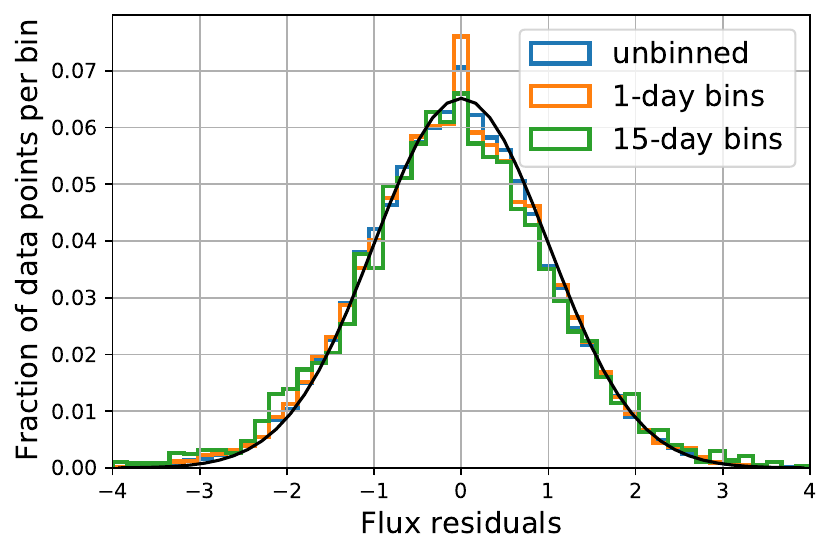}
\caption{\label{fig:flux_residuals} Flux residuals for the background sample of mirrored positions (see Sec.~\ref{sec:fp_bgsample}) compared to a normal distribution. The baselines have been centered around zero and underestimated flux errors have been rescaled as described in Sec.~\ref{sec:fp_rescaling}.}
\end{figure}

We verify that the flux residuals indeed follow a normal distribution with a width of 1 by showing the flux residuals (i.e., the flux divided by its uncertainty) in Fig.~\ref{fig:flux_residuals}. Except for statistical fluctuations, the residuals roughly follow a normal distribution. We expect some deviations from a normal distribution, for example because we do not reduce the size of overestimated flux errors (see Sec.~\ref{sec:fp_rescaling}). For 15-day bins the systematic error in the reference image also becomes relevant, such that we expect a slightly narrower distribution. The distributions in Fig.~\ref{fig:flux_residuals} illustrate that only very few data points deviate from 0 by more than $3\sigma$. This indicates that the forced-photometry pipeline and our cleaning process work well in locations where interacting SNe explode and that the error bars have an appropriate size after the scaling described in Sec.~\ref{sec:fp_rescaling}. For the precursor search we use a $5\sigma$ threshold.

\subsection{Expected Number of False Detections and Astrophysical Backgrounds}
\label{sec:fp_tests}

The empty positions as well as the mirrored positions and historic SNe serve as a quality check of the forced-photometry pipeline. We do not expect any astrophysical precursors at these positions and can therefore use these to calculate the false-alarm rate. Table~\ref{tab:cuts} shows that our actual search (last column) yields 152 $5\sigma$ detections for unbinned light curves even though only a few false detections are expected. This gives us confidence that the majority of the detected precursors are astrophysical. In addition, these precursors are almost exclusively detected among Type IIn SNe and they prefer low-redshift objects, as expected.

Nevertheless, a small number of false detections persist after all cuts. We inspect them to identify possible reasons. For empty and mirrored positions all false detections occur in the $g$-band images that contain SN\,2018bih. A visual check shows that the reference image contains structures that are not astrophysical. In the actual search for this SN, we do not find any precursor candidates, potentially because only few pre-explosion images are available owing to its explosion date in May 2018. Another notable issue is the large number of detections at locations where PTF SNe exploded prior to 2015 (penultimate column in Table~\ref{tab:cuts}). Most of them (97 out of 116 detections before cuts) are at the position of SN\,2011cc and are likely due to AGN activity close to the SN position (see below). Our cuts remove most detections at this position, but a few remain. We conclude that our precursor search might yield a few false detections, for example owing to faulty reference images (1 out of 860 reference images affected) and background AGNs (1 out of 100 SN positions affected).

A large number of false detections is found at the position of SN\,2011cc, which exploded at a distance of $2.4''$ from the center of its host galaxy. The host, IC\,4612, is classified as a star-forming galaxy in the SDSS catalog \citep{ahumada2019} and as a narrow-line AGN by \citet{liu2011}. We therefore hypothesize that the variability observed in ZTF data is caused by AGN activity. Requiring a reduced $\chi^2$ of $<1.4$ at the SN location removes most detections, because the background AGN is slightly offset from the SN position. No variability or precursors were detected in the PTF pre-explosion light curve of SN\,2011cc \citep{ofek2014}, likely owing to the relatively small number of observations. Another case of apparent variability due to potential background AGN activity is detected in the pre-explosion light curve of SN\,2019meh when searching the actual SN locations (see Sec.~\ref{sec:pre_detections}).

In addition, we find that light from the Type Ia SN SN\,2018big contaminated the pre-explosion light curve of the flash-spectroscopy object SN\,2019nvm. Both SNe happened in the same host galaxy with a separation of $3.6''$, so SN\,2018big is just at the edge of the $7\times7$ pixel region for which the PSF fit is done (see Sec.~\ref{sec:fp_pipeline}). Since we require a small reduced $\chi^2$ at the SN position (step 9 in Table~\ref{tab:cuts}), all detections of SN\,2018big are rejected, such that the object does not show up as a potential precursor in the search described in Sec.~\ref{sec:pre_detections}. 
These coincidences serve as reminders that pre-explosion activity does not necessarily originate from the progenitor star, but could be related to bright, variable objects within $\lesssim4''$. 

It is also possible that a different star close to the progenitor produces precursor eruptions. It could even be the progenitor of a SN that might explode at a later time. We consider this scenario relatively rare, as no further precursors are detected in ZTF data at 100 positions where PTF detected Type IIn SNe before 2015 (see Sec.~\ref{sec:fp_tests}). Nevertheless, this possibility cannot be ruled out in individual cases.

Another challenge is distinguishing between a precursor and the rising SN light curve. Double peaks or early plateaus, likely powered by shock cooling (see e.g.~\citealt{sapir2017}), have been observed for several SNe of Type Ib, Ibn and IIb (see e.g. \citealt{gal-yam2017}). \citet{piro2013} estimate that SNe powered by radioactivity can undergo a dark phase of up to several days. After this time emission from centrally located radioactive nickel-56 is able to diffuse outwards and the SN starts to rise to its main peak. We find such early detections for several SNe (e.g. for SN\,2019fci). If the detection is separated by less than a week from the observed rise, we assume conservatively that the SN has already exploded at this time and adjust the discovery date $t_0$ accordingly. As a consequence, we might miss short-lived precursors immediately prior to the SN detection. This is especially true for objects for which the rise of the light curve is not well observed.

\section{Precursor properties}
\label{sec:precursors}

After developing and testing our analysis in the previous section, here we apply it to the actual data. The detected precursors and additional tests are described in Sec.~\ref{sec:pre_detections}, the precursor absolute magnitude light curves and radiated energies are calculated in Sec.~\ref{sec:pre_energy}, and their $g-r$ colors are presented in Sec.~\ref{sec:pre_colors}.

\subsection{Detected Precursors}
\label{sec:pre_detections}

To search for precursors, we produce forced photometry light curves at the SN positions and apply the cuts and corrections described in Sec.~\ref{sec:fp}. Any pre-explosion data points that are significant at the $5\sigma$ level are considered detections. To gain sensitivity to fainter precursors, we search in addition the binned light curves (see Sec.~\ref{sec:fp_binned}). The precursor durations are unknown and moreover depend on the detection threshold. To cover a wide range of timescales we use six different bin sizes with lengths of 1, 3, 7, 15, 30, and 90 days. The bin sizes are chosen such that the amount of data approximately doubles or triples when going to the next larger bin size.

\begin{figure}[tb]
\centering
\includegraphics[width=\columnwidth]{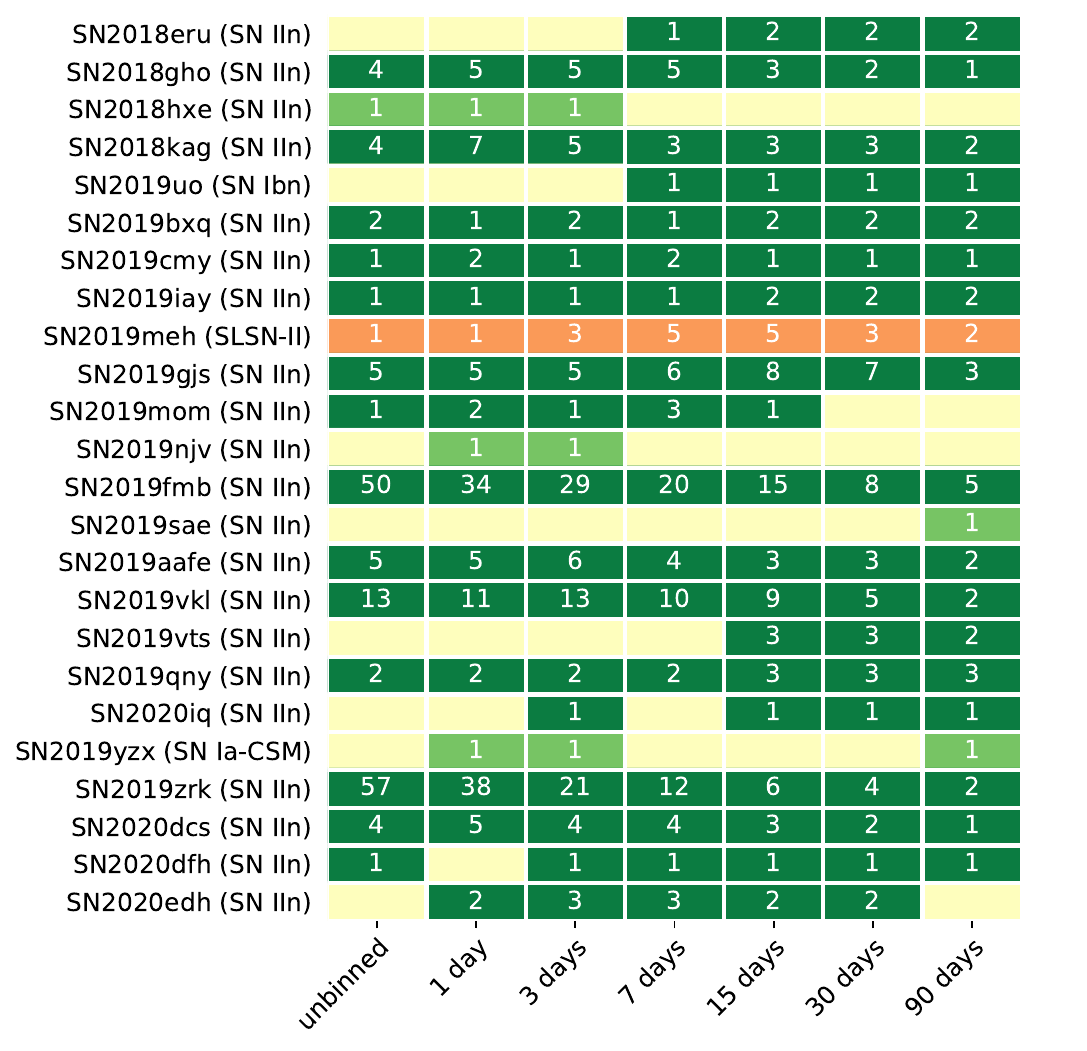}
\caption{Number of pre-explosion detections above the $5\sigma$ threshold for each bin size. Observations in different photometric bands are binned separately and the white numbers indicate the total number of detections in all three bands. Dark-green fields represent securely detected precursors, while light-green fields mark unconfirmed precursors that are only seen in a single bin and band (see text). Orange fields signal that the flux variability is likely caused by AGN activity in the host galaxy and not by the progenitor star.}
\label{fig:precursor_overview}
\end{figure}

\begin{figure*}[h!]
\centering
\includegraphics[width=8.7cm]{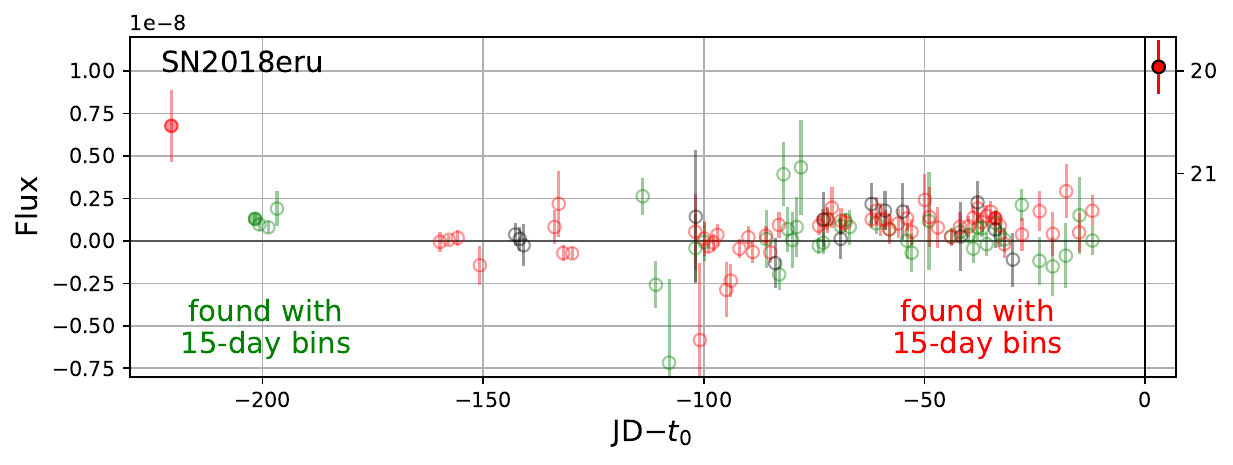} \hfill 
\includegraphics[width=8.7cm]{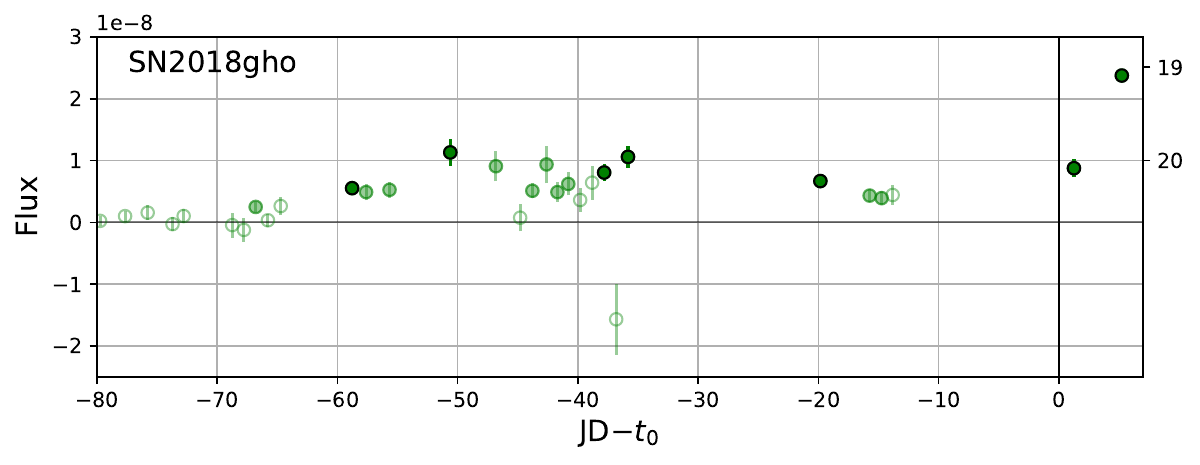} \\
\includegraphics[width=8.7cm]{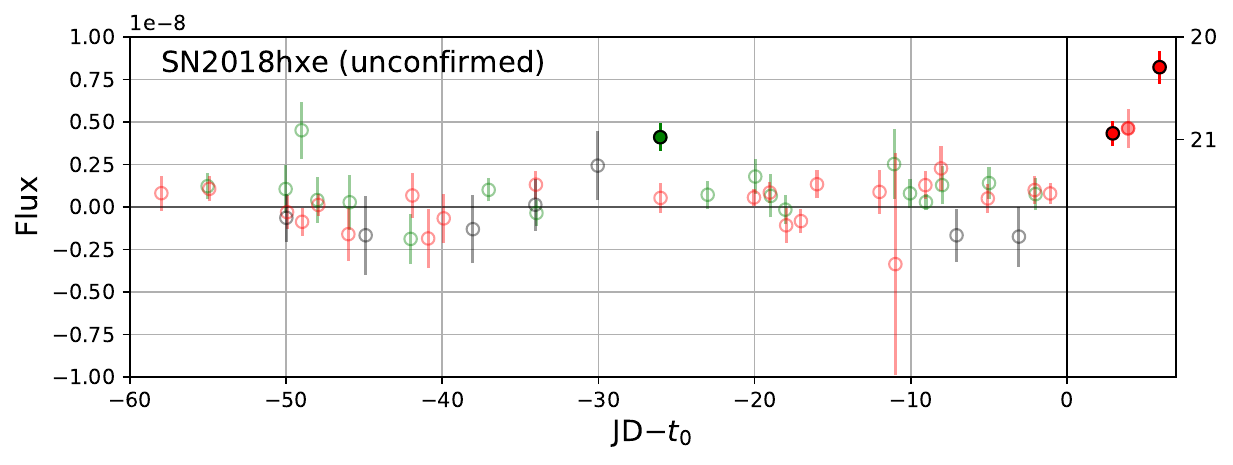} \hfill
\includegraphics[width=8.7cm]{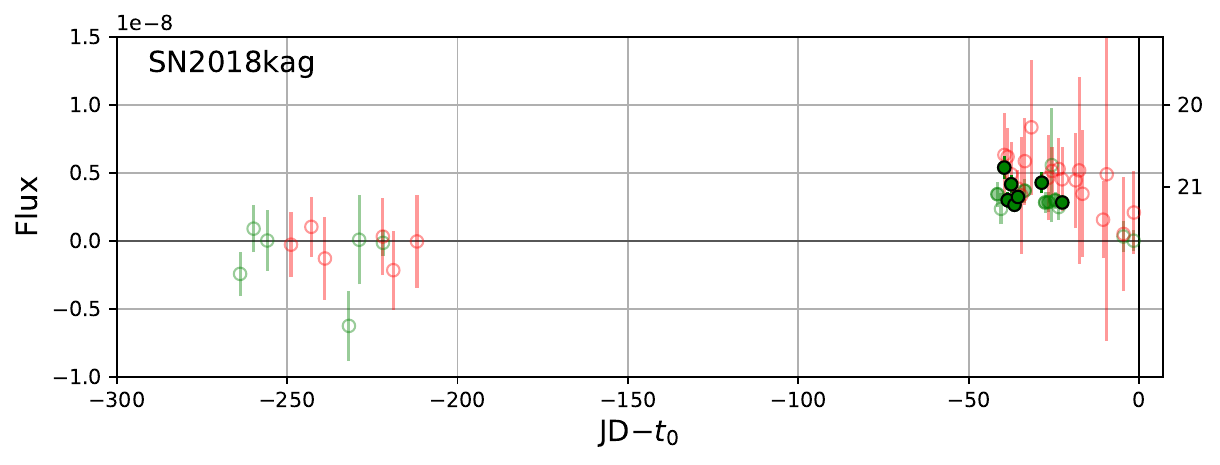} \\
\includegraphics[width=8.7cm]{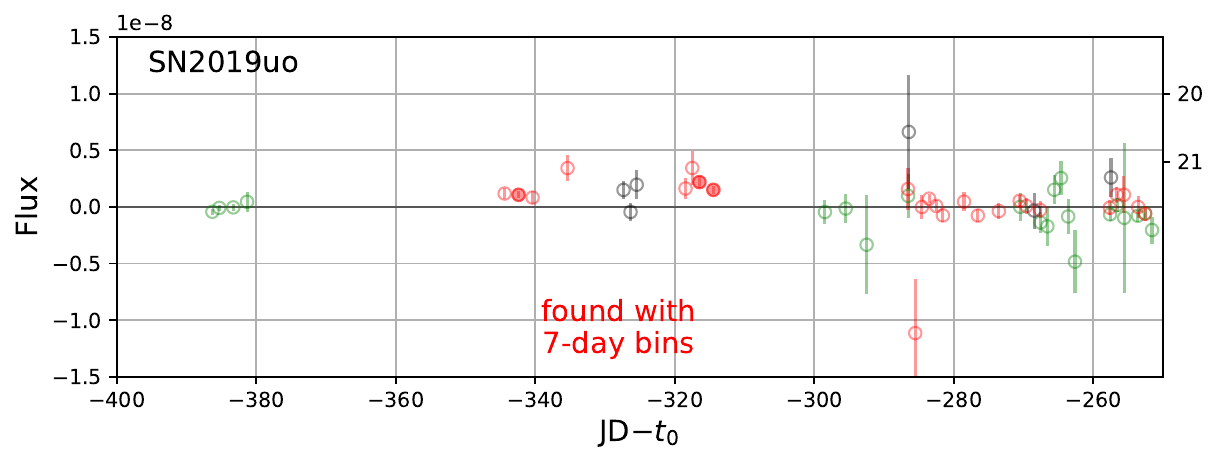} \hfill
\includegraphics[width=8.7cm]{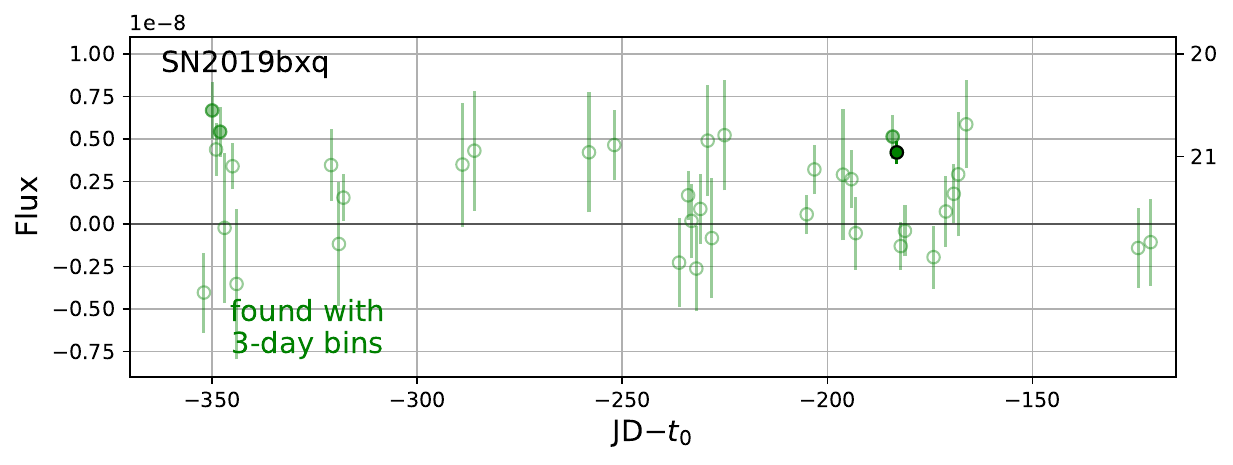} \\
\includegraphics[width=8.7cm]{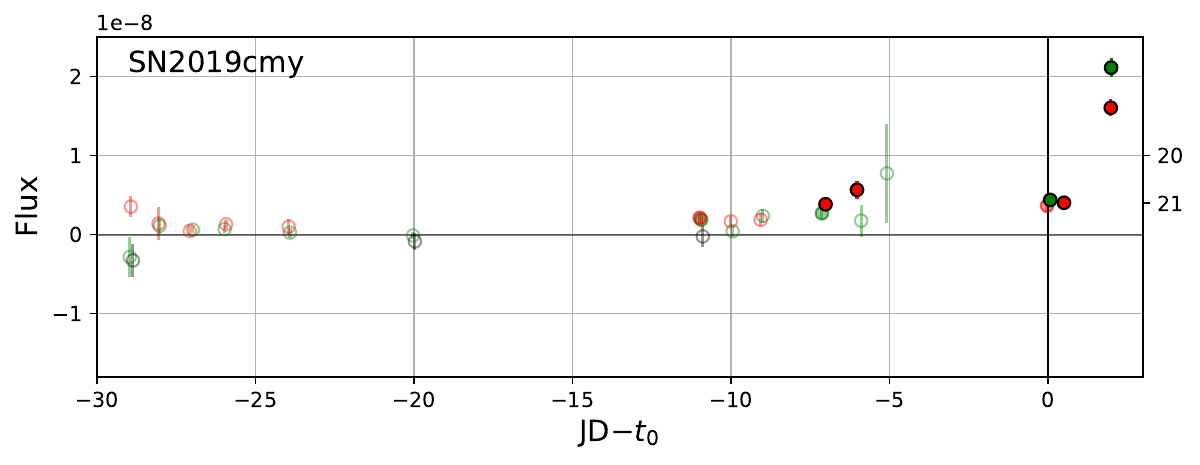} \hfill
\includegraphics[width=8.7cm]{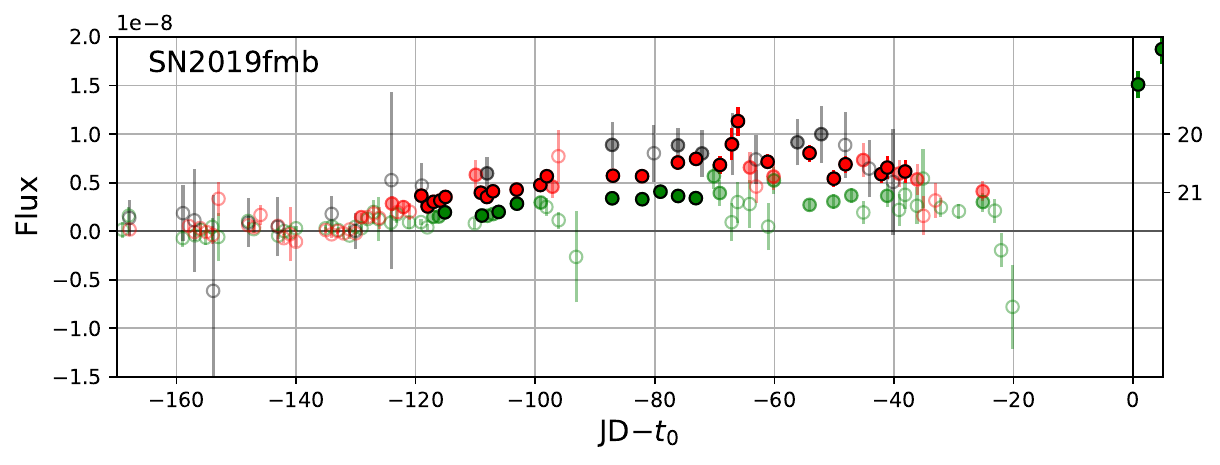} \\
\includegraphics[width=8.7cm]{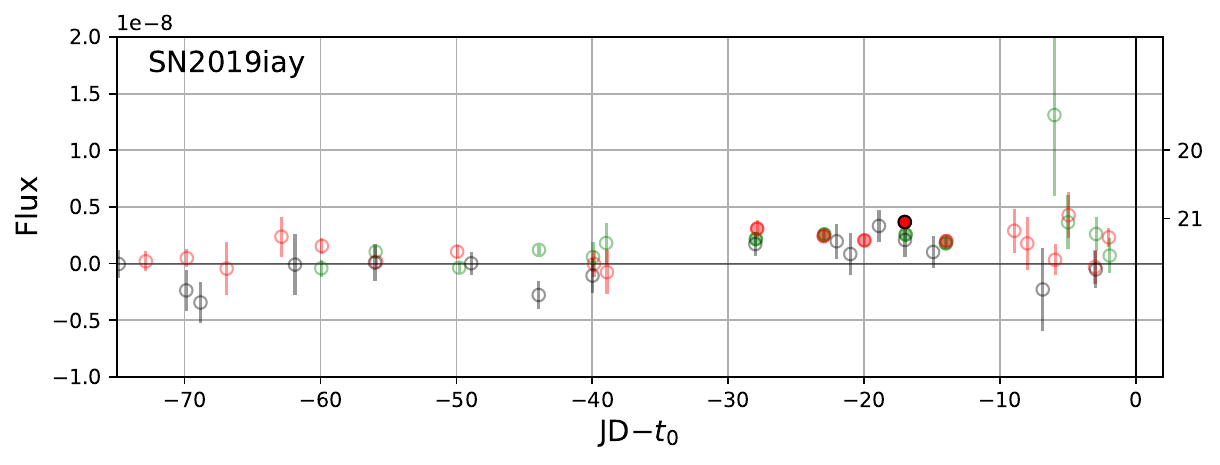} \hfill
\includegraphics[width=8.7cm]{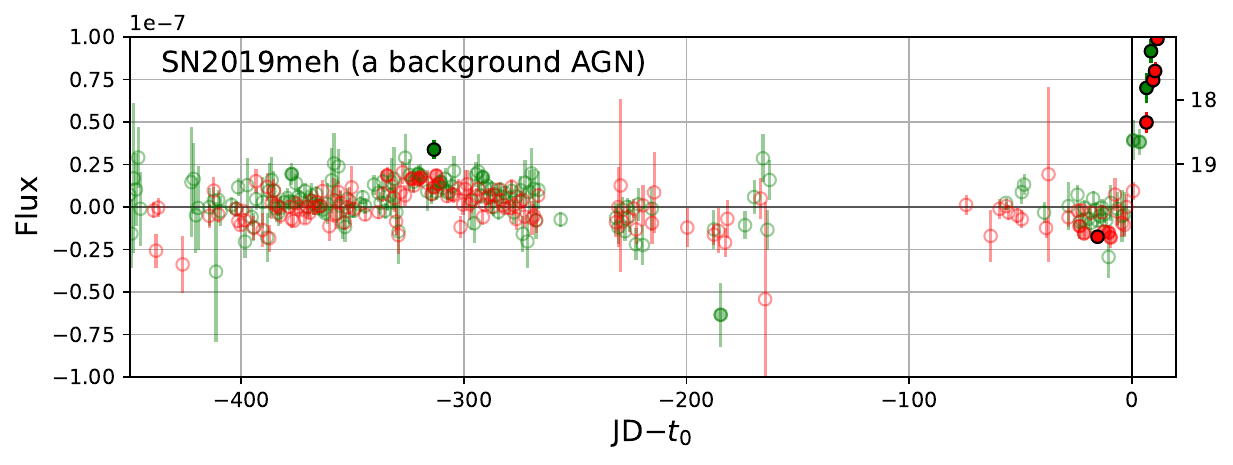} \\
\includegraphics[width=8.7cm]{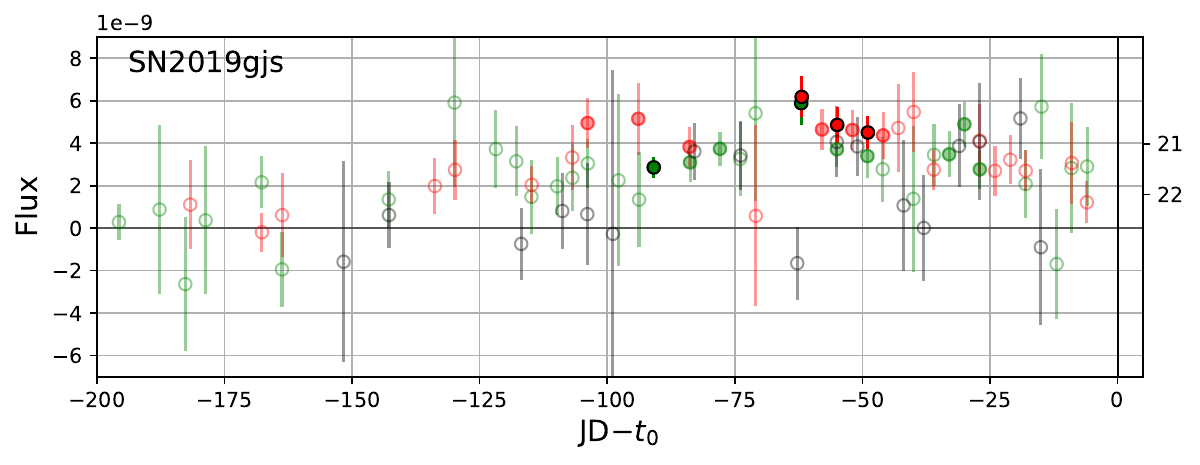} \hfill
\includegraphics[width=8.7cm]{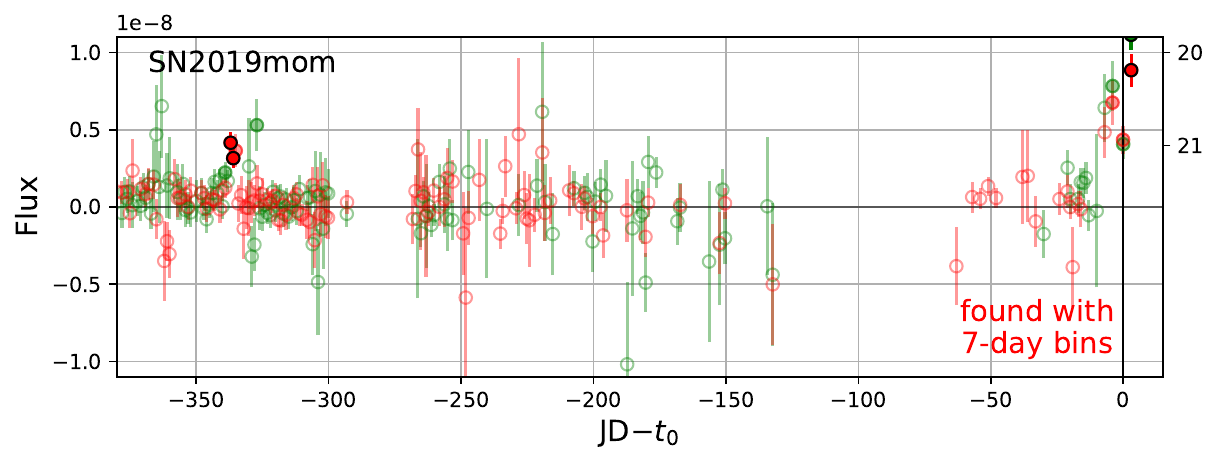}
\caption{Precursor light curves in 1-day bins. Solid circles mark $5\sigma$ detections, circles filled with a lighter shade and without a black edge have significance between $3\sigma$ and $5\sigma$, and open points are less significant. Green, red, and black data points were obtained in the $g$, $r$, and $i$ bands, respectively. The vertical black line indicates $t_0$, a rough estimate for the explosion date, and colored areas identify precursors that are detected significantly when using larger bins. The flux $f$ is given as a unitless ratio relative to the zeropoint which is equivalent to the unit ``maggie" used in SDSS catalogs \citep{finkbeiner2004}. Corresponding AB magnitudes are given on the right-hand ordinate axis and are calculated as $m_{\text{AB}}=-2.5 \log_{10}(f)$.
}
\label{fig:precursor_lcs}
\end{figure*}

\begin{figure*}[h!]
\centering
\includegraphics[width=8.7cm]{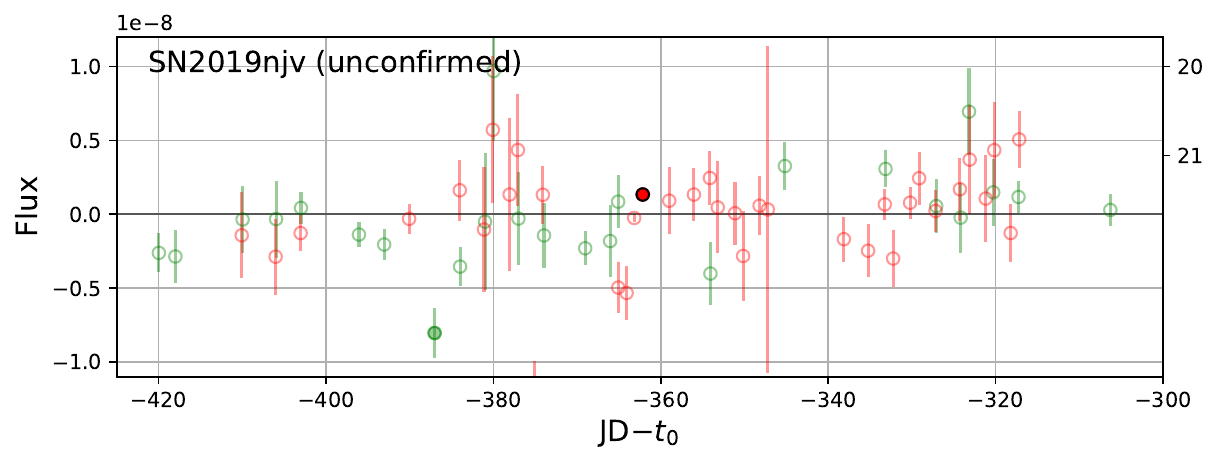} \hfill
\includegraphics[width=8.7cm]{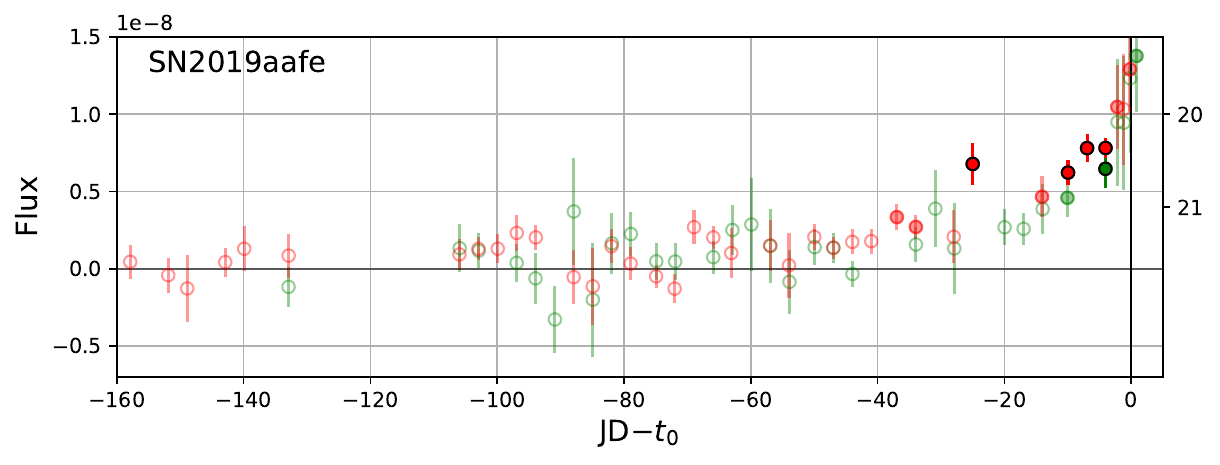} \\
\includegraphics[width=8.7cm]{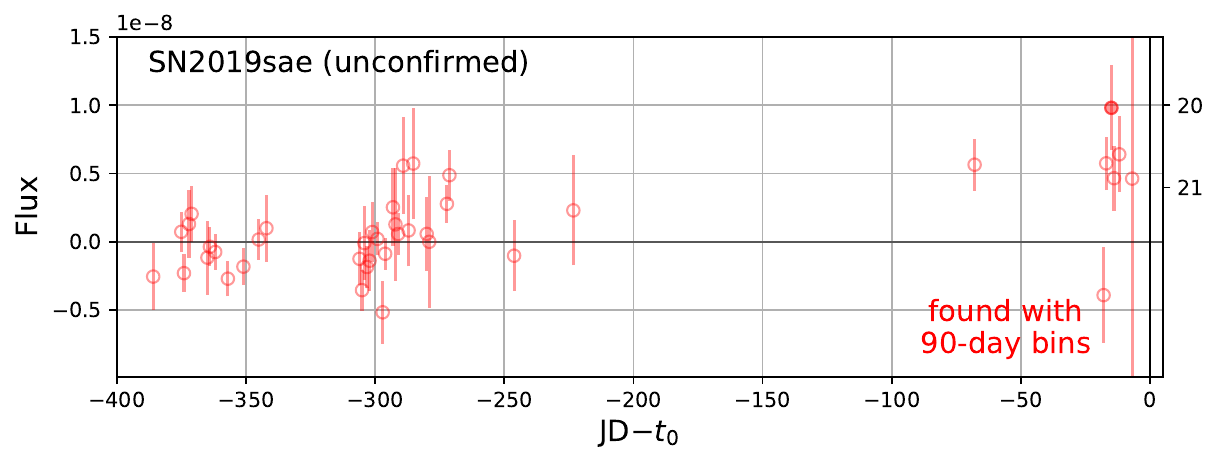} \hfill
\includegraphics[width=8.7cm]{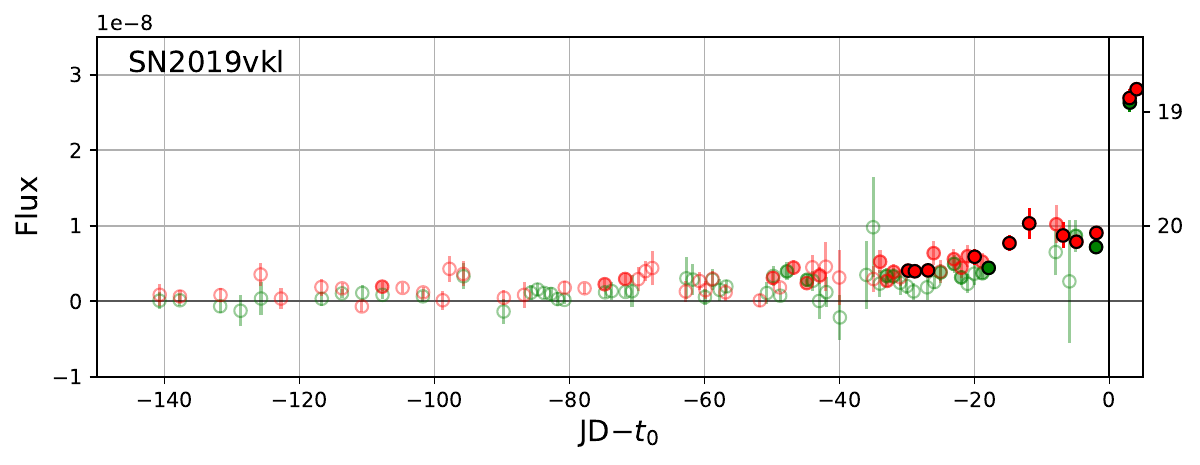} \\
\includegraphics[width=8.7cm]{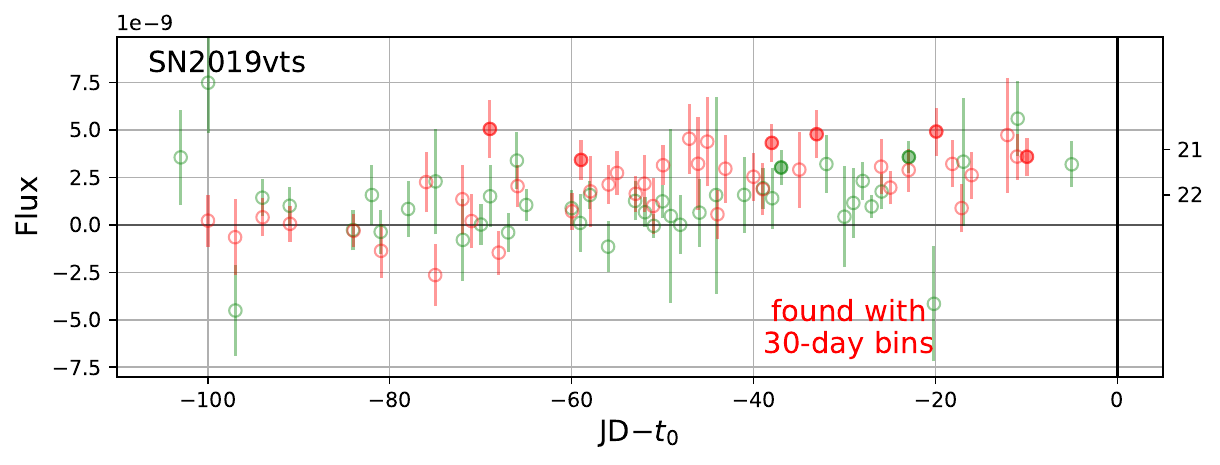} \hfill
\includegraphics[width=8.7cm]{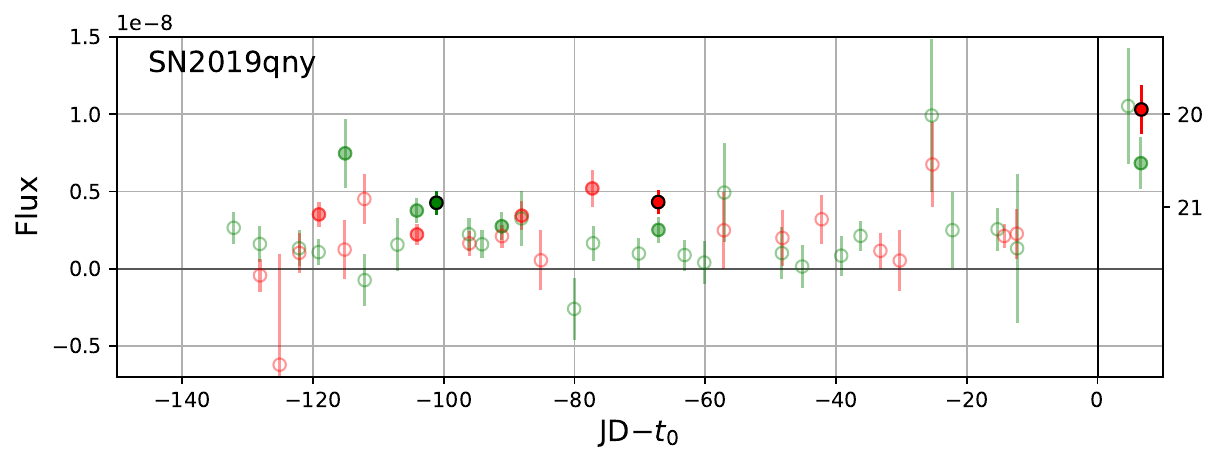} \\
\includegraphics[width=8.7cm]{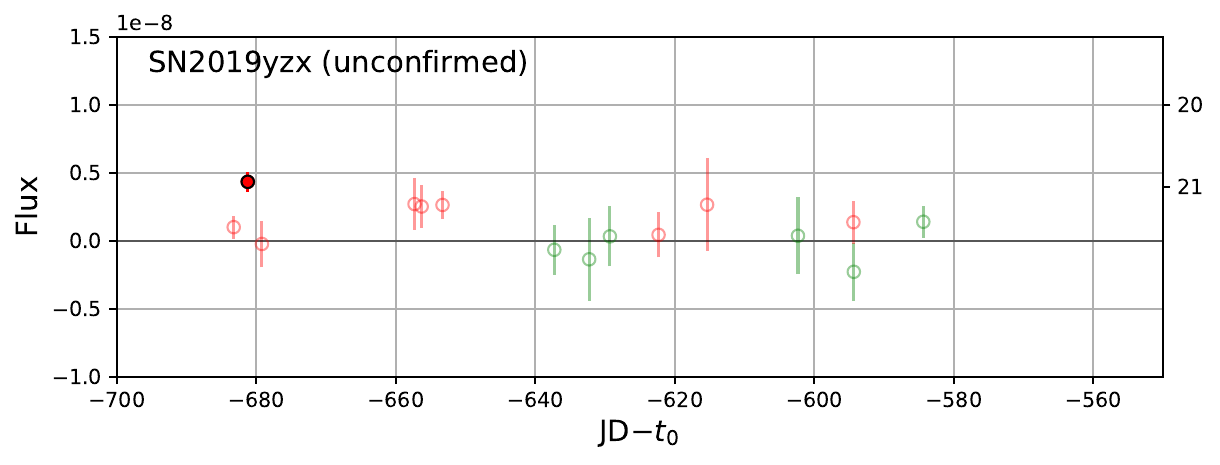} \hfill
\includegraphics[width=8.7cm]{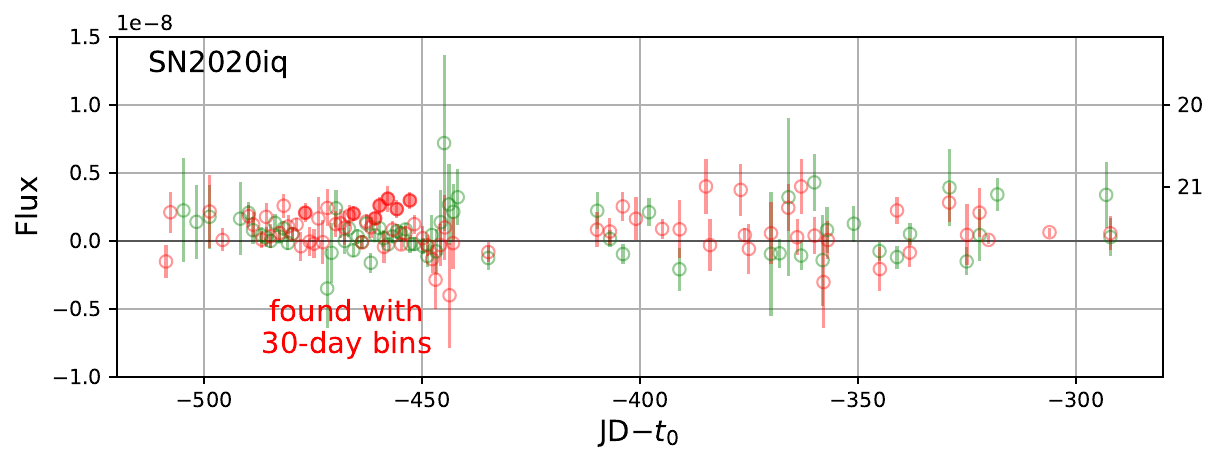} \\
\includegraphics[width=8.7cm]{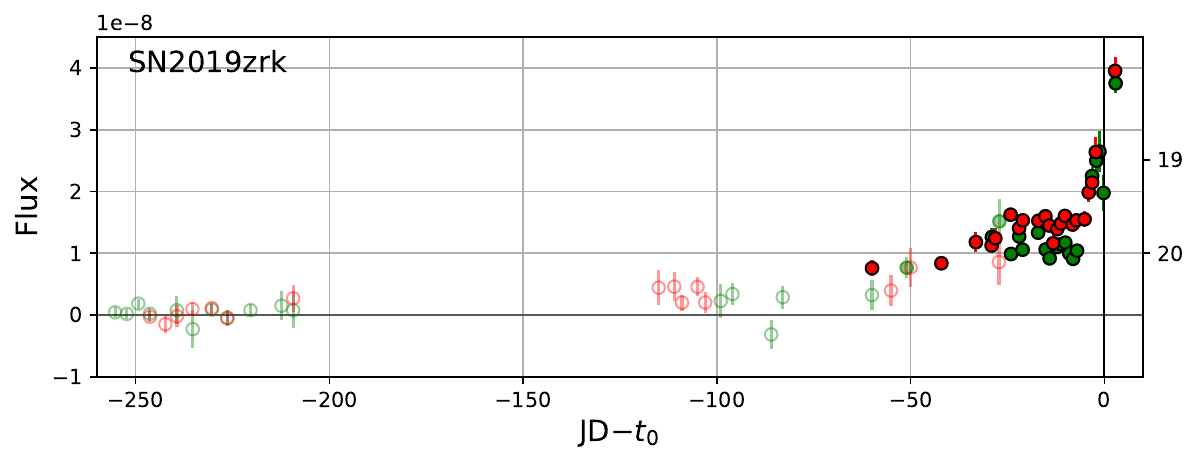}  \hfill
\includegraphics[width=8.7cm]{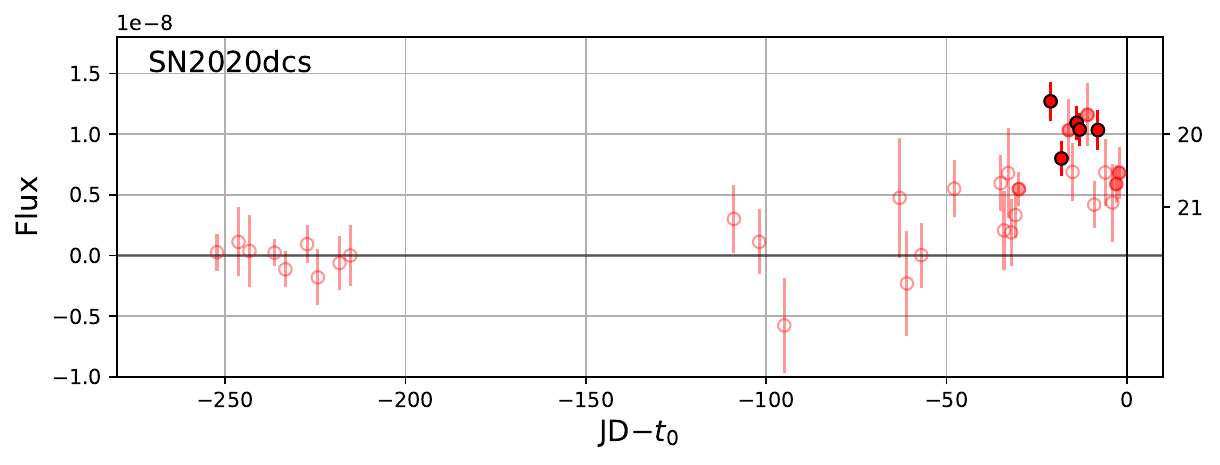} \\
\includegraphics[width=8.7cm]{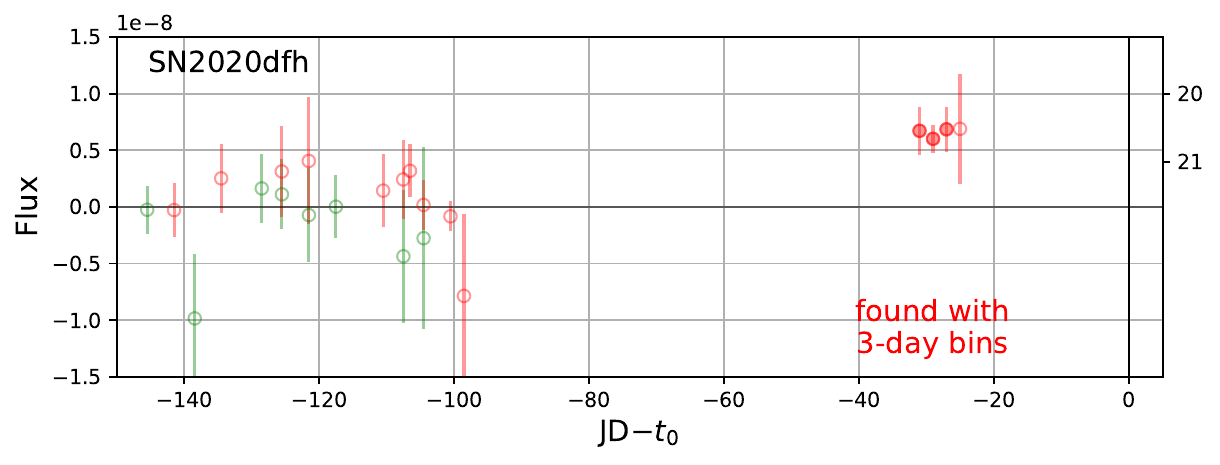} \hfill
\includegraphics[width=8.7cm]{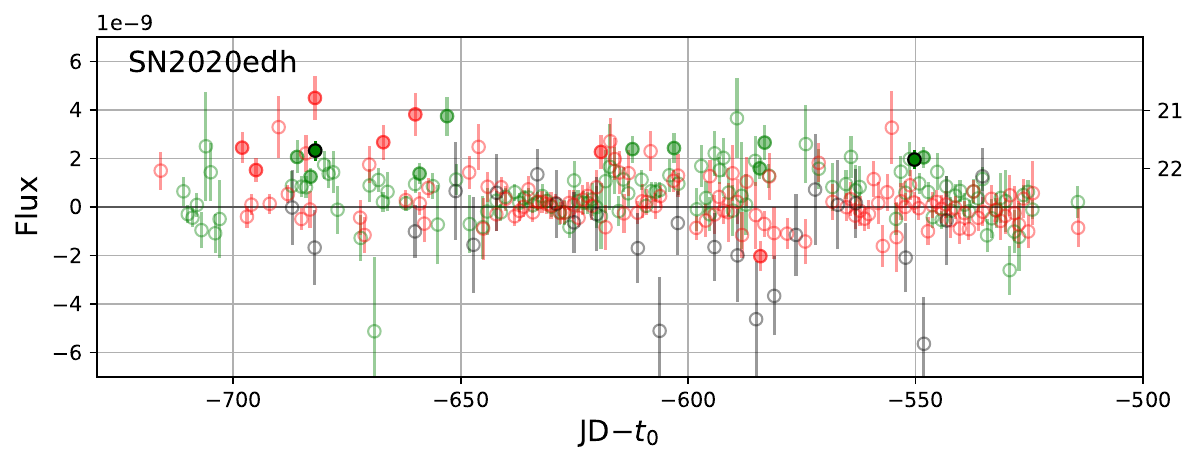}
\caption{Precursor light curves -- continuation of Fig.~\ref{fig:precursor_lcs}}
\label{fig:precursor_lcs2}
\end{figure*}

In our search of the pre-explosion data of 196 SNe, we find precursor candidates prior to 24 SNe, mostly of Type IIn; Fig.~\ref{fig:precursor_overview} indicates the number of $5\sigma$ detections in each search channel. Most precursors are detected using several different bin sizes, indicating that they are both bright and long-lasting. 
The precursor light curves in 1-day bins are shown in Figs.~\ref{fig:precursor_lcs} and~\ref{fig:precursor_lcs2}, and their properties are summarized in Table~\ref{tab:precursors}. In addition, we show coadded difference images of the precursors in Appendix~\ref{sec:images}. They demonstrate that the detections are indeed due to point sources at the SN location with the exception of SN\,2019sae, which might be spurious.

Marginally detected precursor candidates are inspected in more detail, to test whether they are genuine. For precursors that are only detected in a single bin, we check whether fluxes in the three bins before or after the detection surpass the $3\sigma$ significance threshold or whether reducing the bin size leads to at least two data points above the $3\sigma$ threshold. If we do not find any additional $3\sigma$ detections, we conclude that the detection is driven by data collected within a single night and band and refer to these detections as \emph{unconfirmed precursors}. The light-green color in Fig.~\ref{fig:precursor_overview} highlights the four precursors that do not pass this test (the Type IIn SN\,2018hxe, SN\,2019njv, and SN\,2019sae, and the Type Ia-CSM SN\,2019yzx). The fact that we only found few false $5\sigma$ detections when searching the background samples in Sec.~\ref{sec:fp_bgsample} suggests that at least some of the unconfirmed precursors are astrophysical nonetheless. In the following we only focus on the 19 securely detected precursors.


One pre-explosion light curve, prior to SN\,2019meh, shows long-term up-and-down fluctuations as expected for AGNs (see Fig.~\ref{fig:precursor_lcs}). Indeed, the SN is located within $1''$ of the center of its host galaxy and we therefore conclude that the variability is likely due to nuclear activity and is not caused by the progenitor star, as described for SN\,2011cc in Sec.~\ref{sec:fp_tests}. SN\,2019meh is a SLSN of Type II located at a relatively high (for our sample) redshift of 0.0935. Such a distant progenitor star would have to reach an extreme luminosity to be detectable, which supports the hypothesis that we are seeing AGN activity rather than stellar flares. We therefore remove this object from the sample.

\begin{figure*}[tb]
\centering
\includegraphics[width=\textwidth]{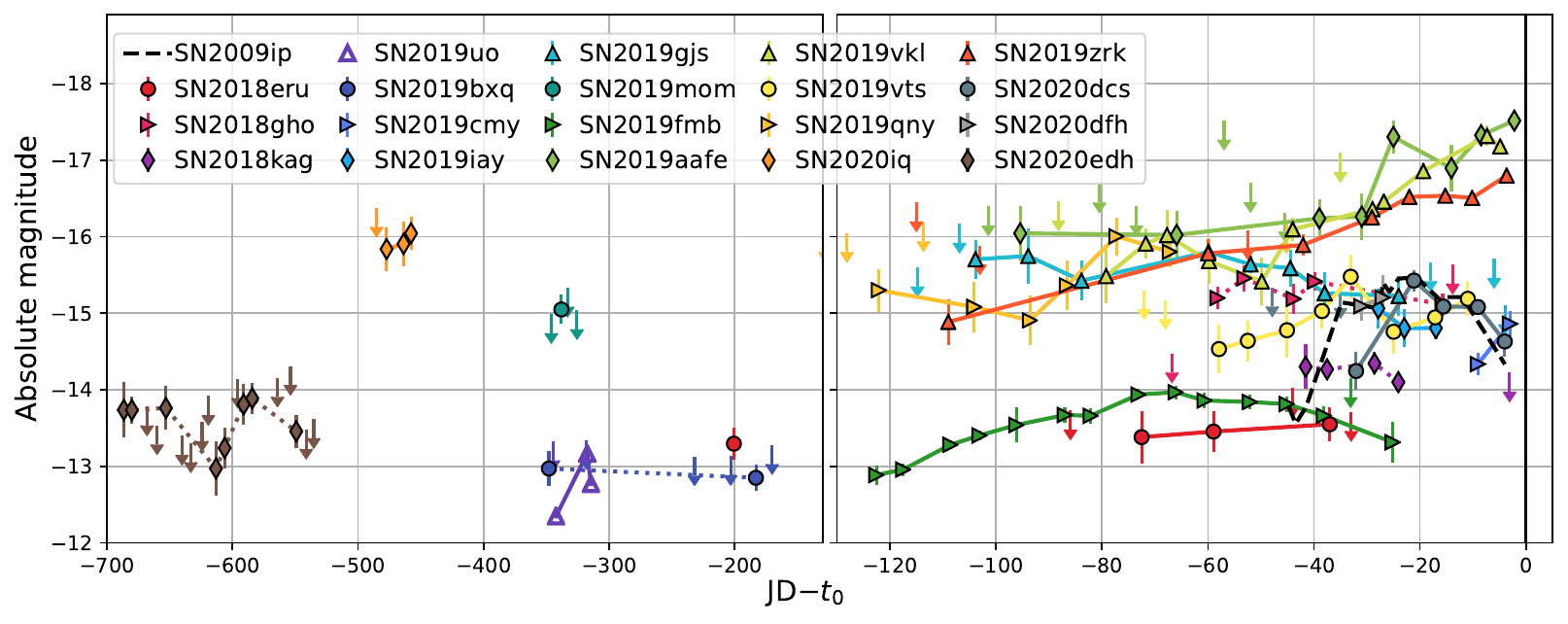} \hfill
\caption{\label{fig:shapes} Absolute magnitude precursor light curves in 7-day bins. All SNe are of Type IIn, except for the Type Ibn SN\,2019uo (open markers). If available, we show the $r$-band light curve, and dotted lines indicate that the $g$-band light curves are depicted for SN\,2018gho, SN\,2018kag, SN\,2019bxq, SN\,2020edh, and the early detection of SN\,2018eru. For clarity, only the most relevant nondetections are displayed. The 2012a event prior to the likely final explosion of SN\,2009ip is shown as a black dashed line for comparison (light curve taken from \citealt{margutti2013}, including data from \citealt{prieto2013} and \citealt{pastorello2013}).
}
\end{figure*}

We also check whether shifting the bin positions leads to the detection of additional precursors. For this purpose, we repeat the search with 7-day bins six times while moving the bin edges by one day for each new search. We detect precursors prior to a few SNe that are not found with the original 7-day bins, but all additional SNe already have detected precursors when using smaller or larger bins (see Fig.~\ref{fig:precursor_overview}). We thus conclude that the bin positions only have a minor influence on the results.

We summarize that we securely detect pre-explosion outbursts prior to 18 different SNe of Type IIn and prior to the Type Ibn SN\,2019uo (see Fig.~\ref{fig:precursor_overview}). Figures~\ref{fig:precursor_lcs} and~\ref{fig:precursor_lcs2} show that some SNe, such as SN\,2018eru, SN\,2019bxq, SN\,2019mom, or SN\,2020edh, might undergo several separate precursor eruptions. It is, however, also possible that the detections are part of a single flaring episode that lasts for several hundred days. 

\subsection{Precursor Energy}
\label{sec:pre_energy}

To put the precursor eruptions into context, we calculate the absolute magnitude light curves and estimate the radiated energies of the securely detected precursors found in Sec.~\ref{sec:pre_detections}.
Fluxes are converted to ``asinh magnitudes," also called ``luptitudes," with a softening parameter of $10^{-10}$ as defined by \citet{lupton1999}. Magnitude errors are given as $\sigma_{\text{mag}} = 1.0857 \log(f/\sigma_{\text{flux}})$, where $f$ and $\sigma_{\text{flux}}$ are the dimensionless normalized fluxes and uncertainties shown in Figs.~\ref{fig:precursor_lcs} and ~\ref{fig:precursor_lcs2}. The $5\sigma$ limiting magnitude is calculated as $m_{\text{lim}} = -2.5 \log(5\times\sigma_{\text{flux}})$ and the significance of a detection is given as $f/\sigma_{\text{flux}}$.
All calculations are here done for $7$-day bins and we consider $3\sigma$ detections significant if they are part of a previously detected precursor. For data points that do not reach the $3\sigma$ threshold we calculate $5\sigma$ upper limits.

Figure~\ref{fig:shapes} shows the resulting absolute magnitude $r$-band light curves (dashed lines indicate that the $g$ band was used instead for SN\,2018gho, SN\,2018kag, SN\,2019bxq, SN\,2020edh, and the early detection of SN\,2018eru). For clarity we omit nondetections that do not directly constrain the precursor duration. Most precursors are detectable for several weeks and some of them start more than 100 days before the explosion. The peak magnitudes vary between $-13$ and $-17.5$ as also summarized in Table~\ref{tab:precursors}. 
For comparison we add the $r$-band light curve measured for the 2012a event observed immediately prior to the likely final explosion of SN\,2009ip (data taken from \citealt{margutti2013}, \citealt{prieto2013}, and \citealt{pastorello2013}). Its duration, peak magnitude, and shape are similar to those of several of the less energetic precursors found in this search. We hence conclude that bright and long-lasting precursors are common in the last months before the explosion of Type IIn SNe. Their rate is quantified in Sec.~\ref{sec:rates}.

Next, we calculate the precursor energies by integrating the fluxes per bin from the first to the last detection, even if individual data points in between are not significant at the $3\sigma$ level. The calculation is done for each band separately and gaps in the data are interpolated if one or two 7-day bins are empty, such as for SN\,2018gho (see Fig.~\ref{fig:precursor_lcs}). This interpolation increases the total energy by at most 30\% and thus does not have a major impact on the results. We obtain similar results for 3-day bins and thus conclude that the energy estimates in Table~\ref{tab:precursors} roughly describe the observed precursor energy. These are lower limits on the true radiated energies of the precursors, which are often only partially detected, and also radiate outside of the visible-light bands which we cover. The brightest precursors reach radiative energies close to $10^{49}\,\text{ergs}$, about 10\% of the total radiative energy in a typical SN explosion.

\begin{deluxetable*}{l c c c c c c c c c}
\tablecaption{Detected precursors\label{tab:precursors}}
\tablewidth{0pt}
\tablehead{
\colhead{} & \colhead{band} & \colhead{start phase} & \colhead{end phase} & \colhead{median flux} & \colhead{energy} & \colhead{$v_{\text{CSM}}$} & \colhead{$\epsilon M_{\text{CSM, pre.}}$} & \colhead{$t_{\text{rise}}$} & \colhead{$M_{\text{CSM, diff.}}$}\\
\colhead{} & \colhead{} & \colhead{(days)} & \colhead{(days)} & \colhead{(mag)} & \colhead{($10^{46}$ ergs)} & \colhead{($\text{km}\,\text{s}^{-1}$)} & \colhead{(M$_\odot$)} & \colhead{(days)} & \colhead{(M$_\odot$)} 
}
\startdata
SN\,2018eru & $g$ & $-202.2$ & $-195.2$ & $-13.3$ & $4$ & 1100 & 0.02 & $-$ & $-$\\
& $g$ & $-69.2$ & $-62.2$ & $-13.3$ & $4$ & & & & \\
& $r$ & $-76.2$ & $-34.2$ & $-13.4$ & $23$ & & & & \\
\rule{0pt}{3ex}
SN\,2018gho & $g$ & $-63.0$ & $-14.0$ & $-15.2$ & $160$ & 210 & 4 & 13 & $<1.3$ \\
\rule{0pt}{3ex}
SN\,2018kag & $g$ & $-48.1$ & $-20.1$ & $-14.3$ & $40$ & 1100 & 0.04 & $-$ & $-$ \\
& $r$ & $-41.1$ & $-20.1$ & $-14.7$ & $50$ & & & & \\
\rule{0pt}{3ex}
SN2019uo & $r$ & $-342.8$ & $-307.8$ & $-13.0$ & $17$ & 880 & 0.007 & $8$ & $<0.8$\\
\rule{0pt}{3ex}
SN\,2019bxq & $g$ & $-349.4$ & $-342.4$ & $-13.0$ & $3$ & 330 & 0.06 & 18 & $<1.8$ \\
& $g$ & $-188.4$ & $-181.4$ & $-12.9$ & $3$ & & & & \\
\rule{0pt}{3ex}
SN\,2019cmy & $g$ & $-13.4$ & $-6.4$ & $-13.9$ & $7$ & 150 & 1.1 & 8 & $<0.8$ \\
& $r$ & $-13.4$ & $-6.4$ & $-14.6$ & $30$ & & & & \\
\rule{0pt}{3ex}
SN\,2019fmb & $g$ & $-125.3$ & $-27.3$ & $-12.9$ & $40$ & 990 & 0.08 & $-$ & $-$ \\
& $r$ & $-174.3$ & $-20.3$ & $-13.3$ & $80$ & & & & \\
& $i$ & $-111.3$ & $-41.3$ & $-14.1$ & $80$ & & & & \\
\rule{0pt}{3ex}
SN\,2019iay & $g$ & $-34.2$ & $-13.2$ & $-14.7$ & $40$ & 340 & 0.4 & 9 & $<0.9$ \\
& $r$ & $-34.2$ & $-13.2$ & $-14.8$ & $50$ & & & & \\
\rule{0pt}{3ex}
SN\,2019gjs & $g$ & $-97.2$ & $-20.2$ & $-15.3$ & $300$ & 320 & 3 & 7 & $<0.7$ \\
& $r$ & $-104.2$ & $-20.2$ & $-15.6$ & $300$ & & & & \\
& $i$ & $-55.2$ & $-48.2$ & $-15.5$ & $30$ & & & & \\
\rule{0pt}{3ex}
SN\,2019mom & $g$ & $-342.5$ & $-335.5$ & $-14.7$ & $13$ & 590 & 0.19 & $-$ & $-$ \\
& $r$ & $-342.5$ & $-335.5$ & $-15.1$ & $19$ & & & & \\
& $g$ & $-6.5$ & $0$ & $-16.0$ & $50$ & & & & \\
& $r$ & $-6.5$ & $0$ & $-16.0$ & $50$ & & & & \\
\rule{0pt}{3ex}
SN\,2019aafe & $g$ & $-20.4$ & $-6.4$ & $-16.9$ & $300$ & 1100 & 0.9 & 4 & $<0.4$ \\
& $r$ & $-97.4$ & $-6.4$ & $-16.1$ & $1000$ & & & & \\
\rule{0pt}{3ex}
SN\,2019vkl & $g$ & $-62.2$ & $-6.2$ & $-16.0$ & $500$ & 770 & 1.3 & 10 & $<1.0$ \\
& $r$ & $-83.2$ & $-6.2$ & $-16.1$ & $800$ & & & & \\
\rule{0pt}{3ex}
SN\,2019vts & $g$ & $-41.5$ & $-20.5$ & $-14.6$ & $40$ & 340 & 1.2 & $-$ & $-$ \\
& $r$ & $-62.5$ & $-6.5$ & $-14.9$ & $140$ & & & & \\
\rule{0pt}{3ex}
SN\,2019qny & $g$ & $-104.6$ & $-90.6$ & $-15.4$ & $50$ & 350 & 2 & 25 & $<2.5$ \\
& $r$ & $-125.6$ & $-62.6$ & $-15.4$ & $300$ & & & & \\
\rule{0pt}{3ex}
SN\,2020iq & $r$ & $-482.2$ & $-454.2$ & $-15.9$ & $170$ & 160 & 7 & $-$ & $-$ \\
\rule{0pt}{3ex}
SN\,2019zrk & $g$ & $-55.5$ & $-6.5$ & $-16.2$ & $500$ & 350 & 5 & 7 & $<0.7$ \\
 & $r$ & $-111.5$ & $-6.5$ & $-16.1$ & $700$ & & & & \\
\rule{0pt}{3ex}
SN\,2020dcs & $r$ & $-34.4$ & $-6.4$ & $-15.1$ & $90$ & 180 & 3 & 14 & $<1.4$ \\
\rule{0pt}{3ex}
SN\,2020dfh & $r$ & $-34.6$ & $-20.6$ & $-15.2$ & $40$ & 200 & 1.1 & $-$ & $-$ \\
\rule{0pt}{3ex}
SN\,2020edh & $g$ & $-692.5$ & $-545.5$ & $-13.0$ & $80$ & 600 & 0.2 & $-$ & $-$ 
\enddata
\tablecomments{Properties of the detected precursors and the SNe. The first columns list the beginning and end of precursors with respect to $t_0$, the median magnitudes and precursor energies. The CSM velocity is derived from the median width of narrow lines and P~Cygni profiles (see Sec.~\ref{sec:sne_history}) and is used to estimate the CSM mass multiplied by an unknown efficiency factor $\epsilon$. $t_{\text{rise}}$ quantifies how many days it takes the SN to rise by a factor of $e$ (1.086 mag) to its peak in the $r$ band ($g$ band used for SN\,2018gho and SN\,2019iay). The rise time provides a rough upper limit on the total CSM mass given in the last column (see Sec.~\ref{sec:sne_corr}).}
\end{deluxetable*}

\subsection{Precursor Colors}
\label{sec:pre_colors}

\begin{figure*}[tb]
    \centering
\includegraphics[width=\textwidth]{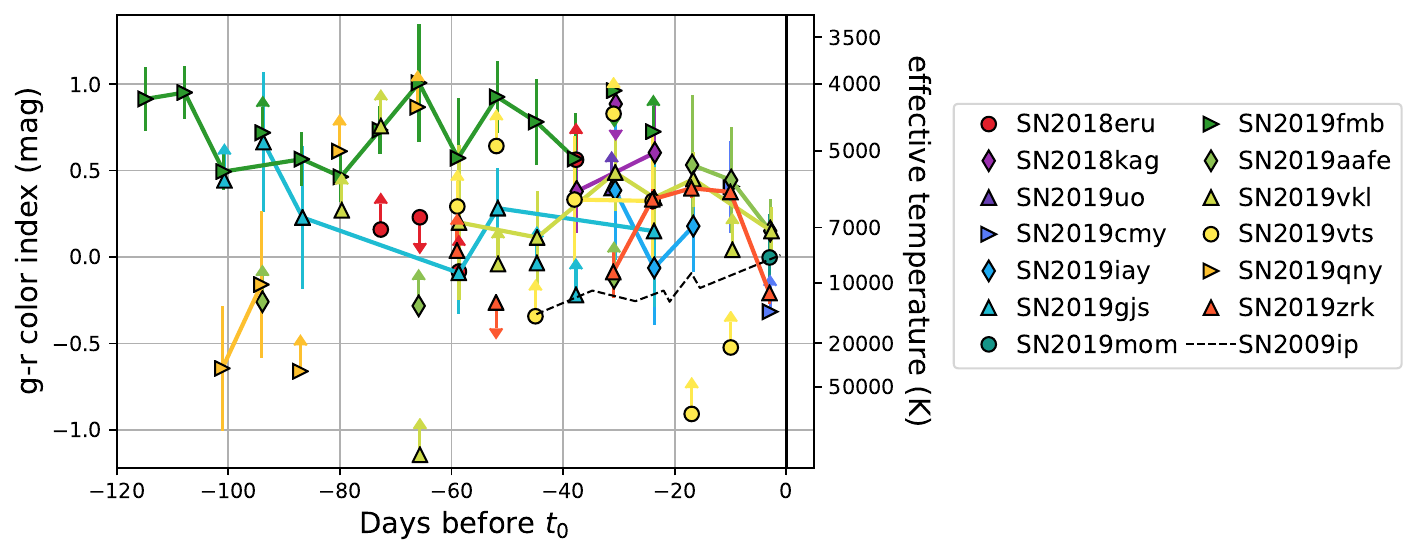}
\caption{\label{fig:pre_colors} Color index of precursors observed in the $g$ and $r$ bands. The light curves are binned in 7-day bins and we quote $3\sigma$ lower or upper limits instead of detections if the $g$-band or $r$-band flux is not significant at the $3\sigma$ level.}
\end{figure*}

Here, we calculate the $g-r$ color index for precursors that have observations in both bands. For this purpose we select all bins in which a significance of $3\sigma$ is reached in at least one band. If the detection in the second band is less significant we quote lower or upper limits accordingly. The resulting colors are shown in Fig.~\ref{fig:pre_colors}.
Compared to young SNe, the precursors exhibit quite red colors, which correspond to lower effective temperatures. 
We caution, however, that the $\text{H}\alpha$ line falls within the $r$ band. The spectrum of the precursor prior to PTF\,13efv showed relatively strong, narrow hydrogen lines \citep{ofek2016}, and the same is true for Type IIn SNe and LBV outbursts. A red color could therefore be mimicked by a blue continuum flux with a strong $\text{H}\alpha$ line. The precursor prior to SN\,2019fmt is also detected in the $i$ band and shows a mean $g-i$ color of 1.1 mag. This corresponds to an effective temperature of $\sim4300\,\text{K}$, similar to the result from the $g-r$ color index shown in Fig.~\ref{fig:pre_colors}. For this object, at least, we conclude that the rather low effective temperature is not primarily due to a strong $\text{H}\alpha$ line.

We also show the effective temperatures of the 2012a outburst of SN\,2009ip in Fig.~\ref{fig:pre_colors}. They were obtained by fitting a blackbody continuum to the multiband photometry \citep{margutti2013} and are therefore less susceptible to line fluxes. The precursor of SN\,2009ip is slightly hotter than most precursors observed in our sample, and we find that the precursors detected here typically do not cool down as observed for the 2012a event prior to the final explosion of SN\,2009ip \citep{margutti2013}.

If the precursor's bolometric luminosity $L$ and temperature $T$ are known, photospheric radii can be estimated via the Stefan-Boltzmann law $R=({L}/{4\pi\sigma_{\text{B}}})^{0.5} T^{-2}$, where $\sigma_B$ is the Stefan-Boltzmann constant.
A faint and hot precursor (with a temperature of $8,000\,\text{K}$ and a bolometric magnitude of $-13$) would have a photosphere with a small radius of $\sim4\times10^{13}\,\text{cm}$, while a bright and cool precursor ($4,000\,\text{K}$ and a magnitude of $-17$) would have a radius of $\sim10^{15}\,\text{cm}$. When using $r$-band luminosities (shown in Fig.~\ref{fig:shapes}) as order of magnitude estimates for the precursor bolometric luminosity and the $g-r$ color index as a crude temperature estimate, we find that most detected precursors have photospheric radii of a few times $10^{14}\,\text{cm}$. These large radii suggest that we cannot see down to the surface of the progenitor star. 

\section{Precursor Rates}
\label{sec:rates}

Here we focus on the whole sample of pre-explosion light curves and use it to calculate precursor rates. Except for one, all confirmed precursors are found prior to Type IIn SNe and we therefore first describe the rate for this SN class in Sec.~\ref{sec:rates_iin} and Sec.~\ref{sec:rates_timedep}. Precursor rates for other types of possibly interacting SNe are presented in Sec.~\ref{sec:rates_subsamples}.

\subsection{Precursor Rates for Type IIn SNe}
\label{sec:rates_iin}

\begin{figure*}[tb]
\centering
\includegraphics[width=\textwidth]{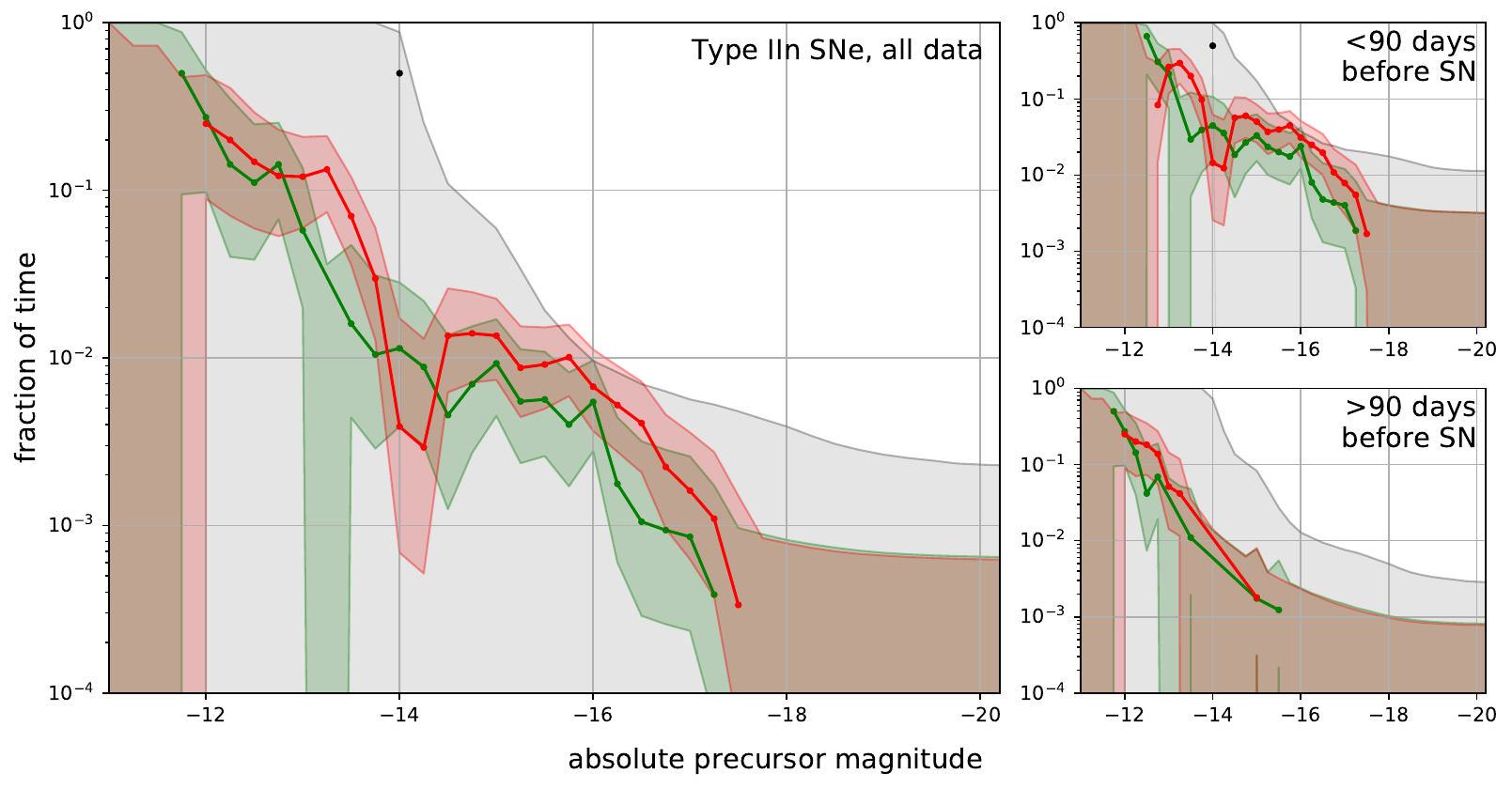}
\caption{Fraction of time during which precursors brighter than the respective absolute magnitude are observed for Type IIn SNe. The lines show the measured rates and the shaded area indicates the 95\% confidence region. If it reaches down to zero, its upper edge should be interpreted as an upper limit; otherwise it corresponds to the 95\% error bar on the rate. The calculation was done for $7$-day bins; the green, red, and black colors show the rates for the $g$, $r$, and $i$ bands, respectively. The $i$-band precursor rate is only measured at magnitude $-14$, as indicated by the black dot. In the main panel, the median phase of the observations is $\sim6$ months prior to the explosion, while it is $1.4$ months and $10.4$ months (respectively) for the two right panels. The $r$-band precursor rate measured in the last 90 days before the SN explosion (upper small panel) is typically 6 times larger than the 95\% upper limit on the rate measured at earlier times (lower small panel).}
\label{fig:rates_vs_time}
\end{figure*}

The rate calculation is done for 7-day bins, because this search channel is sensitive to faint precursors without losing short precursors (see Fig.~\ref{fig:precursor_overview}). Another advantage of using 7-day bins is that they partly compensate for differences between light curves obtained by the private and public surveys, which have typical cadences of one day and three days, respectively. None of the unconfirmed precursors is detected for 7-day bins, so they do not enter the rate calculation.

The precursor rate is here defined as the fraction of time during which precursors are observed above a certain limiting magnitude. As a result, we do not distinguish between two 1-week long precursors and a single precursor that lasts for two weeks. The rate depends on the absolute magnitude of the precursors and we calculate it in steps of $0.25$ mag. For each absolute magnitude we select all pre-explosion bins with a deeper limiting magnitude. We then calculate which fraction of these bins have precursor detections. Consequently, precursors detected with a high significance (i.e., a large difference between its magnitude and the limiting magnitude of the bin) may contribute in several magnitude bins. On the other hand, detections just at the $5\sigma$ threshold may not contribute at all, if they fall in between the magnitude steps\footnote{For example, a precursor detected with an absolute magnitude of $-14.2$ and with a limiting magnitude of $-14.1$ would not count as a detection in the bin at magnitude $-14.25$ because it is not bright enough. In the next fainter bin at a magnitude $-14$ it also does not contribute because the limiting magnitude is not sensitive enough.}. The resulting rate is cumulative, as we search for precursors that are brighter than the corresponding magnitude threshold.
The 95\% uncertainty associated with the rate is calculated using the Wilson binomial confidence interval \citep{wilson1927, wallis2013} as implemented in the \emph{astropy} package \citep{astropy2013, astropy2018}.

The main panel of Fig.~\ref{fig:rates_vs_time} shows the fraction of time during which precursors are detected for Type IIn SNe as a function of the absolute magnitude. The green, red, and gray shaded regions correspond to the parameter space that is allowed at the 95\% confidence level for the $g$, $r$, and $i$ bands (respectively), and the solid lines depict the cumulative precursor rate. For bins without detections (e.g., for bright absolute magnitudes) the colored area reaches down to zero and its upper edge corresponds to a 95\% upper limit.

As shown in the main panel of Fig.~\ref{fig:rates_vs_time}, $g$-band and $r$-band precursors are detected with absolute magnitudes ranging from $-17$ to $-12$. The rate is slightly lower in the $g$ band, because of the red precursor colors observed in Sec.~\ref{sec:pre_colors}. The fraction of time during which we observe bright precursors with an absolute $r$-band magnitude of $-16$ or brighter is $\sim0.7\%$ with a 95\% confidence range of 0.4\% to 1.2\%. For fainter magnitudes, the rate increases and reaches $\sim12\%$ (6--23\%) for precursors brighter than magnitude $-13$. The measured rates are also summarized in Table~\ref{tab:rates}.

We caution that $r$-band precursors fainter than magnitude $-14$ are only detected for SN\,2019fmb, so the rate of such faint precursors is determined by this object and by the fact that few other SNe have as constraining observations. The $g$-band rate is more robust, since such faint precursors are detected for four different SNe. The dip in the $r$-band rate at magnitude $-14$ is likely a statistical fluctuation caused by the relatively small number of SNe with precursors. The gray shaded region indicates that the $i$-band observations are typically not sensitive enough to detect precursors. The reason is that fewer observations were obtained and they have in addition larger error bars, in part owing to the lower quantum efficiency in this wavelength range for the ZTF CCD \citep{bellm2019}. A black dot marks the only $i$-band detection, a precursor with magnitude $-14$ prior to SN\,2019fmb.

\begin{deluxetable*}{l c c c c c}
\tablecaption{Precursor rates\label{tab:rates}}
\tablewidth{0pt}
\tablehead{
\colhead{sample} & \colhead{band} & \colhead{number of SNe} & \colhead{median phase} & \colhead{rate of bright pre. ($\leqslant-16$ mag)} & \colhead{rate of faint pre. ($\leqslant-13$ mag)} \\
\colhead{} & \colhead{} & \colhead{} & \colhead{(months)} & \colhead{(\%)} & \colhead{(\%)}
}
\startdata
Type IIn, all data & $g$ & 122 & $-8.7$ & $0.5\ (0.3 - 1.1)$ & $6\ (2-15)$ \\
& $r$ & 126 & $-8.8$ & $0.7\ (0.4 - 1.2)$ & $12\ (6-23)$ \\
& $i$ & 49 & $-9.0$ & $<0.9$ & $-$ \\
\rule{0pt}{4ex}Type IIn, $\leqslant90$ days before SN & $r$ & 107 & $-1.3$ & $3\ (1.7 - 6)$ & $26\ (12-49)$ \\
Type IIn, $>90$ days before SN & $r$ & 121 & $-10.4$ & $<0.2$ & $5\ (1.4-17)$ \\
\rule{0pt}{4ex}bright Type IIn (peak mag. $<-18.5$) & $r$ & 84 & $-8.5$ & $2\ (1.2 - 4)$ & $<73$ \\
faint Type IIn (peak mag. $>-18.5$) & $r$ & 33 & $-8.4$ & $<0.3$ & $12\ (6-12)$ \\
\rule{0pt}{4ex}Type Ibn & $r$ & 11 & $-8.6$ & $<1.0$ & $<31$ \\
SLSNe-II & $r$ & 24 & $-8.4$ & $<72$ & $-$ \\
flash-spectroscopy SNe & $r$ & 20 & $-4.4$ & $<0.5$ & $<2.7$ \\
Type Ia-CSM & $r$ & 7 & $-7.2$ & $<5.0$ & $-$
\enddata
\tablecomments{Fraction of time during which bright or faint precursors are observed with the $95\%$ confidence range given in parentheses. If no precursors are detected the $95\%$ upper limit is quoted instead. The calculation was done for 7-day bins and the numbers are taken from Figs.~\ref{fig:rates_vs_time}, \ref{fig:rates_snlumi}, and \ref{fig:rates_subsample}. The number of SNe with data is given in the third column, and the fourth column lists the median phase of the pre-explosion observations which is close to nine months for most subsamples. Dashes indicate that no data are available, so the rate remains unconstrained (e.g., the rate of faint precursors in the $i$ band).}
\end{deluxetable*}

\subsection{Time Dependence of the Precursor Rate for SNe IIn}
\label{sec:rates_timedep}

The rate calculation in the left-hand panel of Fig.~\ref{fig:rates_vs_time} was done using all pre-explosion data that were collected over a period of up to $2.5$\,yr before each SN explosion. The median phase of the pre-explosion light curves is 267 days (nearly nine months) before the discovery date $t_0$. The precursor light curves in Fig.~\ref{fig:shapes} show that most precursors are detected in the final few months before the SN explosion. To quantify the time dependency, we split the dataset into two parts: observations collected within 90\,days before the estimated explosion date (with a median of 42\,days) and observations collected earlier (at a median time of 317\,days before the SN). We then repeat the rate calculation and display the results in the two side panels of Fig.~\ref{fig:rates_vs_time}.

The $r$-band 95\% confidence regions in the two smaller panels of Fig.~\ref{fig:rates_vs_time} do not overlap for absolute magnitudes $>-14.5$ and the precursor rate is significantly larger in the final 90\,days before the explosion. The measured rate in the final months before the explosion is up to 16 times larger than the 95\% upper limit on the precursor rate before that. In the $r$ band, the median difference for all magnitude bins between magnitude $-13$ and $-17.5$ is a factor of 6 (i.e., the precursor rate at early times is typically more than 6 times smaller).
The difference would be even larger when dividing the dataset at 120\,days, because several of the detections in the lower-right panel of Fig.~\ref{fig:rates_vs_time} are part of the $>100$-day long precursors (e.g., prior to SN\,2019fmb and SN\,2019gjs; see Fig.~\ref{fig:shapes}). The large number of precursor detections shortly prior to the explosion is hence not caused by the larger amount of data available at these times, but is a genuine and significant difference.

In the three months before the explosion, faint precursors with an $r$-band magnitude of $>-13$ are observed $26\%$ of the time (with a 95\% confidence range of 12--49\%; see also Table~\ref{tab:rates}), while the rate is $5\%$ (1.4--17\%) at earlier times. The time dependence of the rate is even stronger for brighter precursors with absolute magnitudes $>-16$: their rate is $3\%$ (1.7--6\%) in the three months before explosion, while it is $<0.2\%$ prior to that. We conclude that precursors become brighter and more frequent in the final months leading up to the explosion. 

\begin{figure}[tb]
    \centering
\includegraphics[width=\columnwidth]{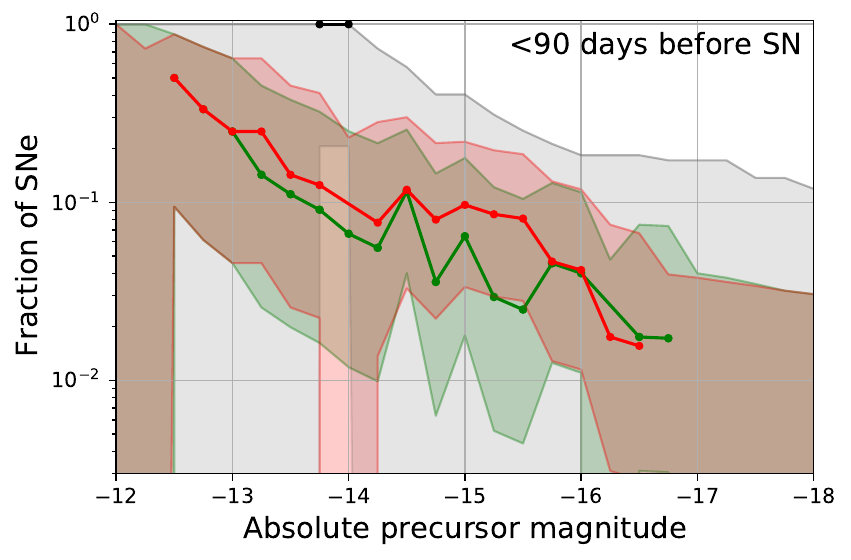}
\caption{\label{fig:rate_per_sn}Fraction of Type IIn SNe with long-lasting ($\gtrsim30$ days) precursors within 90\,days prior to their explosion. The lines mark measured rates and the shaded area indicates the 95\% confidence regions.}
\end{figure}

Early precursors are only observed prior to five SNe (see Fig.~\ref{fig:shapes}) and they appear to be fainter and short-lived compared to the precursors immediately before the explosion that typically last for several months. The rate calculation in Fig.~\ref{fig:rates_vs_time} shows that this effect is real and not caused by a smaller number of observations at early times. The luminosity increase likely continues within the last three months before the explosion as shown in Fig.~\ref{fig:shapes}.

While we so far constrained the fraction of time during which precursors are observed, we here calculate in addition the fraction of progenitor stars that undergo long-lasting precursors within 90\,days before the SN explosion. For this purpose, we compile a sample of SNe for which such precursors are detectable. We bin the light curves in 30-day bins and require that at least two bins contain data (i.e., that observations are available in two out of three months). If this condition is met, we estimate down to which limiting magnitude a precursor can be detected. For this purpose we use the second lowest limiting magnitude --- that is, the median for three data points or the least constraining bin for two data points.
The rate is then calculated for each magnitude bin by dividing the number of detected precursors by the number of light curves for which such a precursor would have been detectable.

The fraction of SNe with long-lasting precursors in the last three months before the explosion is shown in Fig.~\ref{fig:rate_per_sn}. Long-lasting precursors brighter than magnitude $-16$ occur for about $4\%$ (1.1--14\%, 95\% confidence range) of the SNe in the $r$ band, while fainter precursors with an absolute magnitude brighter than $-13$ occur for $25\%$ (5--69\%) of the Type IIn SNe. The $i$-band precursor rate is unity at magnitude $-14$, but it is purely determined by SN\,2019fmb as no other SN has as constraining observations. The rate is here detected in two magnitude bins because the 30-day-long light curve bins yield deeper limiting magnitudes than the 7-day-long bins used in Fig.~\ref{fig:rates_vs_time}.

We hence conclude that precursor eruptions brighter than magnitude $-13$ occur prior to many, but not all Type IIn SNe. This result is in tension with some of the findings by \citet{ofek2014}, who calculate that the average Type IIn progenitor undergoes several precursors brighter than magnitude $-14$ in the last year before its explosion. Based on this they estimate that $>52\%$ of all Type IIn SNe exhibit at least one bright precursor in the final four months before the explosion at a confidence level of $99\%$. \citet{ofek2014} calculate the precursor rate by dividing the number of precursors by the time during which such precursors are detectable, the so-called ``control time." However, if the light curve has gaps, the control time (and thus the rate) depends on the bin size while the number of precursors does not change, as long as the bin size is smaller than their duration. 
To avoid such a dependence on the bin size, we calculate instead the fraction of bins with precursors or the fraction of well-observed SNe with precursors. The rate calculation used by \citet{ofek2014} and \citet{strotjohann2015} are thus only valid if each light-curve bin contains observations.

Our results are likely consistent with the findings of \citet{bilinski2015}, who did not detect any precursors for a sample of five Type IIn SNe and one SN imposter. They report that a precursor similar to the 2012a event prior to the explosion of SN\,2009ip would have been detectable for two of their objects. We measure that $\sim10\%$ of all Type IIn SNe have precursors as bright as magnitude $-15$ (see Fig.~\ref{fig:rate_per_sn}) which is consistent with their non-detections. \citet{bilinski2015} do not quote a control time, so we cannot compare to all of their results.

\begin{figure*}[tb]
\centering
\includegraphics[width=8.5cm]{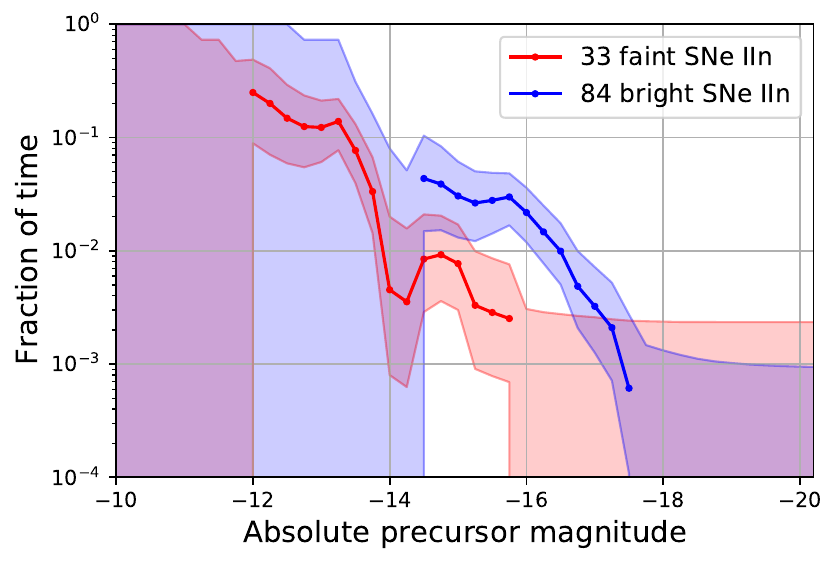} 
\includegraphics[width=8.5cm]{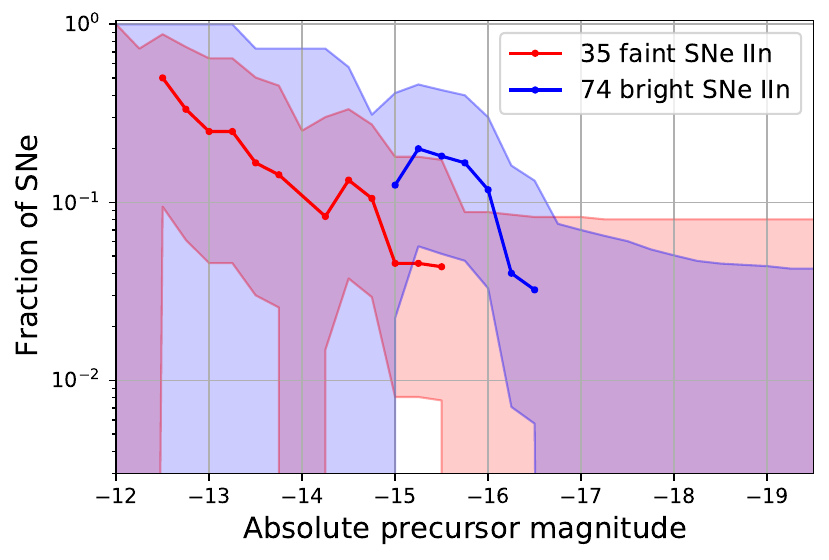}\\
\caption{$r$-band precursor rates for faint and bright (peak magnitude $<-18.5$ in the $g$, $r$, or $i$ band) SNe of Type IIn. The left plot displays the fraction of light-curve bins in which precursors are detected, which means that the rate depends on the precursor duration (like Fig.~\ref{fig:rates_vs_time}). The right figure is only based on well-observed SNe and the rate indicates whether a long-lasting precursor is detected within the last three months before the explosion (like in Fig.~\ref{fig:rate_per_sn}).}
\label{fig:rates_snlumi}
\end{figure*} 

We conclude that the precursor rate increases by a factor of more than six within the last three months before the SN explosion compared to earlier observations obtained on average ten months before the SN. While the rate of faint precursors (with an $r$-band magnitude of $-13$) increases by a factor of $\sim5$, the difference is more than a factor of $10$ for bright precursors with an $r$-band magnitude of brighter than $-16$. Our observations do not constrain the rate of long-lasting precursors that are fainter than magnitude $-13.5$. It is hence possible that all progenitors of Type IIn SNe exhibit precursors if at least one third of them are fainter than this threshold.

\subsection{Precursor Rates for Faint and Bright SNe IIn}
\label{sec:rates_snlumi}

Type IIn SNe can have diverse peak luminosities and SN energies. Here, we split the sample of Type IIn SNe into  bright and  faint subsamples to test whether they have similar precursor rates. We consider a SN bright if it reaches an absolute magnitude of $-18.5$ in any ZTF band. This threshold is chosen such that the measured precursor rates are relatively well constrained in both subsamples. Detections in all three bands are considered, because some SN light curves only have sparse observations, especially if their peak occurred in the year 2020, for which part of the data has not yet been released (see Sec.~\ref{sec:fp_pipeline}).

We compare the $r$-band precursor rates for bright and faint SNe in Fig.~\ref{fig:rates_snlumi} and the subsample of bright SNe has a higher rate of bright precursors. The rate of faint precursors is not well constrained for the bright SN sample, because most objects in this subsample are located at large distances. The rates could therefore agree below an absolute magnitude of $-14$ (see also Table~\ref{tab:rates}). The difference between the bright and faint sample is relatively strong in the left-hand panel of Fig.~\ref{fig:rates_snlumi}, which shows the rate as the fraction of time during which the progenitor stars undergo precursors and thus depends on the precursor duration (see also Sec.~\ref{sec:rates_iin}). In the right-hand panel of the figure, we show instead the fraction of SNe that undergo a long-lasting precursor immediately before the explosion (like in Fig.~\ref{fig:rate_per_sn}) and the difference is not significant any more. A possible explanation for this change could be that bright precursors have longer durations. Indeed, the three brightest precursors in Fig.~\ref{fig:shapes} are all observed for $\sim100$ days. We hence find indications that luminous SNe typically undergo brighter and longer-lasting precursors. This correlation is quantified in Sec.~\ref{sec:sne_corr}.

\subsection{Precursor Rates for Different Interacting SNe}
\label{sec:rates_subsamples}

\begin{figure*}[tb]
\centering
\includegraphics[width=8.5cm]{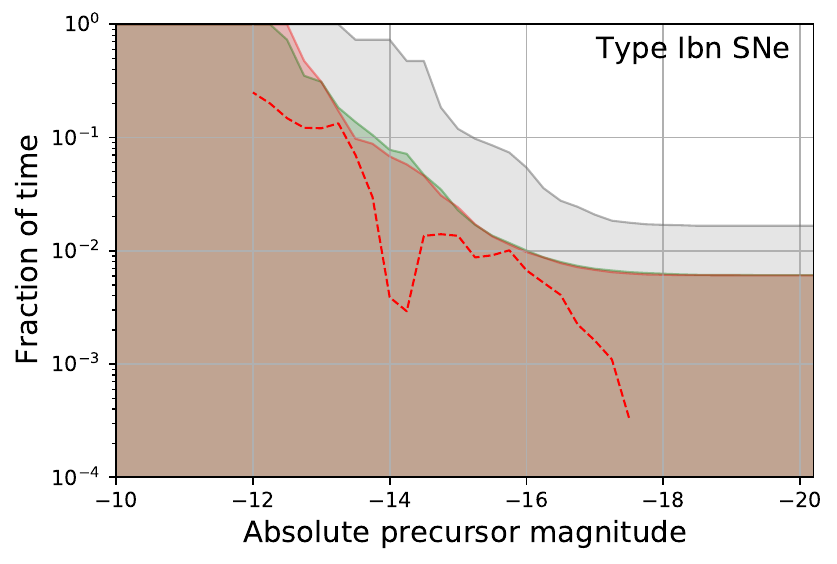} 
\includegraphics[width=8.5cm]{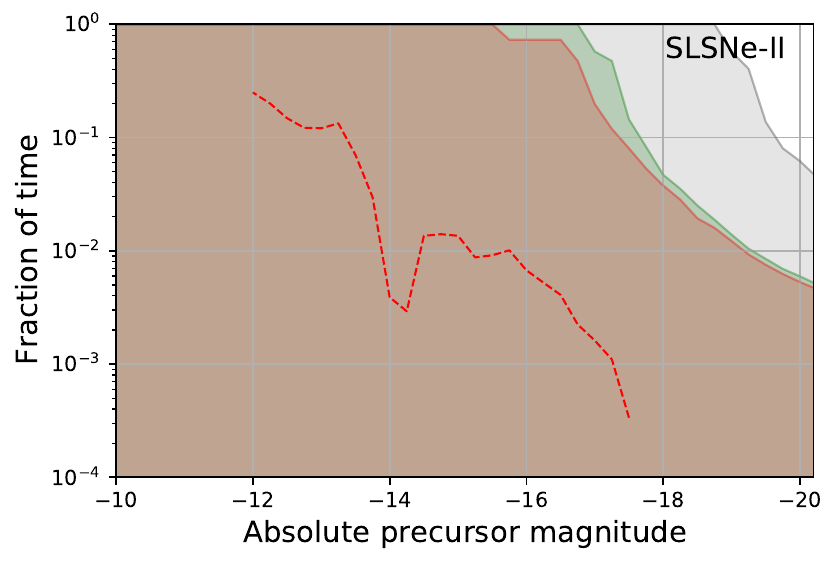}\\
\includegraphics[width=8.5cm]{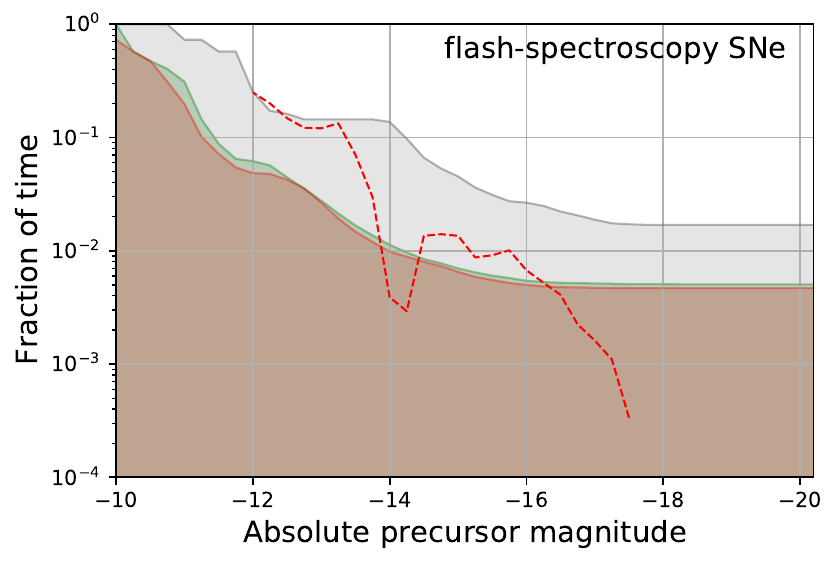}
\includegraphics[width=8.5cm]{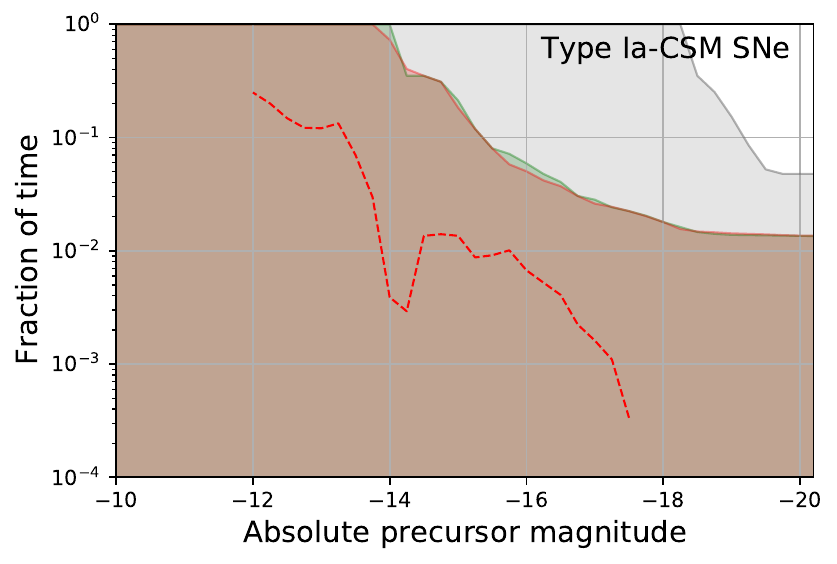}\\
\caption{95\% upper limits on the precursor rates for 7-day bins for Type Ibn SNe, SLSNe-II, flash-spectroscopy events, and Type Ia-CSM SNe. The shaded areas indicate the 95\% confidence area and the area above it excluded by the nondetection of precursors. The $g$ and $r$-band observations are nearly equally sensitive such that the upper limits fall on top of each other, but the $i$-band is less constraining. The red dashed line indicates the $r$-band rate that was measured for Type IIn SNe in the main panel of Fig.~\ref{fig:rates_vs_time}.}
\label{fig:rates_subsample}
\end{figure*}

As described in Sec.~\ref{sec:fp_sample}, our full sample also contains interacting SNe that do not belong to the class of Type IIn SNe. Here, we present precursor rates for SNe of Type Ibn (based on 12 objects for which pre-explosion observations are available; see the online version of Table~\ref{tab:sample}), SLSNe-II (26 objects after excluding SN\,2019meh which falls on top of a background AGN), flash-spectroscopy SNe (20 objects), and Type Ia-CSM SNe (7 objects). The number given in Table~\ref{tab:rates} can be lower as not all SNe have pre-explosion data in the $r$-band.
Flash-spectroscopy events are here defined as objects showing narrow \ion{He}{2} lines in their early-time spectra up to a week after the discovery. There is some overlap between flash-spectroscopy SNe and the other classes: Some flash-spectroscopy SNe show narrow hydrogen lines for several weeks and are here included in the sample of Type IIn SNe (such as SN\,2019cmy). SN\,2019uo, is considered a Type Ibn SN, even though it might show flash-spectroscopy lines at early times \citep{gangopadhyay2020}.

We calculate the fraction of time during which precursors are observed in the same way as in Sec.~\ref{sec:rates_iin} and show the results for each subsample in Fig.~\ref{fig:rates_subsample} for 7-day bins. The precursor detected prior to the Type Ibn SN\,2019uo (described in more detail in Sec.~\ref{sec:sne_sn2019uo}), does not appear because it is marginally above the $5\sigma$ threshold. An unconfirmed precursor is detected $680$\,days before the explosion of the Type Ia-CSM SN\,2019yzx, as shown in Fig.~\ref{fig:precursor_lcs2}. However, its significance is purely driven by observations in a single night while the two neighboring data points are consistent with zero. The location is observed relatively sparsely, so we cannot confirm whether the detection is real. We here conservatively assume that the detection is not astrophysical.

For comparison, the measured $r$-band precursor rate for Type IIn SNe (from the main panel of Fig~\ref{fig:rates_vs_time}) is shown as a dashed red line in Fig.~\ref{fig:rates_subsample}. The Type IIn rate is nearly always in the allowed region of parameter space, which means that we do not expect to detect any precursors even if the rates are as high as for Type IIn SNe. The lower sensitivity is due to the small sample size, or in the case of SLSNe to the fact that the objects are located at large distances (see also Table~\ref{tab:rates}).
The only region where the Type IIn SN rate is higher than the upper limit is for the sample of flash-spectroscopy SNe at faint precursor magnitudes of $>-14$. However, in this region the Type IIn SN rate is completely dominated by SN\,2019fmb and we therefore consider it less reliable.

Thus, we conclude that we only observe a single precursor that was not associated with a Type IIn SN, but with the Type Ibn SN\,2019uo. However, this small number of detections is expected owing to the small sample sizes of the subclasses and to the large distances of SLSNe.

\section{Impact of the Precursors on the SNe}
\label{sec:sne}

In this section we explore the impact of the observed precursors on the SN spectra and light curves. In Sec.~\ref{sec:sne_history} we consider whether the narrow lines in the SN spectra can be produced by material emitted during the observed precursors.
Next, in Sec.~\ref{sec:sne_corr} we test whether SNe with observed precursors are brighter than other SNe in our sample. Finally, in Sec.~\ref{sec:sne_sn2019uo} we describe how the precursor prior to the Type Ibn SN\,2019uo could account for both the SN light curve and the spectral evolution of this object.

\subsection{Progenitor Mass-Loss History}
\label{sec:sne_history}

\begin{figure*}[p]
\centering
\includegraphics[width=0.9\textwidth]{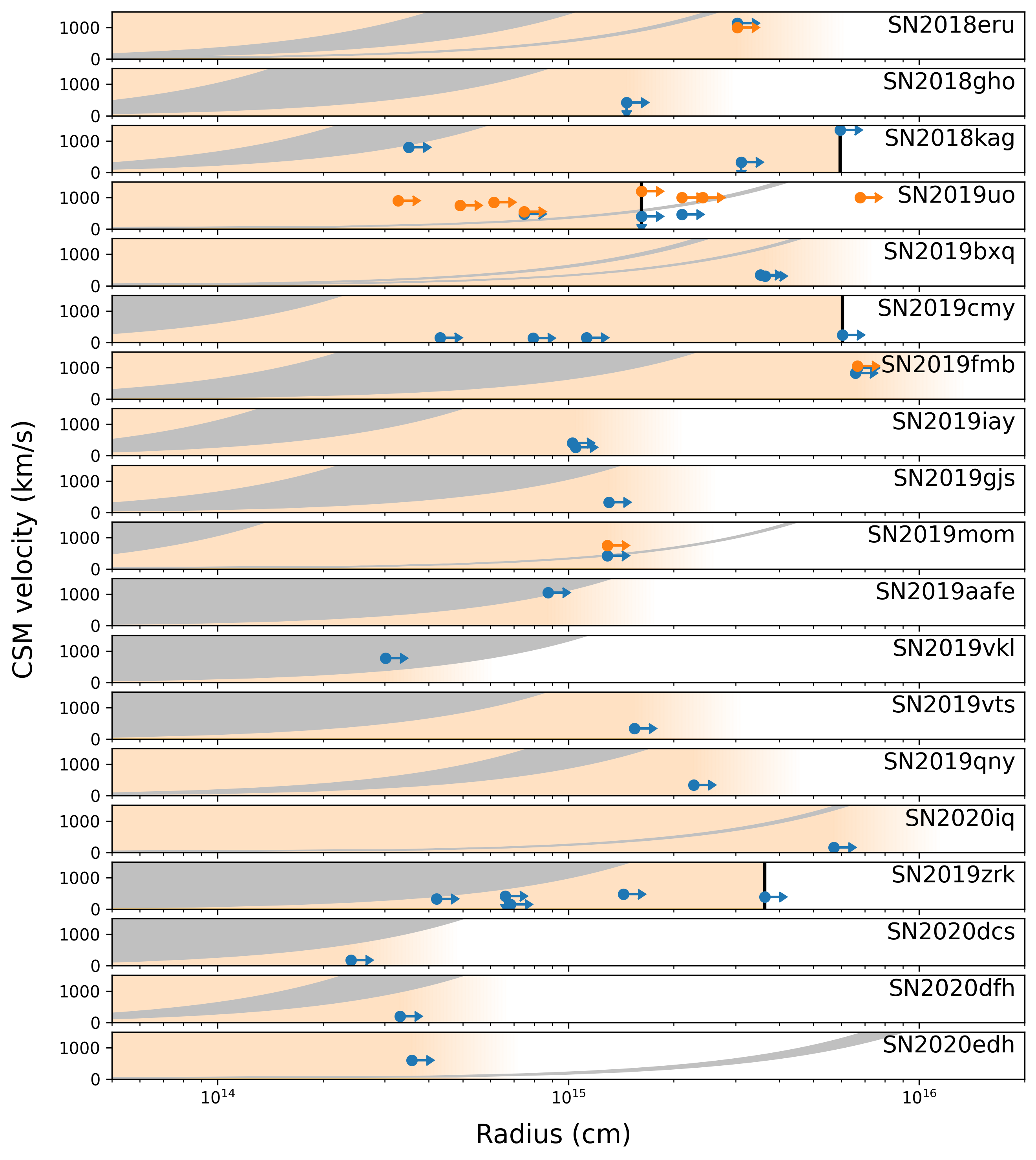} \caption{\label{fig:interaction_sig}Radial CSM distribution for SNe with precursors. The gray regions indicate the possible locations of material ejected during the observed precursors depending on its velocity (shown on a linear scale).
Data points correspond to spectra in which we observe narrow lines (blue points) or narrow P~Cygni profiles (orange points). The ordinate represents the measured velocity while the abscissa is a lower limit on the radius of the material. The unshocked material must be located above the ejecta for which we assume a fiducial mean velocity of $10^4\,\text{km}\,\text{s}^{-1}$. For four SNe, black lines indicate that broad features from shocked material or the SN ejecta emerge, which means that the unshocked CSM is no longer optically thick. All other SNe are still optically thick at the time when the last spectrum is obtained and the shaded area represents a lower limit on the extension of the CSM.
}
\end{figure*}

A massive star of $100\,M_\odot$ reaches its Eddington luminosity when it becomes brighter than $1.3\times10^{40}\,\text{ergs}\,\text{s}^{-1}$. For a hot LBV star with a temperature of $20,000\,\text{K}$ (see, e.g.,~\citealt{smith2004}) this luminosity corresponds to an absolute $r$-band magnitude of $-8.1$, while it is $-10.0$ for a temperature of $5,000\,\text{K}$ which is more similar to the temperatures observed for the precursors in Fig.~\ref{fig:pre_colors}. Figure~\ref{fig:shapes} shows that the luminosities of all detected precursors are clearly above this threshold, so the outbursts are likely accompanied by strong mass-loss events.

By detecting the precursor, we measure the time of the mass ejection; however, the velocity of the material is uncertain (see also Sec.~\ref{sec:nature_interaction}). The gray regions in Fig.~\ref{fig:interaction_sig} indicate out to which radii the material has expanded at the time of the SN explosion, depending on its velocity. CSM velocities between 0 and 1,500\,$\text{km}\,\text{s}^{-1}$ are shown on the linear ordinate axis. We here assume that the material was ejected from a radius of $100\,R_\odot$, $7\times10^{12}\,\text{cm}$. Using a radius that is a factor of a few larger or smaller does not have a major impact on the results, as spectra are usually obtained after the SN ejecta have expanded far beyond this radius.

Most detected precursors occur within the last few months before the explosion, so any ejected material is still located within a radius of $10^{15}\,\text{cm}$ even if it has a velocity of $\lesssim1,000\,\text{km}\,\text{s}^{-1}$. Earlier precursors are only observed for six SNe (SN\,2018eru, SN\,2019uo, SN\,2019bxq, SN\,2019mom, SN\,2020iq, and SN\,2019edh; see Figs.~\ref{fig:precursor_lcs}, \ref{fig:precursor_lcs2}, and~\ref{fig:shapes}). The material ejected in these precursors might be located at radii of a few $\times 10^{15}\,\text{cm}$, but likely below $10^{16}\,\text{cm}$.

Next, we consider the narrow emission lines in the spectra of the SNe. To estimate CSM velocities, we measure the full width at half-maximum intensity (FWHM) of the narrow component of the $\text{H}\alpha$ line. We subtract the approximate resolution of the spectrograph in quadrature or quote upper limits if the result is smaller than half of the resolution. In addition, we look for narrow P~Cygni features in the $\text{H}\alpha$ line (He lines for the Type Ibn SN\,2019uo), as their minimum indicates the typical velocity of material moving toward the observer. The results for all spectra with clear narrow features are listed in Table~\ref{tab:spectroscopy}. The quoted velocities are only order-of-magnitude estimates as we do not fit line profiles, measure the actual resolution of the spectra, or subtract host-galaxy contributions.

The exact location of the material that produces the narrow features is unknown, but the time when the spectrum was obtained provides an order-of-magnitude lower limit on its radius. Narrow features can only originate from unshocked material, which must be located at larger radii than the SN ejecta. In order to estimate these radii, we adopt a fiducial average ejecta velocity of $10^4\,\text{km}\,\text{s}^{-1}$, which is close to the width of the broad hydrogen features observed in the late-time spectra of SN\,2018kag, SN\,2019cmy, and SN\,2019zrk. To estimate out to which radius the ejecta have approximately expanded we multiply this velocity by the time since the explosion. The resulting distances and CSM velocities are represented by the data in Fig.~\ref{fig:interaction_sig}, where blue points indicate velocities measured from the line width while orange points indicate the velocities of narrow P~Cygni profiles. We emphasize that both the radii and velocities are rough estimates. 

For most SNe, the data points are located below or to the right of the gray shaded region which indicates the location of the CSM produced during the observed precursor. This implies that the material ejected during the precursor cannot account for the observed narrow emission lines, because it would be located at smaller radii if it propagates with the observed velocity. Instead, it is more likely that the emission lines are produced by slow-moving material that was expelled earlier. This conclusion is exclusively based on the distance out to which the SN ejecta have expanded at a certain time and is therefore also valid for aspherical CSM distributions (see, e.g.,~\citealt{soumagnac2020}), as long as the SN ejecta expand with an average velocity of at least $10^4\,\text{km}\,\text{s}^{-1}$ in all directions. The only SNe for which the narrow features might originate from CSM produced during the precursor are SN\,2019uo, SN\,2019mom (material from the early precursor), SN\,2019aafe, SN\,2019vkl, and SN\,2020edh. In all other cases, material ejected during the precursor is swept up quickly if it has a low velocity or, if it is faster, it cannot account for the low line velocities.

The measured CSM velocities and the lower limits on the radius allow us to roughly estimate when the material that produces the narrow lines was ejected. In half of the spectra we see matter that was presumably ejected at least 1\,yr before the explosion, while 10\% of the spectra show signatures of material ejected 2.5\,yr or more before the SN. Additional material could be ejected earlier and the resulting CSM shells at larger distances can lead to rebrightenings or bumps in the SN light curve, as observed for example in SN\,2009ip \citep{margutti2013}, PTF\,10tel \citep{ofek2013}, or iPTF\,13z \citep{nyholm2017}.

We typically observe similar line velocities in spectra of the same SNe, perhaps with the exceptions of SN\,2018kag and SN\,2019uo, where the scatter is larger. One explanation is that the narrow lines are produced by the same material that is located at a large radius above the photosphere. Another option is that progenitor stars eject material with a characteristic velocity (see, e.g., \citealt{owocki2019}, who find an equipartition between the gravitational and kinetic energy of material ejected from the surface of an LBV). 
If the CSM velocities are indeed determined by the surface gravity of the progenitor stars, the escape velocity of the progenitor of SN\,2019cmy (and maybe SN\,2020dcs and SN\,2020dfh, for which we only have lower resolution spectra) are relatively low as shown in Table~\ref{tab:spectroscopy}. The escape velocity is determined by the stellar mass and radius, and is given by $v_{\text{escape}} = (2\times G \times M/R)^{0.5}$. For a stellar mass of $30\,M_\odot$, the stars would have large radii of $300$ to $500\,R_\odot$. The highest escape velocities are observed for SN\,2019uo, SN\,2019fmb, and SN\,2019aafe, which would yield radii of only $10\,R_\odot$ to $15\,R_\odot$, again assuming a stellar mass of $30\,M_\odot$. Especially for the Type Ibn SN\,2019uo, this interpretation seems appropriate: the star has already stripped its hydrogen envelope and is therefore likely much more compact than a typical LBV star.

For four SNe, broad emission lines or broad P~Cygni features become visible a few weeks or months after the SN explosion. This suggests that the ejecta have reached the radius where the CSM is optically thin. The corresponding radii are marked by black lines for SN\,2019uo, SN\,2019cmy, SN\,2019aafe, and SN\,2019zrk. The late-time spectra of the first three SNe continue to exhibit narrow features on top of the broad line, indicating that unshocked, optically thin material is still located above the ejecta. Spectroscopic monitoring of SN\,2019zrk continued and about one month after the broad features first emerged, it turned into a Type II SN without any narrow components (as will be described by Fransson et al., in prep.). For all other SNe, the CSM is still optically thick at the time when the last spectrum was obtained, meaning that the dense CSM extends to larger radii as indicated by the shaded area.

We conclude that the material ejected during the observed precursors typically cannot account for the narrow emission lines in the SN spectra (see also \citealt{moriya2014}). The narrow lines that are observed while the SN is bright are instead produced by slow-moving material ejected years before the observed precursors and SN explosion.

\subsection{Correlations with SN Properties}
\label{sec:sne_corr}

\begin{figure}[tb]
    \centering
\includegraphics[width=\columnwidth]{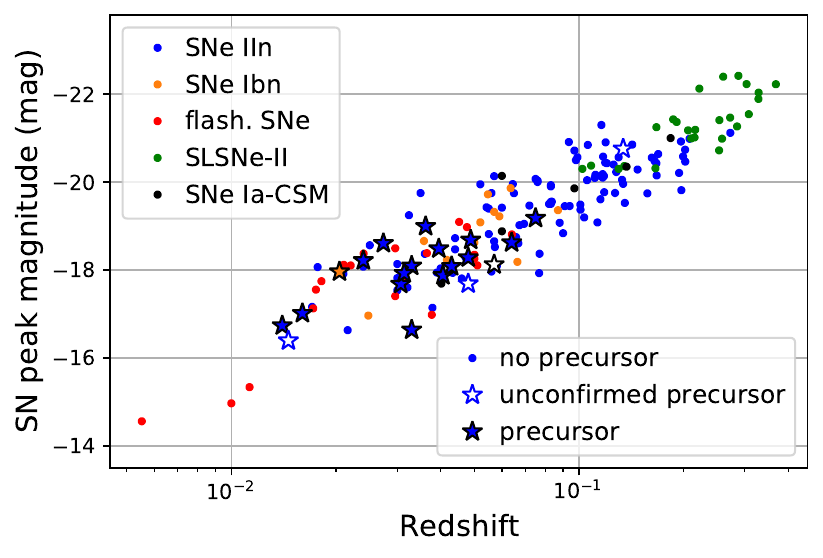}
\caption{\label{fig:sne_with_and_without_pre} SN peak magnitudes for SNe with and without precursors. We use the brightest detection in the $g$, $r$, or $i$ band as a proxy for the peak magnitude. The apparent correlation between the redshift and peak magnitude is caused by the Malmquist bias \citep{malmquist1922}. Most precursors are observed for relatively nearby SNe as they are faint. We find that SNe with detected precursors are not significantly more luminous than the complete sample. We do not apply K corrections.}
\end{figure}

Here, we test whether the observed precursors increase the SN peak brightness or prolong the rise time. \citet{ofek2014} found several marginally significant and weak correlations between the CSM mass estimate and the SN peak luminosity, rise time, and SN energy. All of these correlations are based on a small sample of precursors and require confirmation.
Figure~\ref{fig:sne_with_and_without_pre} shows all SNe with and without precursors and their peak magnitudes. Precursors are detected for many nearby, faint Type IIn SNe, but not for nearby SNe of other types with the exception of the Type Ibn SN\,2019uo. Bright precursors are rare, as demonstrated in Sec.~\ref{sec:rates}, so fewer precursors are detected for distant SNe. The correlation between the redshift and the SN luminosity in Fig.~\ref{fig:sne_with_and_without_pre} is due to the Malmquist bias \citep{malmquist1922}, which describes that faint objects are undetectable at large distances.

To quantify whether SNe with precursors of any luminosity tend to be more luminous, we calculate a partial correlation between the SN peak magnitude and an array which specifies whether or not a precursor is detected. The distance modulus is used as a control variable to correct for the impact of the Malmquist bias. The distance modulus is chosen rather than the redshift or distance, because it is proportional to the apparent SN magnitude and hence to the detection probability. The partial correlation is calculated for 116 Type IIn SNe with $r$-band pre-explosion observations and with measured peak magnitudes, and we find a Pearson correlation coefficient of $0.06$ which corresponds to a $p$-value of $0.53$. We thus do not detect a correlation between the SN peak magnitude and the detection of a precursor in our search. This might indicate that both groups of SNe have massive CSM shells.

In Sec.~\ref{sec:rates_snlumi} we found indications that luminous SNe tend to have more luminous precursors (see Fig.~\ref{fig:rates_snlumi}). To quantify the significance of this observation we calculate a partial correlation between the precursor and the SN peak luminosity while again using the distance modulus as a control variable. The calculation is done for the $r$-band precursor magnitudes of 12 SNe that have long-lasting precursors in the final three months before the SN explosion, the same objects that contribute to the rate measurement in the right-hand panel of Fig.~\ref{fig:rates_snlumi}. We calculate the Spearman rank coefficient, which measures whether brighter precursors are associated with brighter SNe without assuming a linear relation between the two luminosities. We find a positive correlation with a correlation coefficient of 0.84 and a $p$-value of 0.001 or a significance of $3.1\,\sigma$. We can hence confirm that more luminous precursors occur prior to more luminous SNe even after correcting for the impact of the Malmquist bias. There could be many possible explanations for the detected correlation: A more massive CSM might boost both the luminosity of the precursor and of the SN, or an energetic precursor could create a more massive CSM shell which results in stronger interaction and a more luminous SN (see also the simple exploration by \citealt{ofek2014}). Alternatively, stars with certain properties might produce more violent precursors and more energetic SN explosions. We also caution that less luminous SNe tend to be nearby while more luminous SNe are located at larger redshifts. While we corrected for the fact that distant SNe appear fainter on Earth, there could be many other differences between these objects which all might play a role.

A massive, optically thick CSM slows down the rise of the SN light curve because the photons diffuse out to the photosphere, and we thus expect a correlation between the SN rise time and the estimated precursor mass. To estimate the diffusion time we inspect the SN light curves in 1-day bins and quantify how many days it takes the light curve to reach its peak from a flux level that is $1.086$ mag lower, which corresponds to a luminosity increase by a factor of $e$. This is done in the $r$ band if available and in the $g$ band for SN\,2018gho. The $g$ band typically peaks earlier than the $r$ band, so the numbers might not be directly comparable. The rise time should be measured for the bolometric light curve, which typically rises more quickly as a large part of the energy is emitted in the UV. Our estimates are hence upper limits on the actual rise time. The estimated rise times are given in the penultimate column in Table~\ref{tab:precursors}, if the rise and peak are well observed.

With rise times of 4--25 days, all SNe with detected precursors are part of the fast rising subgroup identified by \citet{nyholm2020}\footnote{While we define the rise time as the time it takes the SN to rise by a factor of $e$ to its peak, \citet{nyholm2020} determine the rise time from a power-law fit. Thus, the quoted numbers might not be directly comparable, but the bottom panel of their Fig. 7 allows us to read off rise times that are consistent with our definition.} which includes approximately two thirds of the Type IIn SNe in their sample. A likely explanation is that fast rising SNe IIn are typically fainter \citep{nyholm2020} and our precursor search is most sensitive to nearby, faint SNe (see Fig.~\ref{fig:sne_with_and_without_pre}).
We here assume that energetic precursors eject more material and search for a correlation between the radiative precursor energy and the estimated SN rise time. With a Pearson correlation coefficient $-0.37$ with a $p$-value of 0.25, no significant correlation is found. It is either possible that a relation is washed out owing to the large uncertainties on both quantities or the observed precursors do not dominate the total CSM mass.

In conclusion, we do observe that more luminous precursors are detected prior to more luminous SNe even after correcting for the impact of the SN distance. However, SNe with detected precursors are not systematically more luminous than SNe without detected precursors, and we do not measure a correlation between the precursor energy and the SN rise time. Thus, we do not find that the observed precursors have a major impact on the SN light curve. This seems to fit with our results from Sec.~\ref{sec:sne_history}, where we find that the narrow hydrogen features in the spectra of Type IIn SNe typically do not originate from material ejected during the observed precursor. Together, both results might indicate that a large fraction of the CSM is ejected in earlier mass-loss events that we do not detect here.

\subsection{The Type Ibn SN\,2019uo}
\label{sec:sne_sn2019uo}

\begin{figure}[tb]
\centering
\includegraphics[width=\columnwidth]{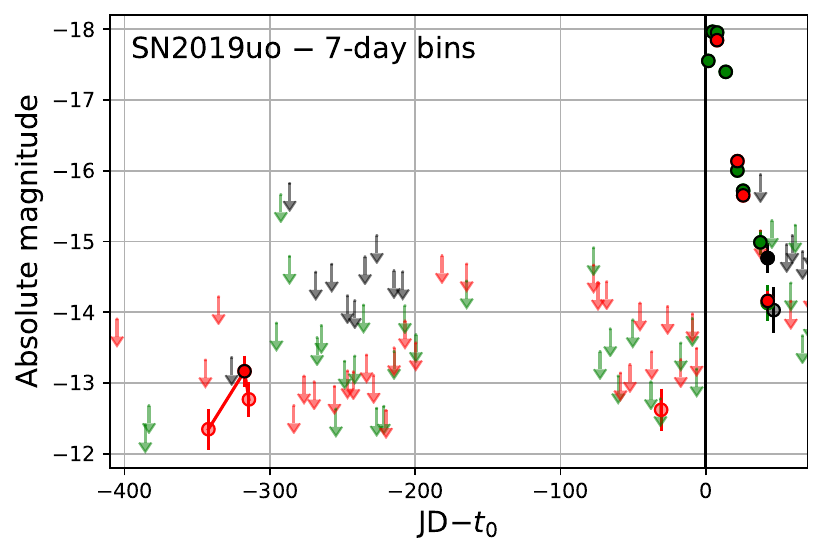} \caption{\label{fig:sn2019uo_lc}Light curve of SN\,2019uo in 7-day bins before the estimated explosion date (JD 2,458,501.3) and in 1-day bins afterward. Solid data points with black edges are $5\sigma$ detections, transparently colored data points have significances larger than $3\sigma$, and upper limits are less significant. Green, red, and black data points correspond to observations in the $g$, $r$, and $i$ bands, respectively. A line connects the three detections that we consider to be part of the precursor. The $3\sigma$ detection at $\sim30$\,days could be a statistical fluctuation and is not discussed here.} 
\end{figure}

One of the precursors discovered in this study occurs prior to SN\,2019uo, making it the second Type Ibn with a detected precursor after SN\,2006jc \citep{pastorello2007,foley2007}. The coadded difference images in Appendix~\ref{sec:images} confirm that a point source is visible $\sim320$\,days before the explosion of SN\,2019uo.
As shown in Fig.~\ref{fig:sn2019uo_lc}, the precursor starts $\sim340$\,days before the explosion of SN\,2019uo and is observed over 35\,days (see also Fig.~\ref{fig:precursor_lcs} and Table~\ref{tab:precursors}). In addition to the $5\sigma$ detection, another two data points surpass the $3\sigma$ threshold when using 7-day bins, and in total 24 individual images contribute to the detection. The mean precursor magnitude is $-13$, making it one of the faintest precursors detected in this search. The observed radiative energy in the $r$ band is $1.7\times10^{47}\,\text{ergs}$. However, the observations shortly before and after the precursor are not very constraining, so it could last longer.

With a redshift of $0.020$, SN\,2019uo is the closest Type Ibn SN out of the 12 objects in our sample. It was classified as a Type Ibn based on a spectrum showing narrow helium P~Cygni features that was obtained 9\,days after the estimated explosion date \citep{fremling2019}. 
SN\,2019uo was studied by \citet{gangopadhyay2020} in detail, and they find that SN\,2019uo is slightly underluminous and evolves quickly, typical for Type Ibn SNe. They model the bolometric light curve with CSM interaction and their best-fitting models require the presence of 0.4-- 0.7\,$\text{M}_\odot$ of material located at a radius of $\gtrsim$ (0.2--2) $\times 10^{15}\,\text{cm}$.

SN\,2019uo is one of the few objects for which the narrow P~Cygni profiles in the spectra might originate from the CSM ejected during the observed precursor (see Fig.~\ref{fig:interaction_sig}). We measure typical velocities of $\sim900\,\text{km}\,\text{s}^{-1}$ (see Table~\ref{tab:spectroscopy}), which would imply that the CSM has propagated to a radius of a $\sim2\times10^{15}\,\text{cm}$ when the SN explodes $320$\,days after the precursor. This radius is approximately consistent with the inner CSM radius required by the modelling of \citet{gangopadhyay2020}. Moreover, it roughly coincides with the radius at which the CSM turns optically thin shown in Fig.~\ref{fig:interaction_sig}: broad features, likely from the shocked CSM, first appear in a spectrum obtained 18\,days after the explosion date (see Fig.~3 by \citealt{gangopadhyay2020}). A CSM shell located at a radius of $2\times10^{15}\,\text{cm}$ is swept up by the SN ejecta if they have an average velocity of $13,000\,\text{km}\,\text{s}^{-1}$. Narrow helium features remain visible in later spectra and might originate from an optically thin wind located above the CSM. The same material could also produce the flash-ionization features detected by \citet{gangopadhyay2020} before the SN reaches its peak.

The observed precursor energy can be produced via interaction if the mass lost during the precursor is $\gtrsim0.007\,\text{M}_\odot$ (see Fig.~\ref{fig:precursor_velo} or Table~\ref{tab:precursors}). However, the SN light curve modelled by \citet{gangopadhyay2020} requires a CSM mass a factor of 50--100 times larger. A possible explanation is that the material ejected during the precursor propagates through a low-density environment such that only 1--2\% of its kinetic energy is converted to radiation in the $r$ band. This would imply that the progenitor star was not surrounded by a massive CSM before the observed precursor. Alternatively, only a small fraction of the CSM is emitted during the observed precursor and the rest of the required material is expelled during earlier or fainter precursors which we do not detect. In this case the radiative efficiency would be larger. We conclude that a relatively simple picture can explain the characteristics of the precursor and subsequent SN: the observed precursor could produce the complete CSM of $\sim0.5\,\text{M}_\odot$ if its radiative efficiency is low with $\epsilon\approx1\%$. The resulting dense CSM shell is confined to a radius of $\sim2\times10^{15}\,\text{cm}$ and can account for the SN bolometric light curve (shown by \citealt{gangopadhyay2020}) as well as for the spectroscopic development.

Until now, the only precursor observed prior to a Type Ibn SN was observed for SN\,2006jc \citep{pastorello2007, foley2007}. The 9-day-long precursor with a peak magnitude of $-14.1$ was detected 2\,yr  before the SN explosion. The two precursors are hence similar to each other, as they are both relatively faint and happen hundreds of days before the SN explosion. In both cases, the CSM is helium-rich, so we do not witness the stripping of the hydrogen envelope. 
Based on 11 Type Ibn SNe with pre-explosion data in the $r$ band, we show in Fig.~\ref{fig:rates_subsample} and Table~\ref{tab:precursors} that precursors with an $r$-band magnitude of $<-16$ happen $< 1.0\%$ of the time ($95\%$ confidence level), while faint precursors with magnitudes brighter than $-13$ might happen up to $31\%$ of the time. These limits are based on observations collected up to $2.5$\,yr before the SN explosion and the median observation time is $8.6$ months before the explosion. We thus do not have very strong constraints on faint outbursts and they might be relatively common.

Thus, we confirm that the progenitor stars of some Type Ibn SNe are able to produce relatively bright flares in the last years before their explosion. Except for the two precursor detections, no Type Ibn SN progenitor has been identified in archival observations and their nature is debated. The most commonly suggested progenitors are Wolf-Rayet stars that have shed their hydrogen envelopes or massive stars that are stripped by a binary partner (see, e.g., \citealt{smith2017, hosseinzadeh2017}). Alternatively, the progenitors of some Type Ibn SNe were hypothesized to be white dwarf binaries \citep{sanders2013, hosseinzadeh2019}, or very massive stars that undergo pulsational pair instability events \citep{woosley2017, karamehmtoglu2019} which would imply that no core collapse occurs and the star is likely still present.
Contrary to LBVs, classical Wolf-Rayet stars are not known to undergo giant eruptions, so bright flares cannot be common during the lifetime of the star. There is an intermediate class of Ofpe/WN9 stars which have stripped most of their hydrogen envelope, but undergo LBV-like outbursts (see, e.g., \citealt{smith2020}). Such stars have been suggested as progenitors for two Type Ibn SNe that also show relatively strong hydrogen lines \citep{smith2012, kool2020}. However, the hydrogen features in the spectra of SN\,2006jc are much weaker and SN\,2019uo might not show any hydrogen. It is therefore unclear under which conditions stripped-envelope stars can produce as bright eruptions. If such flares are related to the late stages of nuclear burning (see Sec.~\ref{sec:nature_waves}), they would only occur shortly before the SN explosion, which would explain why no such flares are observed for Wolf-Rayet stars in the Milky Way or in nearby galaxies.

\section{Nature of the precursors}
\label{sec:nature}

We emphasize that the nature of the observed transients is ambiguous. Owing to the optically thick CSM, we cannot directly observe the expanding SN ejecta, but only see light diffusing out through the photosphere. We therefore cannot determine whether core collapse occurred or at what time it happened (see, e.g.,~\citealt{moriya2015, tartaglia2016b, woosley2017} for possible scenarios). It is conceivable that some of the precursors are already part of the SN light curve, rather than preceding the explosion. Nevertheless, in the following we adopt the interpretation that core collapse occurs shortly before the SN rises to its main peak and that the precursors are produced by the progenitor star before it explodes.

In Sec.~\ref{sec:nature_lumi} we explore whether the precursor luminosity could be produced via interaction or by a continuum wind. The underlying energy source is unknown and suggested mechanisms include unstable nuclear burning phases \citep{smith2014}, shell burning \citep{arnett2011, arnett2011b}, interaction with a binary companion \citep{smith2014, mcley2014, danieli2019, owocki2019}, or reduced gravity due to high neutrino luminosities \citep{moriya2014b}. Specific predictions exist for wave-driven mass loss triggered by instabilities during the neon and oxygen burning phases, and we compare our observations to the model described by \citet{shiode2014} in Sec.~\ref{sec:nature_waves}.

\subsection{What Powers the Precursor Luminosity?}
\label{sec:nature_lumi}

In this section we explore possible mechanisms that could produce the observed precursor luminosity. First, we point out in Sec.~\ref{sec:nature_duration} that the long precursor durations likely require a persistent energy source. This could, for example, be interaction of the ejected material with pre-existing CSM (described in Sec.~\ref{sec:nature_interaction}) or a brightening of the star (discussed in Sec.~\ref{sec:nature_wind}). In Sec.~\ref{sec:nature_diff}, we measure the SN rise times to derive upper limits on the CSM mass.


\subsubsection{Precursor Durations}
\label{sec:nature_duration}

The densely sampled light curves of the ZTF survey allow us to measure the durations for a sample of precursors. As shown in Fig.~\ref{fig:shapes}, the outbursts are typically observed over several months. Their true durations are likely even longer, if the fainter parts of the precursors remain undetected. 
With rise times of only $\sim 1$\,mag over $\sim50$--100\,days, most precursors develop much more slowly than the subsequent SNe. This implies that the diffusion time in the CSM does not dominate the precursor duration. Consequently, the precursors are likely not powered by a single short-lived eruption from the stellar surface, but require a long-lived energy source, such as ongoing interaction, a stellar wind, potentially a series of explosions, or maybe a short-lived event deep within the stellar envelope, where the diffusion time is much larger than within the CSM. 
We thus conclude that the long duration of several detected precursors is likely intrinsic and not due to diffusion.

\subsubsection{Interaction-Powered Precursors}
\label{sec:nature_interaction}

The light curves of Type IIn SNe are mainly powered by interaction between the ejecta and the circumstellar material surrounding the star. It thus might seem logical that the same is true for the precursors. In this scenario, a large amount of material is ejected from the stellar surface and a fraction of the kinetic energy $\epsilon<1$ is converted to radiative energy when the ejected material is slowed down by pre-existing CSM. If the velocity of the newly ejected material is known, its mass can be estimated using
\begin{equation}
\label{eq:interaction}
E_{\text{rad}} = \epsilon E_{\text{kin}} = \epsilon \frac{1}{2} M_{\text{CSM, pre.}} v_{\text{CSM, pre.}}^2\quad,
\end{equation}
where $M_{\text{CSM, pre.}}$ and $v_{\text{CSM, pre.}}$ are respectively the mass and velocity of the material ejected during the precursor.

\begin{figure}[tb]
\centering
\includegraphics[width=\columnwidth]{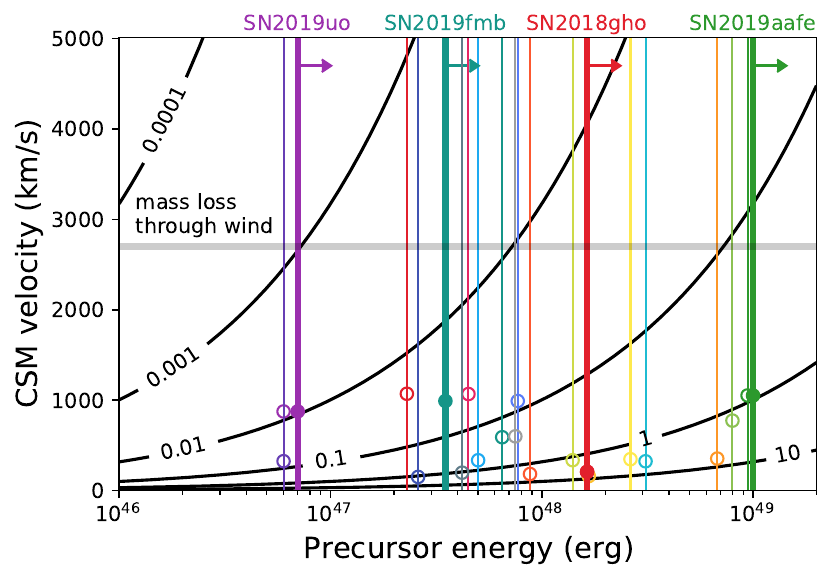}
\caption{\label{fig:precursor_velo} CSM masses required to produce the observed precursor energies, if they are powered by interaction. The black contour lines show the amount of CSM, $\text{M}_{\text{CSM, pre.}}$ in solar masses, ejected during the precursor if the kinetic energy is completely converted to radiation (i.e., $\epsilon=1$). The observed precursor energies (taken from Table~\ref{tab:precursors}) are indicated by vertical lines. They are lower limits on the actual radiative energy as we only detect the brightest part of the precursors and because we did not apply a bolometric correction when calculating the energy. CSM velocities measured from the spectra are marked by dots, but they do not necessarily correspond to the velocities of the precursor material. The gray horizontal line shows the approximate amount of CSM that a continuum-driven wind would eject (see Sec.~\ref{sec:nature_wind}).}
\end{figure}

Equation~\ref{eq:interaction} shows that the required CSM mass strongly depends on the velocity. CSM velocities can be estimated from the narrow hydrogen (or helium) features, however; as demonstrated in Sec.~\ref{sec:sne_history}, these lines are likely produced by material ejected at earlier times. It is hence possible that the matter expelled during the observed precursors has larger velocities which would result in lower CSM mass estimates. The observed velocities and corresponding CSM masses are given in the fourth and third to last columns of Table~\ref{tab:precursors} (the quoted velocity is the median of all spectra given in Table~\ref{tab:spectroscopy}). 
The required CSM masses for different velocities are illustrated in Fig.~\ref{fig:precursor_velo} for some of the precursors.

Most observed precursors develop quite slowly as pointed out in Sec.~\ref{sec:nature_duration}. \citet{nyholm2020} measured the rise and decline rates for $\sim30$ Type IIn SNe from the PTF/iPTF sample and the precursors we detect rise more slowly than any Type IIn SNe in their sample. A possible explanation could be that the precursor shock front propagates with a substantially lower velocity compared to the SN ejecta. As a consequence, the rate at which CSM is swept up is lower and kinetic energy is converted to radiation more slowly. 
According to the model by \citet{svirski2012} (see also \citealt{ofek2014b}), the radiated luminosity produced during CSM interaction is $L_{\text{pre.}}\propto r^2 \rho v_{\text{CSM, pre.}}^3$, which simplifies to $L\propto v_{\text{CSM, pre.}}^3$ for a wind-like density profile with $\rho \propto r^{-2}$. The luminosity difference between precursors and SNe on the order of 100 suggests that the average shock velocity is a factor of $\sim5$ lower (i.e., $v_{\text{CSM, pre.}}\approx 2000\,\text{km}\,\text{s}^{-1}$ compared to $\sim10^{4}$ for SN ejecta). However, this assumes that the radiative efficiency $\epsilon$ (i.e., the fraction of kinetic energy that is converted to radiation) is similar for the precursor and the SN. These higher velocities would reduce the amount of ejected material by about one order of magnitude as shown in Fig.~\ref{fig:precursor_velo}.

Interaction stops either when the shock reaches the edge of the CSM or when it is slowed down, which happens when the mass of the swept-up CSM is comparable to the mass in the shock front. Several precursors (e.g., prior to SN\,2019vkl, SN\,2019aafe, or SN\,2019zrk) continue to rise for more than $100$ days. If their luminosity is dominated by interaction, the associated mass loss must be substantial, such that the ejected material is not slowed down considerably within this time (i.e., $\epsilon \ll 1$). If it expands with an average velocity of $2,000\,\text{km}\,\text{s}^{-1}$ it reaches a radius of $\sim2\times10^{15}\,\text{cm}$ within 100\,days. Several other precursors (e.g., prior to SN\,2019fmb, SN\,2019gjs, or SN\,2020dcs) fade in the weeks to months before the SN explosion, as was also observed for the 2012a event prior to the final explosion of SN\,2009ip. This might indicate that the material has slowed down or that it only collides with a thin CSM shell and then continues to expand through a lower density environment. This qualitative description assumes that each precursor is associated with a single short-lived mass-loss event. It is, however, also possible that material is ejected in a series of eruptions from the stellar surface. If the energy of these eruptions changes with time, this could account for both rising or falling precursor light curves.

As noted in Sec.~\ref{sec:pre_energy} precursors with detections in both the $g$ and $r$ band appear rather red compared to young Type IIn SNe. We suggest that these low temperatures are a consequence of the low precursor luminosity in combination with a relatively large photospheric radius due to the extended CSM.

We conclude that the observed precursors could be powered by interaction if the star undergoes major mass-loss events. The precursor luminosities require that the ejected material has velocities on the order of $\sim2,000\,\text{km}\,\text{s}^{-1}$. Such low-velocity shock fronts could explain the long duration of the precursors compared to the SN light curve. Figure~\ref{fig:precursor_velo} shows that the most energetic precursor requires the ejection of $\sim0.3\,\text{M}_{\odot}$ of material for a CSM velocity of $2,000\,\text{km}\,\text{s}^{-1}$. However, a more realistic radiative efficiency of $\epsilon\lesssim0.3$ would bring the required mass back to $\sim1\,\text{M}_\odot$. CSM envelopes of several solar masses have been observed for some Type IIn SNe, so high-mass estimates are not necessarily unrealistic, and for less energetic precursors the required masses are lower by a factor of up to 100.

\subsubsection{Wind-Driven Precursors}
\label{sec:nature_wind}

In an alternative scenario the observed luminosity originates from the star itself. \citet{shaviv2001b} showed that the outer part of the stellar envelope (from the radius out to which convection is efficient to the ``hydrostatic surface" of the star) becomes unstable when the stellar luminosity approaches the Eddington limit. Local density differences reduce the effective opacity of the star, so it remains quasistable even when it exceeds the Eddington luminosity. However, the instabilities only have an effect as long as the atmosphere is optically thick over a scale height. An optically thick continuum-driven wind is therefore accelerated from this region, with the actual photosphere sitting farther out. The resulting mass loss is smaller than can be expected without the lower effective opacity. The resulting wind has a typical velocity of $(L/L_{\text{Eddington}})^{0.5}\,v_{\text{escape}}$ \citep{shaviv2001} --- that is, usually $v_{\text{CSM, pre.}}\gtrsim1000\,\text{km}\,\text{s}^{-1}$, which is larger than most line velocities measured in Sec.~\ref{sec:sne_history}.

Contrary to the interaction scenario a fraction of the radiative energy is converted into kinetic energy (i.e., $\epsilon>1$). The mass loss can be calculated using
\begin{equation}
\label{eq:continuum_wind}
E_{\text{rad}} = \frac{1}{W} M_{\text{CSM, pre.}} c_{\text{s}} c = \epsilon E_{\text{kin.}} \quad,
\end{equation}
where $W\approx5$ is an empirical factor, $c_{\text{s}}\approx60\,\text{km}\,\text{s}^{-1}$ is the speed of sound at the base of the optically thick wind \citep{shaviv2000, shaviv2001}, and $c$ is the speed of light. We note that the results of Eq.~\ref{eq:continuum_wind} are equal to Eq.~\ref{eq:interaction} for a CSM velocity of $v_{\text{CSM, pre.}}=2,700\,\text{km}\,\text{s}^{-1}$ and $\epsilon=1$. As a result the mass loss is lower by a factor of $\sim10$ compared to the numbers given in the third to last column of Table~\ref{tab:precursors}. The most energetic precursor with a radiative energy of $10^{49}\,\text{ergs}$ only results in a mass loss of $0.1\,\text{M}_{\odot}$ as shown in Fig.~\ref{fig:precursor_velo}.

Once the wind reaches an equilibrium state, it forms a photosphere. \citet{owocki2016} show that a wide range of mass-loss rates and wind velocities result in photospheric temperatures between $5,000\,\text{K}$ and $6,000\,\text{K}$, the temperature at which hydrogen recombines and the opacity drops. These temperatures are consistent with most observed precursor temperatures in Sec.~\ref{sec:pre_colors}. Larger temperatures are expected before the wind reaches its equilibrium state, but lower temperatures are more difficult to explain. We note that the precursor temperatures are also similar to the temperature of the giant eruption of $\eta$ Carina \citep{rest2012}. 


\subsubsection{Constraints on the CSM Mass}
\label{sec:nature_diff}

We also estimate upper limits on the total CSM mass (here material located within $5\times10^{15}\,\text{cm}$) based on the SN rise time. A massive, optically thick CSM slows down the rise of the SN light curve, because the photons diffuse out to the photosphere. Therefore, a quickly rising SN is inconsistent with a large CSM mass, while a slow rise could be intrinsic and does not necessarily imply a massive CSM. SN rise times were estimated in Sec.~\ref{sec:sne_corr} and are given in Table~\ref{tab:precursors}. 
For an infinite wind-like CSM profile (i.e., with a density that decreases with the radius like $\rho \propto r^{-2}$), the diffusion time $t_{\text{diff}}$ is given by
\begin{equation}
\label{eq:tdiff}
\centering
t_{\text{diff.}} \approx \frac{\kappa K}{c} \left(\, \ln(c/v_{\text{shock}}) - 1 \right)\quad\text{,}
\end{equation}
where $\kappa$ is the opacity, $v_{\text{shock}}$ is the velocity of the shock front, and the mass-loading factor $K$ is defined as $K= \dot{M}/(4\pi v_{\text{CSM}})$ for the mass-loss rate $\dot{M}$ and the CSM velocity $v_{\text{CSM}}$ \citep{ginzburg2012}.

The CSM mass between two radii $R_{\text{inner}}$ and $R_{\text{outer}}$ can be obtained by integrating the CSM density profile:
\begin{eqnarray}
M_{\text{CSM, diff.}} &=& \int_{R_{\text{inner}}}^{R_{\text{outer}}} 4 \pi r^{2} K r^{-2} dr = 4 \pi K (R_{\text{outer}}-R_{\text{inner}}) \nonumber \\
&\approx& 4\pi R_{\text{outer}} \frac{t_{\text{diff}} c}{\kappa\left(\, \ln(c/v_{\text{shock}}) - 1 \right)}  \nonumber \\
&\approx& 0.10 \, R_{\text{outer, 5e15\,\text{cm}}} \,\, t_{\text{diff, 1d}} \, \text{M}_{\odot} \quad.
\end{eqnarray}
In the second line, we assume that the inner radius is much smaller than the outer one and insert Eq.~\ref{eq:tdiff}. We then adopt a typical value for $\kappa=0.34 \,\text{cm}^2\,\text{g}^{-1}$,  appropriate for a medium that consists of 70\% hydrogen (see, e.g., \citealt{ginzburg2012, ofek2013b}) and a shock velocity of $v_{\text{shock}}=10^4\,\text{km}\,\text{s}^{-1}$. The CSM mass is hence proportional to the diffusion time as well as to the outer radius of the CSM.

We integrate the CSM density out to a radius of $5\times10^{15}\,\text{cm}$, which is approximately consistent with the observed spectral evolution shown in Fig.~\ref{fig:interaction_sig}. The resulting upper limits on the CSM masses are given in the last column of Table~\ref{tab:precursors}. While many of the mass upper limits are consistent with the mass estimates for interaction-powered precursors, there are discrepancies for a few objects (e.g., SN\,2019zrk, SN\,2019gjs, or SN\,2018gho). We emphasize that a difference of a factor of a few does not necessarily imply an inconsistency, as both CSM masses are order-of-magnitude estimates owing to the assumed wind-like density profile, the adopted shock velocity, and the outer CSM radius of $5\times10^{15}\,\text{cm}$. Furthermore, the diffusion time might be different for nonspherical CSM geometries \citep{soumagnac2020}. Another alternative would be considerably higher CSM velocities of a few $\times 1,000\,\text{km}\,\text{s}^{-1}$ (see Sec.~\ref{sec:nature_interaction}). We summarize that several SNe, such as SN\,2019zrk, rise quickly even though they experienced powerful precursor eruptions shortly before the explosion. This might imply that the precursors are not associated with mass-loss events of several solar masses. Careful modeling of a well-observed SN with an energetic precursor would be required to establish whether fast SN rise times can be reconciled with extensive mass-loss episodes in the last months before the explosion.

\subsection{Can Wave-Driven Mass Loss Trigger the Detected Precursors?}
\label{sec:nature_waves}

Unbinding a solar mass of material requires a substantial energy deposition in the stellar envelope. The fact that the precursor rate increases in the last months before the SN explosion could imply that the precursors are associated with late nuclear burning stages which last from a few years to a few months (see, e.g.,~\citealt{shiode2014}).

When fusing carbon or heavier elements, both the energy production and neutrino cooling rates increase dramatically in the stellar core (see, e.g., \citealt{woosley2002, arnett2011, quataert2012}). While the two processes are in equilibrium on average, local imbalances cause vigorous convection within the core and \citet{shiode2014} estimate that $\lesssim10\%$ of the fusion energy is carried by convection. The convection excites gravity waves which typically remain confined to the core \citep{meakin2006, quataert2012} and do not affect the envelopes of most massive stars.
However, the internal structure of some progenitor stars may allow part of the wave energy to tunnel out of the core and excite acoustic waves in the stellar envelope. The energy deposited in the envelope can be as large as a few times $10^{40}\,\text{ergs}\,\text{s}^{-1}$ or $\sim10^{47}\,\text{ergs}$ over a year. It might trigger strong adiabatic mass-loss events or inflate the stellar envelope \citep{quataert2012, shiode2014, quataert2016, fuller2017, fuller2018}.

The neon and oxygen burning phases last a few years to months with more massive cores burning out more quickly (see, e.g., \citealt{shiode2014}). Silicon fusion only occurs in the last few days to hours before the explosion and therefore cannot account for the observed months-long precursors. \citet{shiode2014} modelled different progenitor stars with the \emph{MESA} stellar code \citep{paxton2011} and found that it takes the waves about several weeks to about a year to reach the stellar surface and that progenitor stars with a wide range of initial conditions can fulfill the requirements for matter outflow.
We observe that the brightest precursors start several months before the SN explosion. They hence might be powered by energy produced during the neon or oxygen burning phases. According to the model by \citet{shiode2014}, the time of the precursor eruption is inversely related to the core mass. A precursor that occurs one month before the SN explosion would imply a helium core mass of $\lesssim 15\,\text{M}_\odot$, while a precursor one year before the explosion requires a lower core mass of $\lesssim5\,\text{M}_\odot$.

The most energetic precursor, detected prior to the explosion of SN\,2019aafe, released an $r$-band energy of $10^{49}\,\text{ergs}$ over 100\,days. If this luminosity is interaction-powered, the kinetic energy would be even larger. For progenitors with $\sim15\,\text{M}_\odot$ helium cores, \citet{shiode2014} calculate wave energies of up to a few times $10^{47}\,\text{ergs}$, about $50$ times lower than the observed radiative energy. Ten times higher energies could be reached by stars with $\gtrsim30\,\text{M}_\odot$ helium cores, but their oxygen burning phase only lasts for about a month, such that the produced waves only reach the surface days before the SN explosion (similar precursor energies and timescales were calculated by \citealt{wu2020}). Fusing $\sim1\,\text{M}_\odot$ of material releases an energy of $\sim10^{51}\,\text{ergs}$ during the neon and oxygen burning phases each \citep{quataert2012}, so the wave transport would have to be extremely efficient to produce as energetic precursors as observed. The precursors observed prior to other SNe, such as SN\,2019fmb or SN\,2020dcs, are at least 10 times less energetic and are more easily explained by the model.

We summarize that wave-driven mass loss, powered by instabilities during the neon and oxygen burning phases, could explain why precursor eruptions occur in the last few months before the SN explosion. According to the model by \citet{shiode2014}, less energetic early precursors are produced by lower mass stars, while stars with massive cores produce more powerful precursors that occur only days before the explosion. This could at least qualitatively explain our observation that brighter precursors become more common in the last months before the SN explosion (see Sec.~\ref{sec:rates_timedep}). However, the brightest detected precursor is about two orders of magnitude more energetic than predicted by \citet{shiode2014}, and it is unclear whether the model could account for such events.
We emphasize that our observations cannot confirm that wave-driven mass loss triggers precursor eruptions, and that other mechanisms such as interaction with a binary companion star might also be able to explain the observations (see, e.g., \citealt{quataert2016, owocki2019}).

\section{Conclusions}
\label{sec:conclusion}

One main finding of this study is that bright precursors are relatively common immediately before the explosion of Type IIn SNe (see Fig.~\ref{fig:shapes}) and that most of them last for one or several months. Long-lasting precursors that are brighter than $-13$ mag in the $r$ band are observed immediately prior to the explosion of about $25\%$ of all Type IIn SNe (with a 95\% confidence range of 5--69\%). Some of the brightest precursors are better described as a continuous brightening rather than a discrete flare (see, e.g., the precursor light curves prior to SN\,2019zrk or SN\,2019aafe in Fig.~\ref{fig:shapes}). The most powerful precursor found here releases an energy of $10^{49}\,\text{ergs}$ over 100\,days ($\sim10$\% of the radiative energy released in a typical SN explosion), but most precursors are an order of magnitude less energetic.

All precursors are much brighter than the Eddington luminosity of a massive star, such that they likely involve extensive mass-loss events. The mass of the ejected CSM is difficult to quantify: if the kinetic energy of the CSM is similar to the radiative energy of the precursor, the mass loss would typically amount to one or a few solar masses with large uncertainties owing to the unknown CSM velocity. For wind-driven precursors, the expected mass loss is typically ten times lower. Several SNe with detected precursors rise to their peak luminosity within a few days, which might imply that the mass of their CSM shell is typically $\lesssim1\,\text{M}_{\odot}$.

To our knowledge, such bright and long-lasting precursors have so far only been detected prior to Type IIn SNe and precursors prior to other SNe are typically less energetic. Nevertheless, material ejected during these events cannot account for the characteristic narrow hydrogen emission lines in the spectra of these SNe. Within the short time before the SN explosion the ejected material cannot expand to radii larger than $10^{14}\,\text{cm}$, if its velocity is as low as the line widths indicate. Hydrogen emission lines are, on the other hand, observed for many weeks or even years and are thus produced by material located at larger radii. For the average spectrum, this material must have been ejected at least $\sim1$\,yr before the explosion. We only find few, rather faint and short, precursors at these times. This might indicate that earlier mass-loss events are likely substantially fainter than the observed precursors and therefore remain undetected. SNe with precursors are not significantly brighter than SNe without observed precursors at similar redshifts, which likely supports our hypothesis that the precursors detected immediately before the explosion do not dominate the total CSM mass.

The Type Ibn SN\,2019uo is an exception: the observed precursor might have ejected $\sim0.5\,\text{M}_{\odot}$ of helium $\sim320$\,days before the SN explosion. The low precursor luminosity is expected if the material propagates through a low-density environment. If the material propagates with the CSM velocity observed in the spectra it would remain confined to a rather small radius. This would explain both the fast light-curve evolution and the appearance of broad spectral features only 18\,days after the explosion that indicate that the ejecta have reached the edge of the optically thick CSM. It is hence possible that the complete CSM of SN\,2019uo is created during the observed precursor, while the spectra and light curves of the Type IIn SNe require earlier mass-loss events (see also \citealt{moriya2016}).

Prior to Type IIn SNe, we only detect five precursors that happen more than three months before core collapse (at phases of 700 to 180\,days before the explosion). At these earlier times the rate (or duration) of bright precursors with $r$-band magnitudes $<-16$ is lower by a factor of more than 10 and the rate of precursors brighter than magnitude $-13$ is a factor of 5 lower. The increasing rate of bright precursors could be explained if the precursors are powered by wave-driven mass loss triggered by instabilities during the neon and oxygen burning phases. \citet{shiode2014} argue that stars with lower core masses undergo fainter precursors about a year before the SN explosion, while more massive stellar cores produce brighter flares only weeks or days before core collapse, because of their shorter nuclear burning phases. However, the most energetic precursor is 100 times more energetic than predicted by \citet{shiode2014}, and it is not clear whether energy transport to the stellar envelope could be as efficient.

The bright outbursts shortly before the SN explosion open the door to the possibility of predicting SN explosions.
Four of the precursors reach an apparent magnitude of $<20$ (see Figs.~\ref{fig:precursor_lcs} and \ref{fig:precursor_lcs2}) and are potentially detectable with the ZTF discovery pipeline. Based on these numbers, we estimate that 1--2 precursors per year are bright enough to allow the prediction of an imminent SN explosion.
Indeed, the precursor prior to SN\,2019fmb was reported as a transient by the Pan-STARRS collaboration \citep{chambers2019}, but it was not realized at the time that this was a pre-explosion outburst. We conclude that the ZTF survey has the potential to predict SN explosions if a dedicated search is implemented.


\begin{deluxetable*}{l c l c c c c l}
\tablecaption{SN Spectra}
\label{tab:spectroscopy}
\tablewidth{0pt}
\tablehead{
\colhead{SN} & \colhead{obs. JD} & \colhead{instrument} & \colhead{time after $t_0$} & \colhead{line width} & \colhead{inst. res.} & \colhead{velocity} & \colhead{comment}\\
\colhead{} & \colhead{} & \colhead{} & \colhead{(days)} & \colhead{($\text{km}\,\text{s}^{-1}$)} & \colhead{($\text{km}\,\text{s}^{-1}$)} & \colhead{($\text{km}\,\text{s}^{-1}$)} & \colhead{}
}
\startdata
SN\,2018eru	&	2458351.7	&	P200/DBSP	&	35.0	&	1150	&	130	&	1143, 1000$^*$	&	\\
\rule{0pt}{3ex}
SN\,2018gho	&	2458372.7	&	P200/DBSP	&	6.2	&	400	&	130	&	378	&		\\
	&	2458383.4	&	LT/SPRAT	&	16.9	&	800	&	830	&	$<415$	&		\\
\rule{0pt}{3ex}
SN\,2018kag	&	2458470.5	&	LT/SPRAT	&	4.0	&	1150	&	830	&	796	&		\\
	&	2458502.5	&	NOT/ALFOSC	&	35.9	&	750	&	640	&	391	&		\\
	&	2458535.5	&	WHT/ACAM	&	68.9	&	1500	&	670	&	1342	&	broad	\\
\rule{0pt}{3ex}
SN\,2019uo	&	2458505.0	&	FTN/FLOYDS	&	3.7	&	$-$	&	$-$	&	990$^*$	&	flash, He line	\\
	&	2458506.9	&	FTN/FLOYDS	&	5.6	&	$-$	&	$-$	&	750$^*$	&	\\
	&	2458508.3	&	LJT/YFOSC	&	7.0	&	$-$	&	$-$	&	850$^*$	&	\\
	&	2458509.9	&	P200/DBSP	&	8.6	&	$-$	&	$-$	&	550$^*$	&	\\
	&	2458519.9	&	FTN/FLOYDS	&	18.6	&	$-$	&	$-$	&	1200$^*$	&	broad	\\
	&	2458525.7	&	NOT/ALFOSC	&	24.4	&	$-$	&	$-$	&	1000$^*$	&	broad	\\
	&	2458529.2	&	LJT/YFOSC	&	27.9	&	$-$	&	$-$	&	1000$^*$	&	broad	\\
	&	2458579.9	&	Keck-1/LRIS	&	78.6	&	$-$	&	$-$	&	1000$^*$	&	broad	\\
\rule{0pt}{3ex}
SN\,2019bxq	&	2458596.7	&	LT/SPRAT	&	40.8	&	900	&	830	&	$<415$	&		\\
	&	2458598.0	&	P200/DBSP	&	42.1	&	330	&	130	&	303	&		\\
\rule{0pt}{3ex}
SN\,2019cmy	&	2458572.8	&	APO/DIS	&	4.9	&	400	&	370	&	$<185$	&	flash	\\
	&	2458577.0	&	Keck-1/LRIS	&	9.1	&	350	&	320	&	$<160$	&		\\
	&	2458580.9	&	APO/DIS	&	13.0	&	400	&	370	&	$<185$	&		\\
	&	2458637.9	&	Keck-1/LRIS	&	70.0	&	400	&	320	&	240	&	broad	\\
\rule{0pt}{3ex}
SN\,2019iay	&	2458668.5	&	Lick 3-m/Kast	&	11.8	&	500	&	300	&	400	&		\\
	&	2458668.7	&	P200/DBSP	&	12.1	&	300	&	130	&	270	&		\\
\rule{0pt}{3ex}
SN\,2019gjs	&	2458705.7	&	P200/DBSP	&	15.0	&	350	&	130	&	325	&		\\
\rule{0pt}{3ex}
SN\,2019mom	&	2458705.9	&	P200/DBSP	&	14.9	&	450	&	130	&	431, 750$^*$	&	\\
\rule{0pt}{3ex}
SN\,2019fmb	&	2458792.0	&	APO/DIS	&	76.2	&	900	&	370	&	820	&		\\
	&	2458793.0	&	P200/DBSP	&	77.2	&	1000	&	130	&	992, 1050$^*$	&	\\
\rule{0pt}{3ex}
SN\,2019aafe	&	2458750.9	&	Keck-1/LRIS	&	10.1	&	1100	&	320	&	1052	&		\\
\rule{0pt}{3ex}
SN\,2019vkl	&	2458812.1	&	Lijiang-2.4m/YFOSC	&	3.4	&	850	&	350	&	775	&		\\
\rule{0pt}{3ex}
SN\,2019vts	&	2458834.8	&	APO/DIS	&	17.8	&	500	&	370	&	336	&		\\
\rule{0pt}{3ex}
SN\,2019qny	&	2458853.4	&	LT/SPRAT	&	26.3	&	900	&	830	&	$<415$	&		\\
\rule{0pt}{3ex}
SN\,2020iq	&	2458898.8	&	Keck-1/LRIS	&	66.1	&	200	&	320	&	$<160^h$	&		\\
\rule{0pt}{3ex}
SN\,2019zrk	&	2458893.8	&	P200/DBSP	&	4.8	&	350	&	130	&	325$^h$	&		\\
	&	2458896.9	&	APO/DIS	&	7.8	&	400	&	370	&	$<185^h$	&		\\
	&	2458896.6	&	LT/SPRAT	&	7.6	&	750	&	830	&	$<415^h$	&		\\
	&	2458905.6	&	NOT/ALFOSC	&	16.6	&	800	&	640	&	480$^h$	&		\\
	&	2458930.9	&	Keck-1/LRIS	&	41.9	&	500	&	320	&	384$^h$	&	broad	\\
\rule{0pt}{3ex}
SN\,2020dcs	&	2458897.6	&	LT/SPRAT	&	2.7	&	850	&	830	&	$<415$	&		\\
\rule{0pt}{3ex}
SN\,2020dfh	&	2458906.9	&	ESO-NTT/EFOSC2	&	3.8	&	1000	&	980	&	$<490$	&		\\
\rule{0pt}{3ex}
SN\,2020edh	&	2458919.0	&	FTN/FLOYDS N-SIRAH	&	4.1	&	1000	&	800	&	600	&		
\enddata
\tablecomments{List of SN spectra with narrow lines or P~Cygni features. The typical resolution or the spectrograph (third column from the end) is subtracted from the measured width in quadrature, or we quote an upper limit if the result would be smaller than half the resolution. For most SNe, additional low-resolution spectra were obtained with the SED Machine.\\
Asterisks ($^*$) mark velocities measured from P~Cygni profiles and not from line widths.
Spectra marked with an $^h$ have strong host lines such that the measured velocities are less reliable.}
\end{deluxetable*}

\section*{Acknowledgements}
We thank A. Nyholm for his comments on the manuscript.
This work would not have been possible without the spectroscopic follow-up observations carried out by
S. Anand, D. Bektesevic, N. Blagorodnova, M. Bulla, S. B. Cenko, W. Chen, P. Chinchilla, R. Clavero Jimenez, C. Cunningham, A. Dahiwale, L. Dominguez, A. J. Drake, C. Frohmaier, F. J. Galindo-Guil, E. Hammerstein, T. Hung, N. Jannsen, J. Jencson, R. Karjalainen, H. Ko, M. Kuhn, E. McEwen, A. A. Miller, S. Moran, M. C. Ramirez-Tannus, A. Smith, E. Swann, K. Teet, J. Vinko, and J. Viuho.
We would like to thank participating observers on the UW APO ZTF follow-up team, including Brigitta Spi\H{o}cz, Eric Bellm, Zach Golkhou, Keaton Bell, and James Davenport.
In addition, we thank A. Gangopadhyay, H. Ko, and S. Prentice for reducing optical spectra and for sharing their data.

Based on observations obtained with the 48-inch Samuel Oschin Telescope and the 60-inch Telescope at Palomar Observatory as part of the Zwicky Transient Facility project. ZTF is supported by the National Science Foundation (NSF) under grant AST-1440341 and a collaboration including Caltech, IPAC, the Weizmann Institute for Science, the Oskar Klein Centre at Stockholm University, the University of Maryland, the University of Washington, Deutsches Elektronen-Synchrotron and Humboldt University, Los Alamos National Laboratories, the TANGO Consortium of Taiwan, the University of Wisconsin at Milwaukee, and the Lawrence Berkeley National Laboratory. Operations are conducted by COO, IPAC, and UW. 
The SED Machine is based upon work supported by NSF grant 1106171.
This work was supported by the GROWTH project funded by the NSF under PIRE grant 1545949.
Partially based on observations made with the Nordic Optical Telescope, operated by the Nordic Optical Telescope Scientific Association at the Observatorio del Roque de los Muchachos, La Palma, Spain, of the Instituto de Astrofisica de Canarias. Some of the data presented herein were obtained with ALFOSC. 
Some of the data presented herein were obtained at the W. M. Keck Observatory, which is operated as a scientific partnership among the California Institute of Technology, the University of California, and NASA; the observatory was made possible by the generous financial support of the W. M. Keck Foundation.
The authors wish to recognize and acknowledge the very significant cultural role and reverence that the summit of Maunakea has always had within the indigenous Hawaiian community. We are most fortunate to have the opportunity to conduct observations from this mountain.
The Liverpool Telescope is operated on the island of La Palma by Liverpool John Moores University in the Spanish Observatorio del Roque de los Muchachos of the Instituto de Astrofisica de Canarias with financial support from the UK Science and Technology Facilities Council. Research at Lick Observatory is partially supported by a
generous gift from Google.
The ztfquery code was funded by the European Research Council (ERC) under the European Union's Horizon 2020 research and innovation programme (grant agreement n759194 - USNAC, PI Rigault). 
%
E.O.O. is grateful for the support by grants from the Israel Science Foundation, Minerva, Israeli Ministry of Technology and Science, the US-Israel Binational Science Foundation, Weizmann-UK, Weizmann-Yale, and the Weizmann-Caltech grants.
A.G.Y.’s research is supported by the EU via ERC grant 725161, the ISF GW excellence center, an IMOS space infrastructure grant and BSF/Transformative and GIF grants, as well as The Benoziyo Endowment Fund for the Advancement of Science, the Deloro Institute for Advanced Research in Space and Optics, The Veronika A. Rabl Physics Discretionary Fund, Paul and Tina Gardner, Yeda-Sela and the WIS-CIT joint research grant;  A.G.Y. is the recipient of the Helen and Martin Kimmel Award for Innovative Investigation.
N.J.S.\ is grateful for the support by the ISF (grant 1770/19).
A.V.F. acknowledges support from the Christopher R. Redlich Fund,
the TABASGO Foundation, and the Miller Institute for Basic Research
in Science. 
L.T. acknowledges support from MIUR (PRIN 2017 grant 20179ZF5KS).
R.L. is supported by a Marie Sk\l{}odowska-Curie Individual Fellowship within the Horizon 2020 European Union (EU) Framework Programme for Research and Innovation (H2020-MSCA-IF-2017-794467).
M.L.G. acknowledges support from the DiRAC Institute in the Department of Astronomy at the University of Washington. The DiRAC Institute is supported through generous gifts from the Charles and Lisa Simonyi Fund for Arts and Sciences, and the Washington Research Foundation.

\bibliographystyle{aa}
\bibliography{precursors_ref.bib}

\begin{thebibliography}{132}
\expandafter\ifx\csname natexlab\endcsname\relax\def\natexlab#1{#1}\fi

\bibitem[{Ahumada {et~al.}(2019)Ahumada, Prieto, Almeida, Anders, Anderson,
  Andrews, Anguiano, Arcodia, Armengaud, Aubert, Avila, Avila-Reese, Badenes,
  Balland, Barger, Barrera-Ballesteros, Basu, Bautista, Beaton, Beers,
  Benavides, Bender, {et~al.}}]{ahumada2019}
Ahumada, R., Prieto, C.~A., Almeida, A., {et~al.} 2019, The Sixteenth Data
  Release of the Sloan Digital Sky Surveys: First Release from the APOGEE-2
  Southern Survey and Full Release of eBOSS Spectra

\bibitem[{{Arnett} \& {Meakin}(2011{\natexlab{a}})}]{arnett2011}
{Arnett}, W.~D. \& {Meakin}, C. 2011{\natexlab{a}}, \apj, 733, 78

\bibitem[{{Arnett} \& {Meakin}(2011{\natexlab{b}})}]{arnett2011b}
{Arnett}, W.~D. \& {Meakin}, C. 2011{\natexlab{b}}, \apj, 741, 33

\bibitem[{{Bellm} {et~al.}(2019){Bellm}, {Kulkarni}, {Graham}, {Dekany},
  {Smith}, {Riddle}, {Masci}, {Helou}, {Prince}, {Adams}, {Barbarino},
  {Barlow}, {Bauer}, {Beck}, {Belicki}, {Biswas}, {Blagorodnova}, {Bodewits},
  {Bolin}, {Brinnel}, {et~al.}}]{bellm2019}
{Bellm}, E.~C., {Kulkarni}, S.~R., {Graham}, M.~J., {et~al.} 2019, \pasp, 131,
  018002

\bibitem[{{Ben-Ami} {et~al.}(2012){Ben-Ami}, {Konidaris}, {Quimby}, {Davis},
  {Ngeow}, {Ritter}, \& {Rudy}}]{ben-ami2012}
{Ben-Ami}, S., {Konidaris}, N., {Quimby}, R., {et~al.} 2012, in Society of
  Photo-Optical Instrumentation Engineers (SPIE) Conference Series, Vol. 8446,
  Ground-based and Airborne Instrumentation for Astronomy IV, 844686

\bibitem[{{Bertin} {et~al.}(2002){Bertin}, {Mellier}, {Radovich}, {Missonnier},
  {Didelon}, \& {Morin}}]{bertin2002}
{Bertin}, E., {Mellier}, Y., {Radovich}, M., {et~al.} 2002, in Astronomical
  Society of the Pacific Conference Series, Vol. 281, Astronomical Data
  Analysis Software and Systems XI, ed. D.~A. {Bohlender}, D.~{Durand}, \&
  T.~H. {Handley}, 228

\bibitem[{{Bilinski} {et~al.}(2015){Bilinski}, {Smith}, {Li}, {Williams},
  {Zheng}, \& {Filippenko}}]{bilinski2015}
{Bilinski}, C., {Smith}, N., {Li}, W., {et~al.} 2015, \mnras, 450, 246

\bibitem[{{Blagorodnova} {et~al.}(2018){Blagorodnova}, {Neill}, {Walters},
  {Kulkarni}, {Fremling}, {Ben-Ami}, {Dekany}, {Fucik}, {Konidaris}, {Nash},
  {Ngeow}, {Ofek}, {O' Sullivan}, {Quimby}, {Ritter}, \&
  {Vyhmeister}}]{blagorodnova2018}
{Blagorodnova}, N., {Neill}, J.~D., {Walters}, R., {et~al.} 2018, \pasp, 130,
  035003

\bibitem[{{Bruch} {et~al.}(2020){Bruch}, {Gal-Yam}, {Schulze}, {Yaron}, {Yang},
  {Soumagnac}, {Rigault}, {Strotjohann}, {Ofek}, {Sollerman}, {Masci},
  {Barbarino}, {Ho}, {Fremling}, {Perley}, {Nordin}, {Cenko}, {Adams},
  {Adreoni}, {Bellm}, {Blagorodnova}, {Bulla}, {Burdge}, {De}, {Dhawan},
  {Drake}, {Duev}, {Dugas}, {Graham}, {Graham}, {Jencson}, {Karamehmetoglu},
  {Kasliwal}, {Kim}, {Kulkarni}, {Kupfer}, {Mahabal}, {Miller}, {Prince},
  {Riddle}, {Sharma}, {Smith}, {Taddia}, {Taggart}, {Walters}, \&
  {Yan}}]{bruch2020}
{Bruch}, R.~J., {Gal-Yam}, A., {Schulze}, S., {et~al.} 2020, arXiv e-prints,
  arXiv:2008.09986

\bibitem[{{Cardelli} {et~al.}(1989){Cardelli}, {Clayton}, \&
  {Mathis}}]{cardelli1989}
{Cardelli}, J.~A., {Clayton}, G.~C., \& {Mathis}, J.~S. 1989, \apj, 345, 245

\bibitem[{{Chambers} {et~al.}(2019){Chambers}, {Boer}, {Bulger}, {Fairlamb},
  {Huber}, {Lin}, {Lowe}, {Magnier}, {Schultz}, {Wainscoat}, {Willman},
  {Smith}, {Young}, {McBrien}, {Srivastav}, {Smartt}, {O'Neil}, {Clark}, {Sim},
  \& {Wright}}]{chambers2019}
{Chambers}, K.~C., {Boer}, T.~D., {Bulger}, J., {et~al.} 2019, Transient Name
  Server Discovery Report, 2019-796, 1

\bibitem[{{Chevalier} \& {Irwin}(2011)}]{chevalier2011}
{Chevalier}, R.~A. \& {Irwin}, C.~M. 2011, \apjl, 729, L6

\bibitem[{{Cutri} {et~al.}(2013){Cutri}, {Wright}, {Conrow}, {Fowler},
  {Eisenhardt}, {Grillmair}, {Kirkpatrick}, {Masci}, {McCallon}, {Wheelock},
  {Fajardo-Acosta}, {Yan}, {Benford}, {Harbut}, {Jarrett}, {Lake}, {Leisawitz},
  {Ressler}, {Stanford}, {Tsai}, {Liu}, {Helou}, {Mainzer}, {Gettings},
  {Gonzalez}, {Hoffman}, {Marsh}, {Padgett}, {Skrutskie}, {Beck}, {Papin}, \&
  {Wittman}}]{cutri2013}
{Cutri}, R.~M., {Wright}, E.~L., {Conrow}, T., {et~al.} 2013, {Explanatory
  Supplement to the AllWISE Data Release Products}, Explanatory Supplement to
  the AllWISE Data Release Products

\bibitem[{{Danieli} \& {Soker}(2019)}]{danieli2019}
{Danieli}, B. \& {Soker}, N. 2019, \mnras, 482, 2277

\bibitem[{{Dilday} {et~al.}(2012){Dilday}, {Howell}, {Cenko}, {Silverman},
  {Nugent}, {Sullivan}, {Ben-Ami}, {Bildsten}, {Bolte}, {Endl}, {Filippenko},
  {Gnat}, {Horesh}, {Hsiao}, {Kasliwal}, {Kirkman}, {Maguire}, {Marcy},
  {Moore}, {Pan}, {Parrent}, {Podsiadlowski}, {Quimby}, {Sternberg}, {Suzuki},
  {Tytler}, {Xu}, {Bloom}, {Gal-Yam}, {Hook}, {Kulkarni}, {Law}, {Ofek},
  {Polishook}, \& {Poznanski}}]{dilday2012}
{Dilday}, B., {Howell}, D.~A., {Cenko}, S.~B., {et~al.} 2012, Science, 337, 942

\bibitem[{{Duev} {et~al.}(2019){Duev}, {Mahabal}, {Masci}, {Graham},
  {Rusholme}, {Walters}, {Karmarkar}, {Frederick}, {Kasliwal}, {Rebbapragada},
  \& {Ward}}]{duev2019}
{Duev}, D.~A., {Mahabal}, A., {Masci}, F.~J., {et~al.} 2019, \mnras, 489, 3582

\bibitem[{{Elias-Rosa} {et~al.}(2016){Elias-Rosa}, {Pastorello}, {Benetti},
  {Cappellaro}, {Taubenberger}, {Terreran}, {Fraser}, {Brown}, {Tartaglia},
  {Morales-Garoffolo}, {Harmanen}, {Richardson}, {Artigau}, {Tomasella},
  {Margutti}, {Smartt}, {Dennefeld}, {Turatto}, {Anupama}, {Arbour}, {Berton},
  {Bjorkman}, {Boles}, {Briganti}, {Chornock}, {Ciabattari}, {Cortini},
  {Dimai}, {Gerhartz}, {Itagaki}, {Kotak}, {Mancini}, {Martinelli},
  {Milisavljevic}, {Misra}, {Ochner}, {Patnaude}, {Polshaw}, {Sahu}, \&
  {Zaggia}}]{elias-rosa2016}
{Elias-Rosa}, N., {Pastorello}, A., {Benetti}, S., {et~al.} 2016, \mnras, 463,
  3894

\bibitem[{{Filippenko}(1997)}]{filippenko1997}
{Filippenko}, A.~V. 1997, \araa, 35, 309

\bibitem[{{Finkbeiner} {et~al.}(2004){Finkbeiner}, {Padmanabhan}, {Schlegel},
  {Carr}, {Gunn}, {Rockosi}, {Sekiguchi}, {Lupton}, {Knapp}, {Ivezi{\'c}},
  {Blanton}, {Hogg}, {Adelman-McCarthy}, {Annis}, {Hayes}, {Kinney}, {Long},
  {Seljak}, {Strauss}, {Yanny}, {Ag{\"u}eros}, {Allam}, {Anderson}, {Bahcall},
  {Baldry}, {Bernardi}, {Boroski}, {Briggs}, {Brinkmann}, {Brunner},
  {Budav{\'a}ri}, {Castander}, {Covey}, {Csabai}, {Doi}, {Dong}, {Eisenstein},
  {Fan}, {Friedman}, {Fukugita}, {et~al.}}]{finkbeiner2004}
{Finkbeiner}, D.~P., {Padmanabhan}, N., {Schlegel}, D.~J., {et~al.} 2004, \aj,
  128, 2577

\bibitem[{Foley {et~al.}(2011)Foley, Berger, Fox, Levesque, Challis, Ivans,
  Rhoads, \& Soderberg}]{foley2011}
Foley, R.~J., Berger, E., Fox, O., {et~al.} 2011, ApJ, 732, 32

\bibitem[{{Foley} {et~al.}(2007){Foley}, {Smith}, {Ganeshalingam}, {Li},
  {Chornock}, \& {Filippenko}}]{foley2007}
{Foley}, R.~J., {Smith}, N., {Ganeshalingam}, M., {et~al.} 2007, \apjl, 657,
  L105

\bibitem[{{Fraser} {et~al.}(2013){Fraser}, {Magee}, {Kotak}, {Smartt}, {Smith},
  {Polshaw}, {Drake}, {Boles}, {Lee}, {Burgett}, {Chambers}, {Draper},
  {Flewelling}, {Hodapp}, {Kaiser}, {Kudritzki}, {Magnier}, {Price}, {Tonry},
  {Wainscoat}, \& {Waters}}]{fraser2013}
{Fraser}, M., {Magee}, M., {Kotak}, R., {et~al.} 2013, \apjl, 779, L8

\bibitem[{{Fremling} {et~al.}(2019){Fremling}, {Dugas}, \&
  {Sharma}}]{fremling2019}
{Fremling}, C., {Dugas}, A., \& {Sharma}, Y. 2019, Transient Name Server
  Classification Report, 2019-188, 1

\bibitem[{{Fremling} {et~al.}(2020){Fremling}, {Miller}, {Sharma}, {Dugas},
  {Perley}, {Taggart}, {Sollerman}, {Goobar}, {Graham}, {Neill}, {Nordin},
  {Rigault}, {Walters}, {Andreoni}, {Bagdasaryan}, {Belicki}, {Cannella},
  {Bellm}, {Cenko}, {De}, {Dekany}, {Frederick}, {Golkhou}, {Graham}, {Helou},
  {Ho}, {Kasliwal}, {Kupfer}, {Laher}, {Mahabal}, {Masci}, {Riddle},
  {Rusholme}, {Schulze}, {Shupe}, {Smith}, {van Velzen}, {Yan}, {Yao},
  {Zhuang}, \& {Kulkarni}}]{fremling2020}
{Fremling}, C., {Miller}, A.~A., {Sharma}, Y., {et~al.} 2020, \apj, 895, 32

\bibitem[{{Fuller}(2017)}]{fuller2017}
{Fuller}, J. 2017, \mnras, 470, 1642

\bibitem[{{Fuller} \& {Ro}(2018)}]{fuller2018}
{Fuller}, J. \& {Ro}, S. 2018, \mnras, 476, 1853

\bibitem[{Gal-Yam(2017)}]{gal-yam2017}
Gal-Yam, A. 2017, Handbook of Supernovae, 195–237

\bibitem[{{Gal-Yam}(2019)}]{gal-yam2019b}
{Gal-Yam}, A. 2019, in American Astronomical Society Meeting Abstracts, Vol.
  233, American Astronomical Society Meeting Abstracts \#233, 131.06

\bibitem[{Gal-Yam(2019)}]{gal-yam2019}
Gal-Yam, A. 2019, \araa, 57, 305–333

\bibitem[{{Gal-Yam} {et~al.}(2014){Gal-Yam}, {Arcavi}, {Ofek}, {Ben-Ami},
  {Cenko}, {Kasliwal}, {Cao}, {Yaron}, {Tal}, {Silverman}, {Horesh}, {De Cia},
  {Taddia}, {Sollerman}, {Perley}, {Vreeswijk}, {Kulkarni}, {Nugent},
  {Filippenko}, \& {Wheeler}}]{gal-yam2014}
{Gal-Yam}, A., {Arcavi}, I., {Ofek}, E.~O., {et~al.} 2014, \nat, 509, 471

\bibitem[{{Gal-Yam} \& {Leonard}(2009)}]{gal-yam2009}
{Gal-Yam}, A. \& {Leonard}, D.~C. 2009, \nat, 458, 865

\bibitem[{{Gal-Yam} {et~al.}(2007){Gal-Yam}, {Leonard}, {Fox}, {Cenko},
  {Soderberg}, {Moon}, {Sand }, {Caltech Core Collapse Program}, {Li},
  {Filippenko}, {Aldering}, \& {Copin}}]{gal-yam2007}
{Gal-Yam}, A., {Leonard}, D.~C., {Fox}, D.~B., {et~al.} 2007, \apj, 656, 372

\bibitem[{Gangopadhyay {et~al.}(2020)Gangopadhyay, Misra, Hiramatsu, Wang,
  Hosseinzadeh, Wang, Valenti, Zhang, Howell, Arcavi, \&
  et~al.}]{gangopadhyay2020}
Gangopadhyay, A., Misra, K., Hiramatsu, D., {et~al.} 2020, ApJ, 889, 170

\bibitem[{{Ginzburg} \& {Balberg}(2012)}]{ginzburg2012}
{Ginzburg}, S. \& {Balberg}, S. 2012, \apj, 757, 178

\bibitem[{{Glas} {et~al.}(2019){Glas}, {Just}, {Janka}, \&
  {Obergaulinger}}]{glas2019}
{Glas}, R., {Just}, O., {Janka}, H.~T., \& {Obergaulinger}, M. 2019, \apj, 873,
  45

\bibitem[{{Graham} {et~al.}(2019){Graham}, {Kulkarni}, {Bellm}, {Adams},
  {Barbarino}, {Blagorodnova}, {Bodewits}, {Bolin}, {Brady}, {Cenko}, {Chang},
  {Coughlin}, {De}, {Eadie}, {Farnham}, {Feindt}, {Franckowiak}, {Fremling},
  {Gezari}, {Ghosh}, {Goldstein}, {Golkhou}, {Goobar}, {Ho}, {Huppenkothen},
  {et~al.}}]{graham2019}
{Graham}, M.~J., {Kulkarni}, S.~R., {Bellm}, E.~C., {et~al.} 2019, \pasp, 131,
  078001

\bibitem[{{Hamuy} {et~al.}(2003){Hamuy}, {Phillips}, {Suntzeff}, {Maza},
  {Gonz{\'a}lez}, {Roth}, {Krisciunas}, {Morrell}, {Green}, {Persson}, \&
  {McCarthy}}]{hamuy2003}
{Hamuy}, M., {Phillips}, M.~M., {Suntzeff}, N.~B., {et~al.} 2003, \nat, 424,
  651

\bibitem[{{Ho} {et~al.}(2019){Ho}, {Goldstein}, {Schulze}, {Khatami}, {Perley},
  {Ergon}, {Gal-Yam}, {Corsi}, {Andreoni}, {Barbarino}, {Bellm},
  {Blagorodnova}, {Bright}, {Burns}, {Cenko}, {Cunningham}, {De}, {Dekany},
  {Dugas}, {Fender}, {Fransson}, {Fremling}, {Goldstein}, {Graham}, {Hale},
  {Horesh}, {Hung}, {Kasliwal}, {Kuin}, {Kulkarni}, {Kupfer}, {Lunnan},
  {Masci}, {Ngeow}, {Nugent}, {Ofek}, {Patterson}, {Petitpas}, {Rusholme},
  {Sai}, {Sfaradi}, {Shupe}, {Sollerman}, {Soumagnac}, {Tachibana}, {Taddia},
  {Walters}, {Wang}, {Yao}, \& {Zhang}}]{ho2019}
{Ho}, A. Y.~Q., {Goldstein}, D.~A., {Schulze}, S., {et~al.} 2019, \apj, 887,
  169

\bibitem[{Hosseinzadeh {et~al.}(2017)Hosseinzadeh, Arcavi, Valenti, McCully,
  Howell, Johansson, Sollerman, Pastorello, Benetti, Cao, \&
  et~al.}]{hosseinzadeh2017}
Hosseinzadeh, G., Arcavi, I., Valenti, S., {et~al.} 2017, ApJ, 836, 158

\bibitem[{Hosseinzadeh {et~al.}(2019)Hosseinzadeh, McCully, Zabludoff, Arcavi,
  French, Howell, Berger, \& Hiramatsu}]{hosseinzadeh2019}
Hosseinzadeh, G., McCully, C., Zabludoff, A.~I., {et~al.} 2019, ApJ, 871, L9

\bibitem[{Humphreys \& Davidson(1994)}]{humphreys1994}
Humphreys, R.~M. \& Davidson, K. 1994, PASP, 106, 1025

\bibitem[{{Janka} {et~al.}(2016){Janka}, {Melson}, \& {Summa}}]{janka2016}
{Janka}, H.-T., {Melson}, T., \& {Summa}, A. 2016, Annual Review of Nuclear and
  Particle Science, 66, 341

\bibitem[{{Justham} {et~al.}(2014){Justham}, {Podsiadlowski}, \&
  {Vink}}]{justham2014}
{Justham}, S., {Podsiadlowski}, P., \& {Vink}, J.~S. 2014, \apj, 796, 121

\bibitem[{{Karamehmetoglu} {et~al.}(2019){Karamehmetoglu}, {Fransson},
  {Sollerman}, {Tartaglia}, {Taddia}, {De}, {Fremling}, {Bagdasaryan},
  {Barbarino}, {Bellm}, {Dekaney}, {Dugas}, {Giomi}, {Goobar}, {Graham}, {Ho},
  {Laher}, {Masci}, {Neill}, {Perley}, {Riddle}, {Rusholme}, \&
  {Soumagnac}}]{karamehmtoglu2019}
{Karamehmetoglu}, E., {Fransson}, C., {Sollerman}, J., {et~al.} 2019, arXiv
  e-prints, arXiv:1910.06016

\bibitem[{{Kasliwal} {et~al.}(2019){Kasliwal}, {Cannella}, {Bagdasaryan},
  {Hung}, {Feindt}, {Singer}, {Coughlin}, {Fremling}, {Walters}, {Duev},
  {Itoh}, \& {Quimby}}]{kasliwal2019}
{Kasliwal}, M.~M., {Cannella}, C., {Bagdasaryan}, A., {et~al.} 2019, \pasp,
  131, 038003

\bibitem[{{Katz} {et~al.}(2011){Katz}, {Sapir}, \& {Waxman}}]{katz2011}
{Katz}, B., {Sapir}, N., \& {Waxman}, E. 2011, arXiv e-prints, arXiv:1106.1898

\bibitem[{Khazov {et~al.}(2016)Khazov, Yaron, Gal-Yam, Manulis, Rubin,
  Kulkarni, Arcavi, Kasliwal, Ofek, Cao, \& et~al.}]{khazov2016}
Khazov, D., Yaron, O., Gal-Yam, A., {et~al.} 2016, ApJ, 818, 3

\bibitem[{{Kiewe} {et~al.}(2012){Kiewe}, {Gal-Yam}, {Arcavi}, {Leonard},
  {Emilio Enriquez}, {Cenko}, {Fox}, {Moon}, {Sand }, {Soderberg}, \&
  {CCCP}}]{kiewe2012}
{Kiewe}, M., {Gal-Yam}, A., {Arcavi}, I., {et~al.} 2012, \apj, 744, 10

\bibitem[{{Kochanek}(2011)}]{kochanek2011b}
{Kochanek}, C.~S. 2011, \apj, 741, 37

\bibitem[{Kochanek \& Szczygiel(2011)}]{kochanek2011}
Kochanek, C.~S. \& Szczygiel, D.~M. 2011, ApJ, 737, 76

\bibitem[{{Kool} {et~al.}(2020){Kool}, {Karamehmetoglu}, {Sollerman},
  {Schulze}, {Lunnan}, {Reynolds}, {Barbarino}, {Bellm}, {De}, {Duev},
  {Fremling}, {Golkhou}, {Graham}, {Green}, {Horesh}, {Kaye}, {Kim}, {Laher},
  {Masci1}, {Nordin}, {Perley}, {Phinney}, {Porter}, {Reiley}, {Rodriguez},
  {van Roestel}, {Rusholme}, {Sharma}, {Sfaradi}, {Soumagnac}, {Taggart},
  {Tartaglia}, {Williams}, \& {Yan}}]{kool2020}
{Kool}, E.~C., {Karamehmetoglu}, E., {Sollerman}, J., {et~al.} 2020, arXiv
  e-prints, arXiv:2008.04056

\bibitem[{{Law} {et~al.}(2009){Law}, {Kulkarni}, {Dekany}, {Ofek}, {Quimby},
  {Nugent}, {Surace}, {Grillmair}, {Bloom}, {Kasliwal}, {Bildsten}, {Brown},
  {Cenko}, {Ciardi}, {Croner}, {Djorgovski}, {van Eyken}, {Filippenko}, {Fox},
  {Gal-Yam}, {Hale}, {Hamam}, {Helou}, {Henning}, {Howell}, {Jacobsen},
  {Laher}, {Mattingly}, {McKenna}, {Pickles}, {Poznanski}, {Rahmer}, {Rau},
  {Rosing}, {Shara}, {Smith}, {Starr}, {Sullivan}, {Velur}, {Walters}, \&
  {Zolkower}}]{law2009}
{Law}, N.~M., {Kulkarni}, S.~R., {Dekany}, R.~G., {et~al.} 2009, \pasp, 121,
  1395

\bibitem[{{Liu} {et~al.}(2011){Liu}, {Shen}, {Strauss}, \& {Hao}}]{liu2011}
{Liu}, X., {Shen}, Y., {Strauss}, M.~A., \& {Hao}, L. 2011, \apj, 737, 101

\bibitem[{{Lupton} {et~al.}(1999){Lupton}, {Gunn}, \& {Szalay}}]{lupton1999}
{Lupton}, R.~H., {Gunn}, J.~E., \& {Szalay}, A.~S. 1999, \aj, 118, 1406

\bibitem[{{Malmquist}(1922)}]{malmquist1922}
{Malmquist}, K.~G. 1922, Meddelanden fran Lunds Astronomiska Observatorium
  Serie I, 100, 1

\bibitem[{Margutti {et~al.}(2013)Margutti, Milisavljevic, Soderberg, Chornock,
  Zauderer, Murase, Guidorzi, Sanders, Kuin, Fransson, \&
  et~al.}]{margutti2013}
Margutti, R., Milisavljevic, D., Soderberg, A.~M., {et~al.} 2013, ApJ, 780, 21

\bibitem[{{Masci} {et~al.}(2019){Masci}, {Laher}, {Rusholme}, {Shupe}, {Groom},
  {Surace}, {Jackson}, {Monkewitz}, {Beck}, {Flynn}, {Terek}, {Landry},
  {Hacopians}, {Desai}, {Howell}, {Brooke}, {Imel}, {Wachter}, {Ye}, {Lin},
  {Cenko}, {Cunningham}, {Rebbapragada}, {Bue}, {Miller}, {Mahabal}, {Bellm},
  {Patterson}, {Juri{\'c}}, {Golkhou}, {Ofek}, {Walters}, {Graham}, {Kasliwal},
  {Dekany}, {Kupfer}, {Burdge}, {Cannella}, {Barlow}, {Van Sistine}, {Giomi},
  {Fremling}, {Blagorodnova}, {Levitan}, {Riddle}, {Smith}, {Helou}, {Prince},
  \& {Kulkarni}}]{masci2019}
{Masci}, F.~J., {Laher}, R.~R., {Rusholme}, B., {et~al.} 2019, \pasp, 131,
  018003

\bibitem[{{Mauerhan} {et~al.}(2013){Mauerhan}, {Smith}, {Filippenko},
  {Blanchard}, {Blanchard}, {Casper}, {Cenko}, {Clubb}, {Cohen}, {Fuller},
  {Li}, \& {Silverman}}]{mauerhan2013}
{Mauerhan}, J.~C., {Smith}, N., {Filippenko}, A.~V., {et~al.} 2013, \mnras,
  430, 1801

\bibitem[{{Mcley} \& {Soker}(2014)}]{mcley2014}
{Mcley}, L. \& {Soker}, N. 2014, \mnras, 445, 2492

\bibitem[{{Meakin} \& {Arnett}(2006)}]{meakin2006}
{Meakin}, C.~A. \& {Arnett}, D. 2006, \apjl, 637, L53

\bibitem[{{Moriya}(2014)}]{moriya2014b}
{Moriya}, T.~J. 2014, \aap, 564, A83

\bibitem[{Moriya(2015)}]{moriya2015}
Moriya, T.~J. 2015, The Astrophysical Journal, 803, L26

\bibitem[{{Moriya} \& {Maeda}(2016)}]{moriya2016}
{Moriya}, T.~J. \& {Maeda}, K. 2016, \apj, 824, 100

\bibitem[{{Moriya} {et~al.}(2014){Moriya}, {Maeda}, {Taddia}, {Sollerman},
  {Blinnikov}, \& {Sorokina}}]{moriya2014}
{Moriya}, T.~J., {Maeda}, K., {Taddia}, F., {et~al.} 2014, \mnras, 439, 2917

\bibitem[{{M{\"u}ller}(2016)}]{mueller2016}
{M{\"u}ller}, B. 2016, \pasa, 33, e048

\bibitem[{{Murase} {et~al.}(2011){Murase}, {Thompson}, {Lacki}, \&
  {Beacom}}]{murase2011}
{Murase}, K., {Thompson}, T.~A., {Lacki}, B.~C., \& {Beacom}, J.~F. 2011, \prd,
  84, 043003

\bibitem[{{Murase} {et~al.}(2014){Murase}, {Thompson}, \& {Ofek}}]{murase2014}
{Murase}, K., {Thompson}, T.~A., \& {Ofek}, E.~O. 2014, \mnras, 440, 2528

\bibitem[{{Nordin} {et~al.}(2019){Nordin}, {Brinnel}, {van Santen}, {Bulla},
  {Feindt}, {Franckowiak}, {Fremling}, {Gal-Yam}, {Giomi}, {Kowalski},
  {Mahabal}, {Miranda}, {Rauch}, {Reusch}, {Rigault}, {Schulze}, {Sollerman},
  {Stein}, {Yaron}, {van Velzen}, \& {Ward}}]{nordin2019}
{Nordin}, J., {Brinnel}, V., {van Santen}, J., {et~al.} 2019, \aap, 631, A147

\bibitem[{Nyholm {et~al.}(2017)Nyholm, Sollerman, Taddia, Fremling, Moriya,
  Ofek, Gal-Yam, De~Cia, Roy, Kasliwal, \& et~al.}]{nyholm2017}
Nyholm, A., Sollerman, J., Taddia, F., {et~al.} 2017, Astronomy \&
  Astrophysics, 605, A6

\bibitem[{Nyholm {et~al.}(2020)Nyholm, Sollerman, Tartaglia, Taddia, Fremling,
  Blagorodnova, Filippenko, Gal-Yam, Howell, Karamehmetoglu, \&
  et~al.}]{nyholm2020}
Nyholm, A., Sollerman, J., Tartaglia, L., {et~al.} 2020, Astronomy \&
  Astrophysics, 637, A73

\bibitem[{{Ofek}(2019)}]{ofek2019}
{Ofek}, E.~O. 2019, \pasp, 131, 054504

\bibitem[{{Ofek} {et~al.}(2016){Ofek}, {Cenko}, {Shaviv}, {Duggan},
  {Strotjohann}, {Rubin}, {Kulkarni}, {Gal-Yam}, {Sullivan}, {Cao}, {Nugent},
  {Kasliwal}, {Sollerman}, {Fransson}, {Filippenko}, {Perley}, {Yaron}, \&
  {Laher}}]{ofek2016}
{Ofek}, E.~O., {Cenko}, S.~B., {Shaviv}, N.~J., {et~al.} 2016, \apj, 824, 6

\bibitem[{{Ofek} {et~al.}(2013{\natexlab{a}}){Ofek}, {Lin}, {Kouveliotou},
  {Younes}, {G{\"o}{\v{g}}{\"u}{\textcommabelow s}}, {Kasliwal}, \&
  {Cao}}]{ofek2013b}
{Ofek}, E.~O., {Lin}, L., {Kouveliotou}, C., {et~al.} 2013{\natexlab{a}}, \apj,
  768, 47

\bibitem[{{Ofek} {et~al.}(2010){Ofek}, {Rabinak}, {Neill}, {Arcavi}, {Cenko},
  {Waxman}, {Kulkarni}, {Gal-Yam}, {Nugent}, {Bildsten}, {Bloom}, {Filippenko},
  {Forster}, {Howell}, {Jacobsen}, {Kasliwal}, {Law}, {Martin}, {Poznanski},
  {Quimby}, {Shen}, {Sullivan}, {Dekany}, {Rahmer}, {Hale}, {Smith},
  {Zolkower}, {Velur}, {Walters}, {Henning}, {Bui}, \& {McKenna}}]{ofek2010}
{Ofek}, E.~O., {Rabinak}, I., {Neill}, J.~D., {et~al.} 2010, \apj, 724, 1396

\bibitem[{{Ofek} {et~al.}(2013{\natexlab{b}}){Ofek}, {Sullivan}, {Cenko},
  {Kasliwal}, {Gal-Yam}, {Kulkarni}, {Arcavi}, {Bildsten}, {Bloom}, {Horesh},
  {Howell}, {Filippenko}, {Laher}, {Murray}, {Nakar}, {Nugent}, {Silverman},
  {Shaviv}, {Surace}, \& {Yaron}}]{ofek2013}
{Ofek}, E.~O., {Sullivan}, M., {Cenko}, S.~B., {et~al.} 2013{\natexlab{b}},
  \nat, 494, 65

\bibitem[{{Ofek} {et~al.}(2014{\natexlab{a}}){Ofek}, {Sullivan}, {Shaviv},
  {Steinbok}, {Arcavi}, {Gal-Yam}, {Tal}, {Kulkarni}, {Nugent}, {Ben-Ami},
  {Kasliwal}, {Cenko}, {Laher}, {Surace}, {Bloom}, {Filippenko}, {Silverman},
  \& {Yaron}}]{ofek2014}
{Ofek}, E.~O., {Sullivan}, M., {Shaviv}, N.~J., {et~al.} 2014{\natexlab{a}},
  \apj, 789, 104

\bibitem[{{Ofek} {et~al.}(2014{\natexlab{b}}){Ofek}, {Zoglauer}, {Boggs},
  {Barri{\'e}re}, {Reynolds}, {Fryer}, {Harrison}, {Cenko}, {Kulkarni},
  {Gal-Yam}, {Arcavi}, {Bellm}, {Bloom}, {Christensen}, {Craig}, {Even},
  {Filippenko}, {Grefenstette}, {Hailey}, {Laher}, {Madsen}, {Nakar}, {Nugent},
  {Stern}, {Sullivan}, {Surace}, \& {Zhang}}]{ofek2014b}
{Ofek}, E.~O., {Zoglauer}, A., {Boggs}, S.~E., {et~al.} 2014{\natexlab{b}},
  \apj, 781, 42

\bibitem[{{Osborn} {et~al.}(2015){Osborn}, {F{\"o}hring}, {Dhillon}, \&
  {Wilson}}]{osborn2015}
{Osborn}, J., {F{\"o}hring}, D., {Dhillon}, V.~S., \& {Wilson}, R.~W. 2015,
  \mnras, 452, 1707

\bibitem[{{Owocki} {et~al.}(2019){Owocki}, {Hirai}, {Podsiadlowski}, \&
  {Schneider}}]{owocki2019}
{Owocki}, S.~P., {Hirai}, R., {Podsiadlowski}, P., \& {Schneider}, F. R.~N.
  2019, \mnras, 485, 988

\bibitem[{{Owocki} \& {Shaviv}(2016)}]{owocki2016}
{Owocki}, S.~P. \& {Shaviv}, N.~J. 2016, \mnras, 462, 345

\bibitem[{{Pastorello} {et~al.}(2013){Pastorello}, {Cappellaro}, {Inserra},
  {Smartt}, {Pignata}, {Benetti}, {Valenti}, {Fraser}, {Tak{\'a}ts}, {Benitez},
  {Botticella}, {Brimacombe}, {Bufano}, {Cellier-Holzem}, {Costado}, {Cupani},
  {Curtis}, {Elias-Rosa}, {Ergon}, {Fynbo}, {Hambsch}, {Hamuy}, {Harutyunyan},
  {Ivarson}, {Kankare}, {Martin}, {Kotak}, {LaCluyze}, {Maguire}, {Mattila},
  {Maza}, {McCrum}, {Miluzio}, {Norgaard-Nielsen}, {Nysewander}, {Ochner},
  {Pan}, {Pumo}, {Reichart}, {Tan}, {Taubenberger}, {Tomasella}, {Turatto}, \&
  {Wright}}]{pastorello2013}
{Pastorello}, A., {Cappellaro}, E., {Inserra}, C., {et~al.} 2013, \apj, 767, 1

\bibitem[{{Pastorello} {et~al.}(2018){Pastorello}, {Kochanek}, {Fraser},
  {Dong}, {Elias-Rosa}, {Filippenko}, {Benetti}, {Cappellaro}, {Tomasella},
  {Drake}, {Harmanen}, {Reynolds}, {Shappee}, {Smartt}, {Chambers}, {Huber},
  {Smith}, {Stanek}, {Christensen}, {Denneau}, {Djorgovski}, {Flewelling},
  {Gall}, {Gal-Yam}, {Geier}, {Heinze}, {Holoien}, {Isern}, {Kangas},
  {Kankare}, {Koff}, {Llapasset}, {Lowe}, {Lundqvist}, {Magnier}, {Mattila},
  {Morales-Garoffolo}, {Mutel}, {Nicolas}, {Ochner}, {Ofek}, {Prosperi},
  {Rest}, {Sano}, {Stalder}, {Stritzinger}, {Taddia}, {Terreran}, {Tonry},
  {Wainscoat}, {Waters}, {Weiland}, {Willman}, {Young}, \&
  {Zheng}}]{pastorello2018}
{Pastorello}, A., {Kochanek}, C.~S., {Fraser}, M., {et~al.} 2018, \mnras, 474,
  197

\bibitem[{{Pastorello} {et~al.}(2007){Pastorello}, {Smartt}, {Mattila},
  {Eldridge}, {Young}, {Itagaki}, {Yamaoka}, {Navasardyan}, {Valenti}, {Patat},
  {Agnoletto}, {Augusteijn}, {Benetti}, {Cappellaro}, {Boles}, {Bonnet-Bidaud},
  {Botticella}, {Bufano}, {Cao}, {Deng}, {Dennefeld}, {Elias-Rosa},
  {Harutyunyan}, {Keenan}, {Iijima}, {Lorenzi}, {Mazzali}, {Meng}, {Nakano},
  {Nielsen}, {Smoker}, {Stanishev}, {Turatto}, {Xu}, \&
  {Zampieri}}]{pastorello2007}
{Pastorello}, A., {Smartt}, S.~J., {Mattila}, S., {et~al.} 2007, \nat, 447, 829

\bibitem[{{Pastorello} {et~al.}(2016){Pastorello}, {Wang}, {Ciabattari},
  {Bersier}, {Mazzali}, {Gao}, {Xu}, {Zhang}, {Tokuoka}, {Benetti},
  {Cappellaro}, {Elias-Rosa}, {Harutyunyan}, {Huang}, {Miluzio}, {Mo},
  {Ochner}, {Tartaglia}, {Terreran}, {Tomasella}, \&
  {Turatto}}]{pastorello2016}
{Pastorello}, A., {Wang}, X.~F., {Ciabattari}, F., {et~al.} 2016, \mnras, 456,
  853

\bibitem[{{Patat} {et~al.}(2011){Patat}, {Taubenberger}, {Benetti},
  {Pastorello}, \& {Harutyunyan}}]{patat2011}
{Patat}, F., {Taubenberger}, S., {Benetti}, S., {Pastorello}, A., \&
  {Harutyunyan}, A. 2011, \aap, 527, L6

\bibitem[{{Patterson} {et~al.}(2019){Patterson}, {Bellm}, {Rusholme}, {Masci},
  {Juric}, {Krughoff}, {Golkhou}, {Graham}, {Kulkarni}, {Helou}, \& {Zwicky
  Transient Facility Collaboration}}]{patterson2019}
{Patterson}, M.~T., {Bellm}, E.~C., {Rusholme}, B., {et~al.} 2019, \pasp, 131,
  018001

\bibitem[{{Paxton} {et~al.}(2011){Paxton}, {Bildsten}, {Dotter}, {Herwig},
  {Lesaffre}, \& {Timmes}}]{paxton2011}
{Paxton}, B., {Bildsten}, L., {Dotter}, A., {et~al.} 2011, \apjs, 192, 3

\bibitem[{{Perley} {et~al.}(2020){Perley}, {Fremling}, {Sollerman}, {Miller},
  {Dahiwale}, {Sharma}, {Bellm}, {Biswas}, {Brink}, {Bruch}, {De}, {Dekany},
  {Drake}, {Duev}, {Filippenko}, {Gal-Yam}, {Goobar}, {Graham}, {Graham}, {Ho},
  {Irani}, {Kasliwal}, {Kim}, {Kulkarni}, {Mahabal}, {Masci}, {Modak}, {Neill},
  {Nordin}, {Riddle}, {Soumagnac}, {Strotjohann}, {Schulze}, {Taggart},
  {Tzanidakis}, {Walters}, \& {Yan}}]{perley2020}
{Perley}, D.~A., {Fremling}, C., {Sollerman}, J., {et~al.} 2020, \apj, 904, 35

\bibitem[{Piro \& Nakar(2013)}]{piro2013}
Piro, A.~L. \& Nakar, E. 2013, ApJ, 769, 67

\bibitem[{{Price-Whelan} {et~al.}(2018){Price-Whelan}, {Sip{\H{o}}cz},
  {G{\"u}nther}, {Lim}, {Crawford}, {Conseil}, {Shupe}, {Craig}, {Dencheva},
  {Ginsburg}, {VanderPlas}, {Bradley}, {P{\'e}rez-Su{\'a}rez}, {de Val-Borro},
  {Paper Contributors}, {Aldcroft}, {Cruz}, {Robitaille}, {Tollerud},
  {Coordination Committee}, {Ardelean}, {Babej}, {Bach}, {Bachetti}, {Bakanov},
  {Bamford}, {Barentsen}, {Barmby}, {Baumbach}, {Berry}, {Biscani}, {Boquien},
  {Bostroem}, {Bouma}, {Brammer}, {Bray}, {Breytenbach}, {Buddelmeijer},
  {Burke}, {Calderone}, {Cano Rodr{\'\i}guez}, {Cara}, {Cardoso}, {Cheedella},
  {Copin}, {Corrales}, {Crichton}, {D{\textquoteright}Avella}, {Deil},
  {Depagne}, {Dietrich}, {Donath}, {Droettboom}, {Earl}, {Erben}, {Fabbro},
  {Ferreira}, {Finethy}, {Fox}, {Garrison}, {Gibbons}, {Goldstein}, {Gommers},
  {Greco}, {Greenfield}, {Groener}, {Grollier}, {Hagen}, {Hirst}, {Homeier},
  {Horton}, {Hosseinzadeh}, {Hu}, {Hunkeler}, {Ivezi{\'c}}, {Jain}, {Jenness},
  {Kanarek}, {Kendrew}, {Kern}, {Kerzendorf}, {Khvalko}, {King}, {Kirkby},
  {Kulkarni}, {Kumar}, {Lee}, {Lenz}, {Littlefair}, {Ma}, {Macleod},
  {Mastropietro}, {McCully}, {Montagnac}, {Morris}, {Mueller}, {Mumford},
  {Muna}, {Murphy}, {Nelson}, {Nguyen}, {Ninan}, {N{\"o}the}, {Ogaz}, {Oh},
  {Parejko}, {Parley}, {Pascual}, {Patil}, {Patil}, {Plunkett}, {Prochaska},
  {Rastogi}, {Reddy Janga}, {Sabater}, {Sakurikar}, {Seifert}, {Sherbert},
  {Sherwood-Taylor}, {Shih}, {Sick}, {Silbiger}, {Singanamalla}, {Singer},
  {Sladen}, {Sooley}, {Sornarajah}, {Streicher}, {Teuben}, {Thomas},
  {Tremblay}, {Turner}, {Terr{\'o}n}, {van Kerkwijk}, {de la Vega}, {Watkins},
  {Weaver}, {Whitmore}, {Woillez}, {Zabalza}, \& {Contributors}}]{astropy2018}
{Price-Whelan}, A.~M., {Sip{\H{o}}cz}, B.~M., {G{\"u}nther}, H.~M., {et~al.}
  2018, \aj, 156, 123

\bibitem[{Prieto {et~al.}(2013)Prieto, Brimacombe, Drake, \&
  Howerton}]{prieto2013}
Prieto, J.~L., Brimacombe, J., Drake, A.~J., \& Howerton, S. 2013, ApJ, 763,
  L27

\bibitem[{{Prieto} {et~al.}(2008){Prieto}, {Kistler}, {Thompson}, {Y{\"u}ksel},
  {Kochanek}, {Stanek}, {Beacom}, {Martini}, {Pasquali}, \&
  {Bechtold}}]{prieto2008}
{Prieto}, J.~L., {Kistler}, M.~D., {Thompson}, T.~A., {et~al.} 2008, \apjl,
  681, L9

\bibitem[{{Quataert} {et~al.}(2016){Quataert}, {Fern{\'a}ndez}, {Kasen},
  {Klion}, \& {Paxton}}]{quataert2016}
{Quataert}, E., {Fern{\'a}ndez}, R., {Kasen}, D., {Klion}, H., \& {Paxton}, B.
  2016, \mnras, 458, 1214

\bibitem[{{Quataert} \& {Shiode}(2012)}]{quataert2012}
{Quataert}, E. \& {Shiode}, J. 2012, \mnras, 423, L92

\bibitem[{{Rau} {et~al.}(2009){Rau}, {Kulkarni}, {Law}, {Bloom}, {Ciardi},
  {Djorgovski}, {Fox}, {Gal-Yam}, {Grillmair}, {Kasliwal}, {Nugent}, {Ofek},
  {Quimby}, {Reach}, {Shara}, {Bildsten}, {Cenko}, {Drake}, {Filippenko},
  {Helfand}, {Helou}, {Howell}, {Poznanski}, \& {Sullivan}}]{rau2009}
{Rau}, A., {Kulkarni}, S.~R., {Law}, N.~M., {et~al.} 2009, \pasp, 121, 1334

\bibitem[{{Reguitti} {et~al.}(2019){Reguitti}, {Pastorello}, {Pignata},
  {Benetti}, {Cappellaro}, {Turatto}, {Agliozzo}, {Bufano}, {Morrell},
  {Olivares E.}, {Reichart}, {Haislip}, {Kouprianov}, {Smartt}, \&
  {Ciroi}}]{reguitti2019}
{Reguitti}, A., {Pastorello}, A., {Pignata}, G., {et~al.} 2019, \mnras, 482,
  2750

\bibitem[{{Rest} {et~al.}(2012){Rest}, {Prieto}, {Walborn}, {Smith}, {Bianco},
  {Chornock}, {Welch}, {Howell}, {Huber}, {Foley}, {Fong}, {Sinnott}, {Bond},
  {Smith}, {Toledo}, {Minniti}, \& {Mandel}}]{rest2012}
{Rest}, A., {Prieto}, J.~L., {Walborn}, N.~R., {et~al.} 2012, \nat, 482, 375

\bibitem[{Rigault(2018)}]{rigault2018}
Rigault, M. 2018, ztfquery, a python tool to access ZTF data

\bibitem[{{Rigault} {et~al.}(2019){Rigault}, {Neill}, {Blagorodnova}, {Dugas},
  {Feeney}, {Walters}, {Brinnel}, {Copin}, {Fremling}, {Nordin}, \&
  {Sollerman}}]{rigault2019}
{Rigault}, M., {Neill}, J.~D., {Blagorodnova}, N., {et~al.} 2019, \aap, 627,
  A115

\bibitem[{{Robitaille} {et~al.}(2013){Robitaille}, {Tollerud}, {Greenfield},
  {Droettboom}, {Bray}, {Aldcroft}, {Davis}, {Ginsburg}, {Price-Whelan},
  {Kerzendorf}, {Conley}, {Crighton}, {Barbary}, {Muna}, {Ferguson},
  {Grollier}, {Parikh}, {Nair}, {Unther}, {Deil}, {Woillez}, {Conseil},
  {Kramer}, {Turner}, {Singer}, {Fox}, {Weaver}, {Zabalza}, {Edwards}, {Azalee
  Bostroem}, {Burke}, {Casey}, {Crawford}, {Dencheva}, {Ely}, {Jenness},
  {Labrie}, {Lim}, {Pierfederici}, {Pontzen}, {Ptak}, {Refsdal}, {Servillat},
  \& {Streicher}}]{astropy2013}
{Robitaille}, T.~P., {Tollerud}, E.~J., {Greenfield}, P., {et~al.} 2013, \aap,
  558, A33

\bibitem[{{Sanders} {et~al.}(2013){Sanders}, {Soderberg}, {Foley}, {Chornock},
  {Milisavljevic}, {Margutti}, {Drout}, {Moe}, {Berger}, {Brown}, {Lunnan},
  {Smartt}, {Fraser}, {Kotak}, {Magill}, {Smith}, {Wright}, {Huang}, {Urata},
  {Mulchaey}, {Rest}, {Sand}, {Chomiuk}, {Friedman}, {Kirshner}, {Marion},
  {Tonry}, {Burgett}, {Chambers}, {Hodapp}, {Kudritzki}, \&
  {Price}}]{sanders2013}
{Sanders}, N.~E., {Soderberg}, A.~M., {Foley}, R.~J., {et~al.} 2013, \apj, 769,
  39

\bibitem[{{Sapir} \& {Waxman}(2017)}]{sapir2017}
{Sapir}, N. \& {Waxman}, E. 2017, \apj, 838, 130

\bibitem[{{Schlafly} \& {Finkbeiner}(2011)}]{schlafly2011}
{Schlafly}, E.~F. \& {Finkbeiner}, D.~P. 2011, \apj, 737, 103

\bibitem[{{Schlegel} {et~al.}(1998){Schlegel}, {Finkbeiner}, \&
  {Davis}}]{schlegel1998}
{Schlegel}, D.~J., {Finkbeiner}, D.~P., \& {Davis}, M. 1998, \apj, 500, 525

\bibitem[{{Shaviv}(2000)}]{shaviv2000}
{Shaviv}, N.~J. 2000, \apjl, 532, L137

\bibitem[{{Shaviv}(2001{\natexlab{a}})}]{shaviv2001b}
{Shaviv}, N.~J. 2001{\natexlab{a}}, \apj, 549, 1093

\bibitem[{{Shaviv}(2001{\natexlab{b}})}]{shaviv2001}
{Shaviv}, N.~J. 2001{\natexlab{b}}, \mnras, 326, 126

\bibitem[{{Shiode} \& {Quataert}(2014)}]{shiode2014}
{Shiode}, J.~H. \& {Quataert}, E. 2014, \apj, 780, 96

\bibitem[{{Silverman} {et~al.}(2013){Silverman}, {Nugent}, {Gal-Yam},
  {Sullivan}, {Howell}, {Filippenko}, {Arcavi}, {Ben-Ami}, {Bloom}, {Cenko},
  {Cao}, {Chornock}, {Clubb}, {Coil}, {Foley}, {Graham}, {Griffith}, {Horesh},
  {Kasliwal}, {Kulkarni}, {Leonard}, {Li}, {Matheson}, {Miller}, {Modjaz},
  {Ofek}, {Pan}, {Perley}, {Poznanski}, {Quimby}, {Steele}, {Sternberg}, {Xu},
  \& {Yaron}}]{silverman2013}
{Silverman}, J.~M., {Nugent}, P.~E., {Gal-Yam}, A., {et~al.} 2013, \apjs, 207,
  3

\bibitem[{Smith(2017)}]{smith2017}
Smith, N. 2017, Handbook of Supernovae, 403–429

\bibitem[{{Smith}(2017)}]{smith2017b}
{Smith}, N. 2017, Philosophical Transactions of the Royal Society of London
  Series A, 375, 20160268

\bibitem[{{Smith} \& {Arnett}(2014)}]{smith2014}
{Smith}, N. \& {Arnett}, W.~D. 2014, \apj, 785, 82

\bibitem[{{Smith} {et~al.}(2020){Smith}, {E Andrews}, {Moe}, {Milne},
  {Bilinski}, {Kilpatrick}, {Fong}, {Badenes}, {Filippenko}, {Kasliwal}, \&
  {Silverman}}]{smith2020}
{Smith}, N., {E Andrews}, J., {Moe}, M., {et~al.} 2020, \mnras, 492, 5897

\bibitem[{{Smith} {et~al.}(2012){Smith}, {Mauerhan}, {Silverman},
  {Ganeshalingam}, {Filippenko}, {Cenko}, {Clubb}, \& {Kand
  rashoff}}]{smith2012}
{Smith}, N., {Mauerhan}, J.~C., {Silverman}, J.~M., {et~al.} 2012, \mnras, 426,
  1905

\bibitem[{{Smith} {et~al.}(2004){Smith}, {Vink}, \& {de Koter}}]{smith2004}
{Smith}, N., {Vink}, J.~S., \& {de Koter}, A. 2004, \apj, 615, 475

\bibitem[{{Soumagnac} \& {Ofek}(2018)}]{soumagnac2018}
{Soumagnac}, M.~T. \& {Ofek}, E.~O. 2018, \pasp, 130, 075002

\bibitem[{{Soumagnac} {et~al.}(2019){Soumagnac}, {Ofek}, {Gal-Yam}, {Waxman},
  {Ginzburg}, {Linn Strotjohann}, {Schulze}, {Barlow}, {Behar}, {Chelouche},
  {Fremling}, {Ganot}, {Gezari}, {Kasliwal}, {Kaspi}, {Kulkarni}, {Laher},
  {Maoz}, {Martin}, {Nakar}, {Neill}, {Nugent}, {Poznanski}, \&
  {Yaron}}]{soumagnac2019}
{Soumagnac}, M.~T., {Ofek}, E.~O., {Gal-Yam}, A., {et~al.} 2019, \apj, 872, 141

\bibitem[{{Soumagnac} {et~al.}(2020){Soumagnac}, {Ofek}, {Liang}, {Gal-Yam},
  {Nugent}, {Yang}, {Cenko}, {Sollerman}, {Perley}, {Andreoni}, {Barbarino},
  {Burdge}, {Bruch}, {De}, {Dugas}, {Fremling}, {Graham}, {Hankins},
  {Strotjohann}, {Moran}, {Neill}, {Schulze}, {Shupe}, {Sip{\H{o}}cz},
  {Taggart}, {Tartaglia}, {Walters}, {Yan}, {Yao}, {Yaron}, {Bellm},
  {Cannella}, {Dekany}, {Duev}, {Feeney}, {Frederick}, {Graham}, {Laher},
  {Masci}, {Kasliwal}, {Kowalski}, {Kupfer}, {Miller}, {Rigault}, \&
  {Rusholme}}]{soumagnac2020}
{Soumagnac}, M.~T., {Ofek}, E.~O., {Liang}, J., {et~al.} 2020, \apj, 899, 51

\bibitem[{{Stritzinger} {et~al.}(2012){Stritzinger}, {Taddia}, {Fransson},
  {Fox}, {Morrell}, {Phillips}, {Sollerman}, {Anderson}, {Boldt}, {Brown},
  {Campillay}, {Castellon}, {Contreras}, {Folatelli}, {Habergham}, {Hamuy},
  {Hjorth}, {James}, {Krzeminski}, {Mattila}, {Persson}, \&
  {Roth}}]{stritzinger2012}
{Stritzinger}, M., {Taddia}, F., {Fransson}, C., {et~al.} 2012, \apj, 756, 173

\bibitem[{{Strotjohann} {et~al.}(2015){Strotjohann}, {Ofek}, {Gal-Yam},
  {Sullivan}, {Kulkarni}, {Shaviv}, {Fremling}, {Kasliwal}, {Nugent}, {Cao},
  {Arcavi}, {Sollerman}, {Filippenko}, {Yaron}, {Laher}, \&
  {Surace}}]{strotjohann2015}
{Strotjohann}, N.~L., {Ofek}, E.~O., {Gal-Yam}, A., {et~al.} 2015, \apj, 811,
  117

\bibitem[{{Svirski} {et~al.}(2012){Svirski}, {Nakar}, \& {Sari}}]{svirski2012}
{Svirski}, G., {Nakar}, E., \& {Sari}, R. 2012, \apj, 759, 108

\bibitem[{{Szczygie{\l}} {et~al.}(2012){Szczygie{\l}}, {Kochanek}, \&
  {Dai}}]{szcygiel2012}
{Szczygie{\l}}, D.~M., {Kochanek}, C.~S., \& {Dai}, X. 2012, \apj, 760, 20

\bibitem[{{Tartaglia} {et~al.}(2016){Tartaglia}, {Pastorello}, {Sullivan},
  {Baltay}, {Rabinowitz}, {Nugent}, {Drake}, {Djorgovski}, {Gal-Yam},
  {Fabrika}, {Barsukova}, {Goranskij}, {Valeev}, {Fatkhullin}, {Schulze},
  {Mehner}, {Bauer}, {Taubenberger}, {Nordin}, {Valenti}, {Howell}, {Benetti},
  {Cappellaro}, {Fasano}, {Elias-Rosa}, {Barbieri}, {Bettoni}, {Harutyunyan},
  {Kangas}, {Kankare}, {Martin}, {Mattila}, {Morales-Garoffolo}, {Ochner},
  {Rebbapragada}, {Terreran}, {Tomasella}, {Turatto}, {Verroi}, \&
  {Wo{\'z}niak}}]{tartaglia2016b}
{Tartaglia}, L., {Pastorello}, A., {Sullivan}, M., {et~al.} 2016, \mnras, 459,
  1039

\bibitem[{{Th{\"o}ne} {et~al.}(2017){Th{\"o}ne}, {de Ugarte Postigo},
  {Leloudas}, {Gall}, {Cano}, {Maeda}, {Schulze}, {Campana}, {Wiersema},
  {Groh}, {de la Rosa}, {Bauer}, {Malesani}, {Maund}, {Morrell}, \&
  {Beletsky}}]{thoene2017}
{Th{\"o}ne}, C.~C., {de Ugarte Postigo}, A., {Leloudas}, G., {et~al.} 2017,
  \aap, 599, A129

\bibitem[{Wallis(2013)}]{wallis2013}
Wallis, S. 2013, Journal of Quantitative Linguistics, 20, 178

\bibitem[{Wilson(1927)}]{wilson1927}
Wilson, E.~B. 1927, Journal of the American Statistical Association, 22, 209

\bibitem[{{Woosley}(2017)}]{woosley2017}
{Woosley}, S.~E. 2017, \apj, 836, 244

\bibitem[{{Woosley} {et~al.}(2002){Woosley}, {Heger}, \&
  {Weaver}}]{woosley2002}
{Woosley}, S.~E., {Heger}, A., \& {Weaver}, T.~A. 2002, Reviews of Modern
  Physics, 74, 1015

\bibitem[{{Wu} \& {Fuller}(2020)}]{wu2020}
{Wu}, S. \& {Fuller}, J. 2020, arXiv e-prints, arXiv:2011.05453

\bibitem[{{Yao} {et~al.}(2019){Yao}, {Miller}, {Kulkarni}, {Bulla}, {Masci},
  {Goldstein}, {Goobar}, {Nugent}, {Dugas}, {Blagorodnova}, {Neill}, {Rigault},
  {Sollerman}, {Nordin}, {Bellm}, {Cenko}, {De}, {Dhawan}, {Feindt},
  {Fremling}, {Gatkine}, {Graham}, {Graham}, {Ho}, {Hung}, {Kasliwal},
  {Kupfer}, {Laher}, {Perley}, {Rusholme}, {Shupe}, {Soumagnac}, {Taggart},
  {Walters}, \& {Yan}}]{yao2019}
{Yao}, Y., {Miller}, A.~A., {Kulkarni}, S.~R., {et~al.} 2019, \apj, 886, 152

\bibitem[{Yaron {et~al.}(2017)Yaron, Perley, Gal-Yam, Groh, Horesh, Ofek,
  Kulkarni, Sollerman, Fransson, Rubin, \& et~al.}]{yaron2017}
Yaron, O., Perley, D.~A., Gal-Yam, A., {et~al.} 2017, Nature Physics, 13,
  510–517

\bibitem[{{Zackay} {et~al.}(2016){Zackay}, {Ofek}, \& {Gal-Yam}}]{zackay2016}
{Zackay}, B., {Ofek}, E.~O., \& {Gal-Yam}, A. 2016, \apj, 830, 27

\end{thebibliography}

\appendix

\section{Precursor images}
\label{sec:images}

\begin{figure*}[tb]
    \centering
\includegraphics[width=0.75\textwidth]{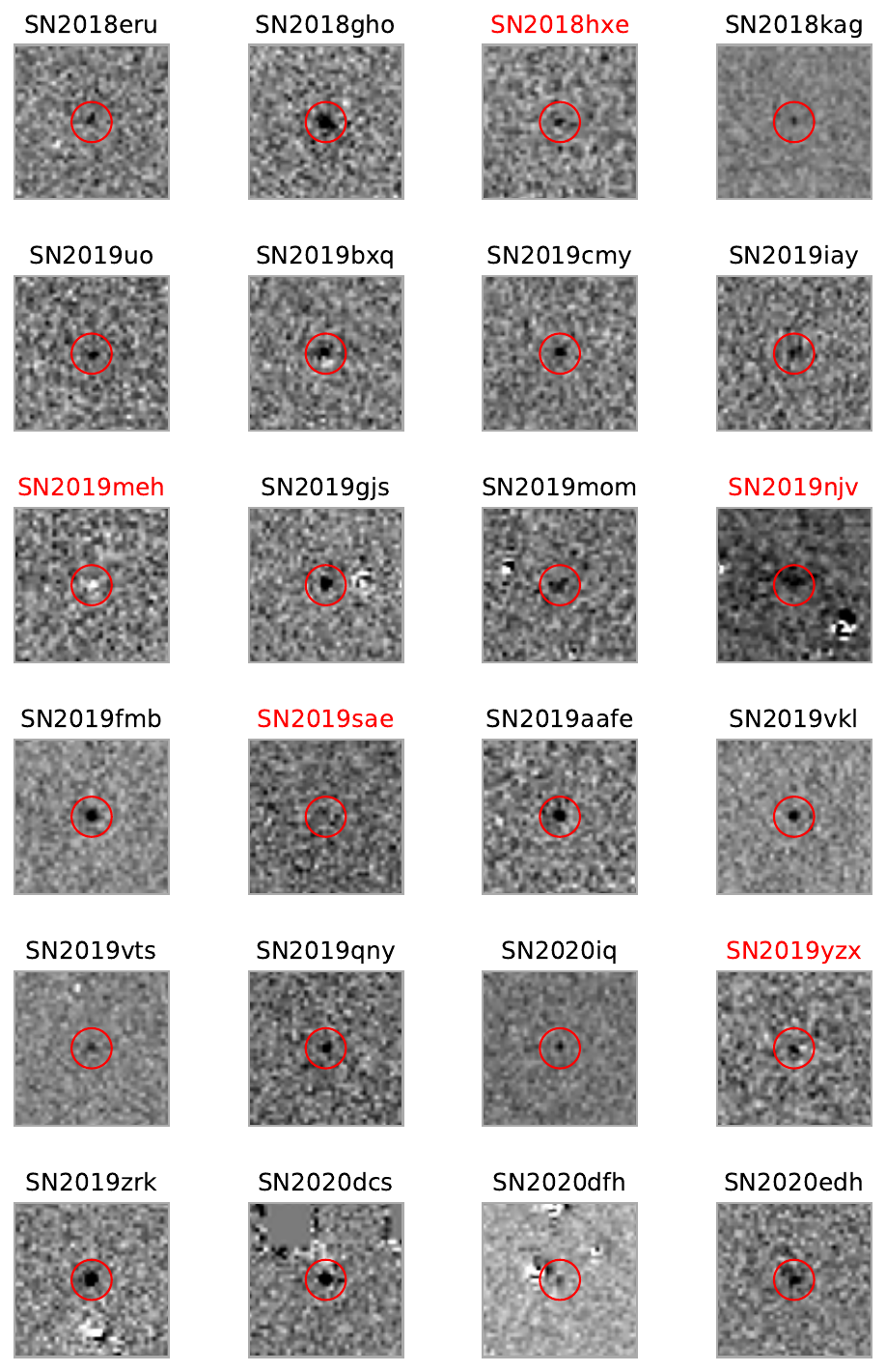}
\caption{\label{fig:skymaps}Coadded ZTF difference images showing the precursors. Red captions indicate that the corresponding precursor is only detected in a single bin and is therefore considered unconfirmed (see Sec.~\ref{sec:pre_detections}). No point source is visible for SN\,2019sae and we conclude that this precursor candidate is not real. A negative source is detected at the position of SN\,2019meh due to the AGN variability in the host galaxy. SN\,2019zrk and SN\,2020dcs are located close to the edge of the CCD.}
\end{figure*}

\begin{figure*}[tb]
    \centering
\includegraphics[width=0.75\textwidth]{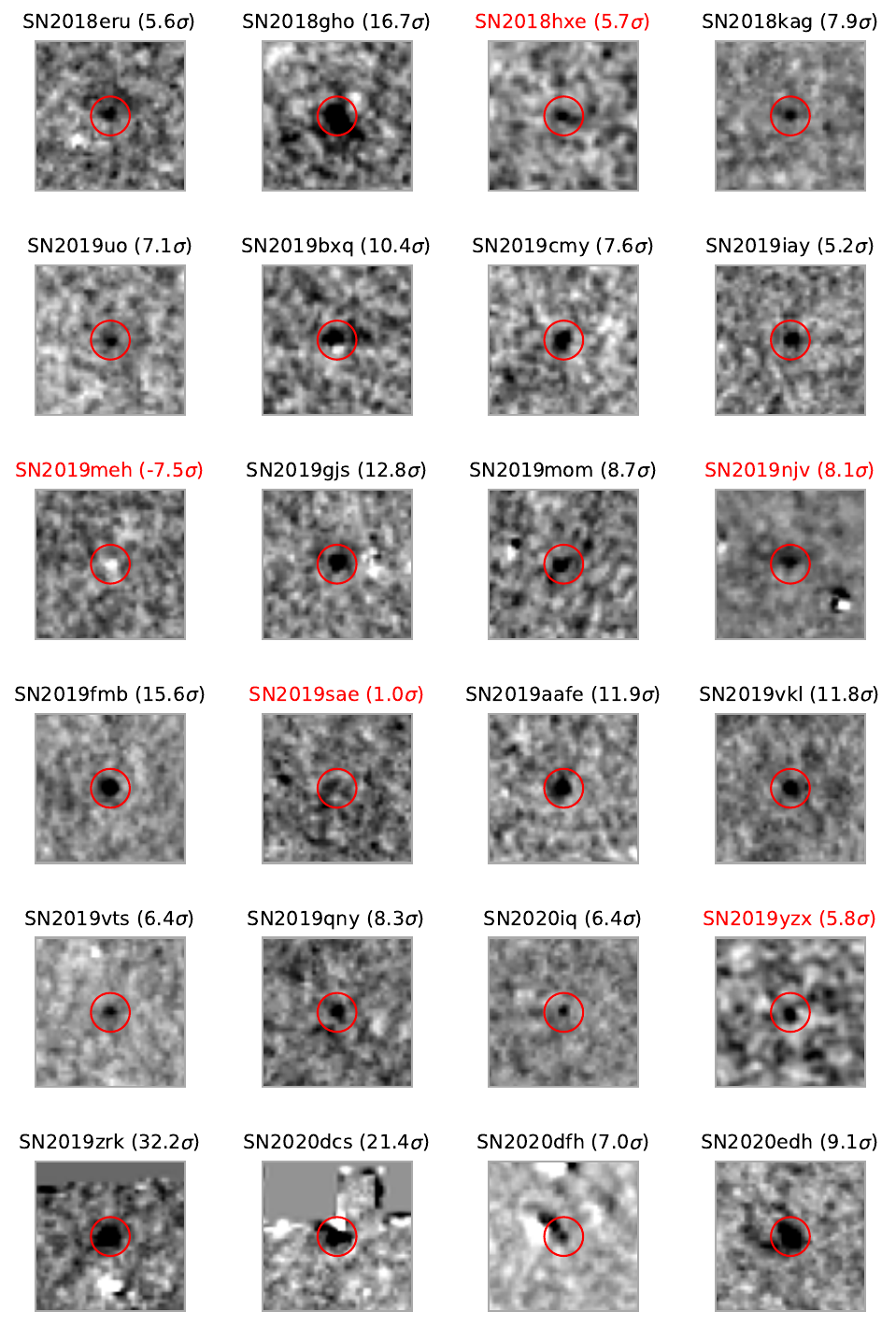}
\caption{\label{fig:skymaps_sig} Significance maps of the precursor candidates.}
\end{figure*}

To test whether the precursor candidates are real, we inspect the difference images. 
We select either the $g$ or $r$ band and coadd the difference images that yield the most significant fluxes in our search (see Figs.~\ref{fig:precursor_lcs} and \ref{fig:precursor_lcs2}). Before the coaddition we subtract the median pixel flux from the image and divide all fluxes by the robust standard deviation (half of the difference between the 15.9\% and 84.1\% percentile). The images are then coadded using the \emph{SWarp} software \citep{bertin2002}. The resulting skymaps are shown in Fig.~\ref{fig:skymaps}. Clear point sources are visible for most SNe.

To quantify the significance of the detections in Fig.~\ref{fig:skymaps}, we calculate the S-image as described by \citet{zackay2016}. For this purpose, we cross correlate the difference image with the PSF of the difference image and coadd the resulting significance maps again using \emph{SWarp}. To normalize the significances we divide each image by the robust standard deviation and show the significance maps in Fig.~\ref{fig:skymaps_sig}.

Most precursors are visible as point sources, with exception of SN\,2019sae. This precursor candidate is considered unconfirmed as only a single bin drives the significance (compare Sec.~\ref{sec:pre_detections}) and we conclude that it is likely spurious. The other three unconfirmed precursors (marked by red titles in Figs.~\ref{fig:skymaps} and \ref{fig:skymaps_sig}) appear as point sources, so we cannot rule out that they are astrophysical. As described in Sec.~\ref{sec:pre_detections} we neglect these marginal detections here and only focus on the 19 confirmed precursors.
In some cases astrometric residuals are visible in the cutouts close to the precursors. Especially affected are SN\,2018gho and SN\,2020dfh. We nevertheless consider them true detections as they also appear to be present in single images without astrometric residuals. We caution however that their light curves could be affected by the missubtractions. They are therefore less reliable than the light curves of other precursors.

\section{Baseline correction of SN\,2019cmy}
\label{sec:sn2019cmy}

\begin{figure*}[tb]
    \centering
\includegraphics[width=\textwidth]{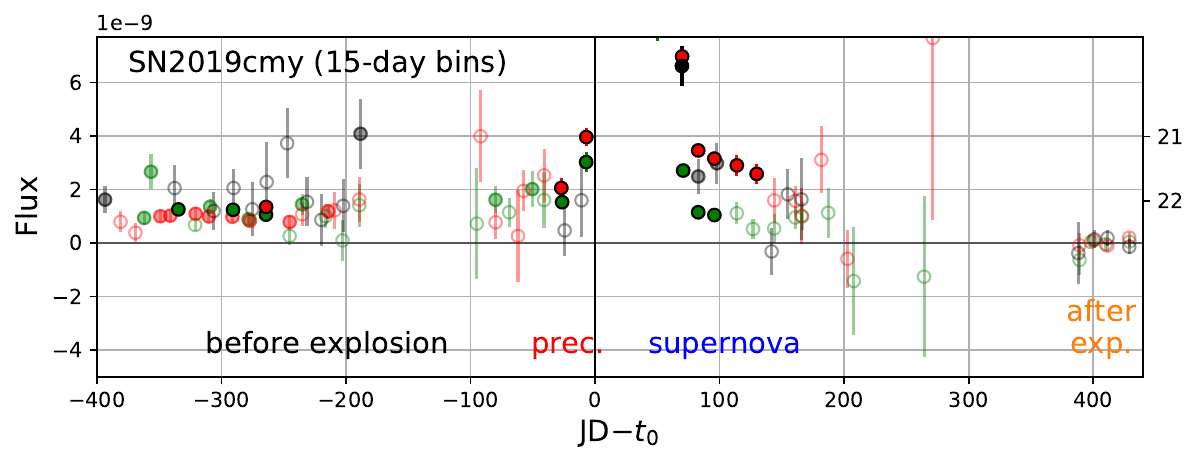}
\caption{\label{fig:baseline_sn2019cmy} Complete light curve for the position of SN\,2019cmy when using the last observations (yellow region) for the baseline correction. The peak of the SN is cut for better visibility. The complete pre-explosion light curve (gray region) is positive, i.e. the fluxes before and after the SN cannot be zero at the same time. We could not identify a systematic effect that might explain this discrepancy. If the effect is astrophysical, it would imply that the progenitor star is detected with an absolute magnitude of $-14.1$. In this paper, we use pre-explosion observations for the baseline correction and only discuss the precursor detected one week before the SN (red region).}
\end{figure*}

When using late-time observations ($>380$ days after the explosion) for the baseline correction of SN\,2019cmy, we find that the complete pre-explosion light curve becomes positive as shown in Fig.~\ref{fig:baseline_sn2019cmy}. The late-time observations include $>60$ observations per filter collected in $>30$ different nights over a period of 50 days. They are consistent with a constant flux, which is however $\sim10^{-9}$ lower than the median flux before the explosion. To probe whether a systematic error could explain this deviation we perform a series of tests: The hour angle and air mass are on average similar before and after the explosion. The seeing is better during the last block of observations collected since March 2020, potentially because of reduced air pollution as a consequence of the lockdown measures due to the Corona pandemic. However, other SNe in our sample do not show a similarly strong flux change, implying that we do not see a global effect caused by changes in the atmosphere.

The pixel coordinates of the SN position drift slowly by in total 100 pixels over three years, but there is no abrupt change consistent with the two different flux levels. We also inspect the mirrored SN position, located at a distance of $11.2\,\text{arcsec}$ on the other side of the host galaxy's center. We observe a constant flux level for the complete light curve, i.e. a similar flux reduction is not observed at late times. Finally, we use the ZUDS pipeline to construct a different reference image, redo the image subtraction and produce a forced photometry light curve. The flux difference between the observations before and after the explosion is similarly large. It is therefore likely not induced by the image subtraction or forced photometry pipeline.

After we could not identify any systematic effect responsible for the lower fluxes after the SN has faded, we here discuss an astrophysical interpretation. If real, the flux drop would imply that the progenitor star is detected for 400 days before the SN explosion. A flux of $10^{-9}$ corresponds to a magnitude of $22.5$, i.e. an absolute magnitude of $-14.1$ for the SN redshift of $0.0314$. The detected progenitor star prior to SN\,2005gl had a $V$ band magnitude of only $-10.4$ \citep{gal-yam2009}, but we cannot rule out that the progenitor underwent a more than year-long outburst with an approximately constant luminosity. Similar outbursts were observed several years prior to the explosion of SN\,2015bh and reached typical $r$-band magnitudes between $-9$ and $-13$ \citep{elias-rosa2016, thoene2017}. It is thus not excluded that we indeed detect the progenitor star.

Since we are not sure whether or not the flux reduction after the SN explosion is astrophysical, we here exclude the late-time observations and do the baseline correction using pre-explosion data. $r$-band observations in two nights, 7 and 6 days before the estimated explosion date surpass the $5\sigma$ detection as shown in Fig.~\ref{fig:precursor_lcs}. They could either be a precursor eruption or a wind-shock breakout peak \citep{ofek2010, chevalier2011, piro2013}.

\end{document}